\documentclass[%draft,
				a4paper,fleqn,usenatbib,useAMS]{mnras}

\usepackage{newtxtext,newtxmath}
\usepackage[T1]{fontenc}
\usepackage{ae,aecompl}

\usepackage{graphicx}	% Including figure files
\usepackage{amsmath}	% Advanced maths commands
\usepackage{amssymb}	% Extra maths symbols

\usepackage[version=3]{mhchem}
\usepackage{datetime}
\usepackage{multirow}
\usepackage{booktabs}
\usepackage{float}
\usepackage{gensymb}
\usepackage{wasysym}
\usepackage{amsfonts}
\usepackage{courier}
\usepackage{caption}
\usepackage{subcaption}
\usepackage{multicol}   % Multi-column entries in tables
\usepackage{pdflscape}	% Landscape pages
\usepackage[para]{threeparttable}

\newdateformat{mydate}{\twodigit{\THEDAY}{ }\monthname[\THEMONTH], \THEYEAR}
\mydate
\newdate{reportdate}{06}{6}{2017}
\newdate{datadate}{02}{2}{2017}

\newcommand{\rom}[1]{%
  \textup{\uppercase\expandafter{\romannumeral#1}}%
}

\title[Stable HZ of single Jovian planet systems]{Stable habitable zones of single Jovian planet systems}

\author[M. T. Agnew et al.]{
Matthew T. Agnew,$^{1}$ %\thanks{E-mail: mn@ras.org.uk (KTS)}
Sarah T. Maddison,$^{1}$
Elodie Thilliez$^{1}$ and
Jonathan Horner$^{2}$
\\
$^{1}$Centre for Astrophysics and Supercomputing, Swinburne University of Technology, Hawthorn, Victoria 3122, Australia\\
$^{2}$University of Southern Queensland, Toowoomba, Queensland 4350, Australia\\
}

\date{Accepted 2017 June 8. Received 2017 June 6; in original form 2016 September 29}

\pubyear{2017}

\begin{document}
\label{firstpage}
\pagerange{\pageref{firstpage}--\pageref{lastpage}}
\maketitle

\begin{abstract}
With continued improvement in telescope sensitivity and observational techniques, the search for rocky planets in stellar habitable zones is entering an exciting era. With so many exoplanetary systems available for follow-up observations to find potentially habitable planets, one needs to prioritise the ever-growing list of candidates. We aim to determine which of the known planetary systems are dynamically capable of hosting rocky planets in their habitable zones, with the goal of helping to focus future planet search programs.\\
We perform an extensive suite of numerical simulations to identify regions in the habitable zones of single Jovian planet systems where Earth mass planets could maintain stable orbits, specifically focusing on the systems in the Catalog of Earth-like Exoplanet Survey Targets (CELESTA). \\
We find that small, Earth-mass planets can maintain stable orbits in cases where the habitable zone is largely, or partially, unperturbed by a nearby Jovian, and that mutual gravitational interactions and resonant mechanisms are capable of producing stable orbits even in habitable zones that are significantly or completely disrupted by a Jovian. \\
Our results yield a list of 13 single Jovian planet systems in CELESTA that are not only capable of supporting an Earth-mass planet on stable orbits in their habitable zone, but for which we are also able to constrain the orbits of the Earth-mass planet such that the induced radial velocity signals would be detectable with next generation instruments. 
\end{abstract}


%%%%%%%%%%%%%%%%%%%%%%%%%%%%%%%%%%%%%%%%%%%%%%%%%%

%%%%%%%%%%%%%%%%% BODY OF PAPER %%%%%%%%%%%%%%%%%%

\section{Introduction}
\label{sec:introduction}
One of the most exciting goals in astrophysics is the discovery of a true, twin Earth: a rocky planet of similar size, structure and composition to Earth on a stable orbit within its host star's habitable zone\footnote{The HZ is a region around a star in which liquid water can be maintained on the surface of a rocky planet that hosts an atmosphere.} (HZ) \citep{Kasting1993,Kopparapu2013}. As a result of biases inherent to observational techniques, the first exoplanets detected were often both massive and close to their host stars \citep[e.g.][]{Mayor1995,Charbonneau2000}. In the decades since, improved technology has allowed for the detection of lower mass planets \citep[e.g.][]{Wright2015,Vogt2015} and planets with greater orbital periods \citep[e.g.][]{Borucki2012,Jenkins2015}. We are only now beginning to discover planets with orbital periods of a decade or more, including Jupiter analogs \citep{Wittenmyer2016}. We now know of over 3400\footnote{As of 2 February 2017.} confirmed exoplanets (NASA Exoplanet Archive, exoplanetarchive.ipac.caltech.edu) with a variety of radii, masses and orbital parameters. In the coming years, we will begin to search for potentially habitable exo-Earths and so in this work we aim to determine how to best focus our future efforts.

Several methods have been used in the past to predict stable regions and the presence of additional exoplanets in confirmed exoplanetary systems. Some methods predict the presence of a planet by simulating observable properties of debris discs \citep[e.g.][]{Thilliez2016}. 
Others utilise dynamical simulations to demonstrate that massless test particles can remain on stable orbits in multiple planet systems, thus identifying potential regions of stability \citep[e.g.][]{Rivera2007,Thilliez2014,Kane2015}. Such stable regions can then be the focus of follow up simulations involving Earth-mass planets \citep{Kane2015}.

Assessing the stability of a system by considering a region of chaos surrounding any known exoplanet has also been used to predict regions of stability in exoplanetary systems \citep{Jones2001,Jones2002,Jones2005,Jones2010,Giuppone2013}. The unstable, chaotic region around a planet is often calculated to be some multiple of its Hill radius \citep{Jones2001,Jones2002}, where the multiplying factor is sometimes derived numerically \citep{Jones2005,Jones2010}.  Alternatively, \cite{Giuppone2013} present a semi-empirical stability criterion to quickly infer the stability of existing systems. They test the validity of the criterion by simulating both single and multiple planet systems, and demonstrate that their criterion is an effective tool for identifying which exoplanetary systems can host additional planets.

In this work, we aim to identify the properties of planetary architectures in single Jovian planet systems that could harbour an Earth-mass planet in the HZ, with a specific focus on those presented in the Catalog of Earth-like Exoplanet Survery Targets (CELESTA) \citep{Chandler2015}.
We first divide the selected systems into three broad classes that indicate their likelihood of hosting stable Earths in their HZ in order to theoretically eliminate systems that almost certainly host stable HZs from our numerical study. Since these HZs are all stable, numerical simulations would not help constrain the locations within the HZ where stable Earths might reside. For the remaining systems, we use the \textsc{swift} N-body software package \citep{Levison1994} to help identify regions where Earth-mass planets could maintain stable orbits by first performing dynamical simulations using massless test particles spread throughout the HZ of each system. We follow these with a suite of dynamical simulations using a $1\ \textrm{M}_\oplus$ planet to ultimately predict which systems could host a stable Earth in their HZ, help constrain where, and determine what the strength of the induced radial velocity signal would be.

In section \ref{sec:jovian_systems} we introduce the motivation for analysing single Jovian planet systems. In section \ref{sec:method}, we describe the method used to select the single Jovian planet systems which we simulate, detail the numerical simulations used to dynamically analyse the systems, and discuss how we interpret the simulation results. We then present and discuss our results in section \ref{sec:results}, and summarise our findings in section \ref{sec:summary}.

% EXISTING JOVIAN SYSTEMS ------------------------------------------------------
\section{Exoplanet Population}
\label{sec:jovian_systems}
Using the Exoplanet Orbit Database \citep[][exoplanets.org]{Han2014a}, we analyse the currently known exoplanet population.\footnote{It should be noted that there are inherent biases in the various observational techniques that may impact the following analysis, but for this work we accept the planetary properties and orbital parameters as they are in the relevant databases.} 
Our analysis %of the exoplanet population 
reveals an interesting feature: the proportion of Jovian planets in single and multiple planetary systems is skewed in favour of single planet systems (see Table~\ref{tab:exoplanet_distribution}). Single Jovian planet systems are an interesting subset of the exoplanet population that could potentially have small rocky planets hidden in their HZs. Jupiter is thought to have played a complicated role in the formation and evolution of the Solar system \citep[e.g.][]{Gomes2005,Horner2009,Walsh2011,Izidoro2013,Raymond2014,Brasser2016,Deienno2016}, although the timing, nature and degree to which it has contributed is a dynamic area of research \citep[e.g.][]{Minton2009,Minton2011,Agnor2012,Izidoro2014,Izidoro2015,Izidoro2016,Levison2015,Kaib2016}. Further to this, it has also been suggested that Jupiter may have had a significant impact on the environment in which life on Earth has developed \citep[e.g.][]{Carter-Bond2010,Carter-Bonda2012,Carter-Bondb2012,Martin2013a,Quintana2014,O'Brien2014}. For this reason, it has been proposed that the presence of a Jupiter analog in an exoplanetary system may be an important indicator for potential habitability \citep{Wetherill1994,Ward2000}, although this hypothesis remains heavily debated \citep{Horner2008,Horner1908,Horner2010,Horner2012,Horner2013,Grazier2016}.

\begin{table}
    \caption{The distribution of exoplanets between Terrestrial planets, Super-Earths, Neptunians and Jovians amongst single and multiple planet systems. The class of each planet is defined by Table~\ref{tab:planet_classification}.}
    \label{tab:exoplanet_distribution}
    \centering
    \begin{tabular}{l c c c c}
        \toprule            
        							& Single		& Multiple	&& Total \\
        \midrule
        Terrestrials				& 320		& 304		&& 624 \\
        Super-Earths				& 458		& 431		&& 889 \\
        Neptunians				& 349		& 308		&& 657 \\
        Jovians					& 601		& 152		&& 753 \\
        	\midrule  
        Total					& 1728		& 1195		&& 2923 \\
        \bottomrule
    \end{tabular}
\end{table}

\begin{table}
    \caption{The radius and mass limits used in this work to classify exoplanets.}
    \label{tab:planet_classification}
    \centering
    \begin{tabular}{l c c c c}
        \toprule                     
        				& $r_{\mathrm{min}}$ 		& $r_{\mathrm{max}}$  		& $m_{\mathrm{min}}$  		& $m_{\mathrm{max}}$ 	\\                     
        				& ($\textrm{r}_\oplus$) 	& ($\textrm{r}_\oplus$) 	& ($\textrm{M}_\oplus$) 	& ($\textrm{M}_\oplus$)	\\				
        \midrule
        Terrestrials	& 0					& $<1.5$						& 0							& $<1.5$					\\
        Super-Earths	& 1.5						& $<2.5$						& 1.5						& $<10$					\\
        Neptunians	& 2.5						& $<6$						& 10							& $<50$					\\
        Jovians		& 6							& $>6$						& 50							& $>50$			       	\\
        \bottomrule
    \end{tabular}
\end{table}

Our analysis using the Exoplanet Orbit Database yields a total of 2923\footnote{Confirmed exoplanets for which good orbital elements and mass and/or radius data is available as of 2 February 2017.} exoplanets, residing in 2208 systems: 1728 single and 480 multiple planet systems. These exoplanets are classified as Terrestrials, Super-Earths, Neptunians or Jovians according to their radius (or according to their mass in lieu of available radius data) as per the ranges defined in Table~\ref{tab:planet_classification}. Analysing all the exoplanet systems, we find the exoplanet classes are distributed amongst the systems as shown in Table~\ref{tab:exoplanet_distribution}. It can be seen that all classes of planet are reasonably well represented within the greater exoplanet population, but also within the single and multiple planet sub-populations.

\begin{figure*}
	\begin{subfigure}[b]{\linewidth}
    		\centering
    		\includegraphics[width=\linewidth]{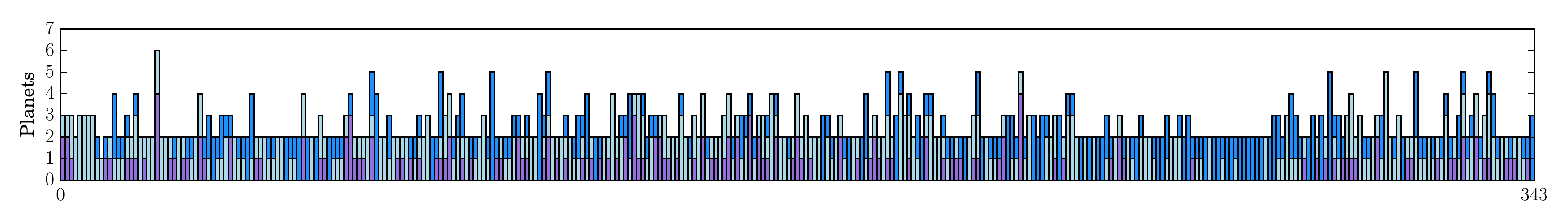}
        \caption{Terrestrial planet or Super-Earth systems without Jovians}\label{fig:multiple_exoplanet_dist_a}
	\end{subfigure}
	
    \begin{subfigure}[b]{\linewidth}
    		\centering
    		\includegraphics[width=\linewidth]{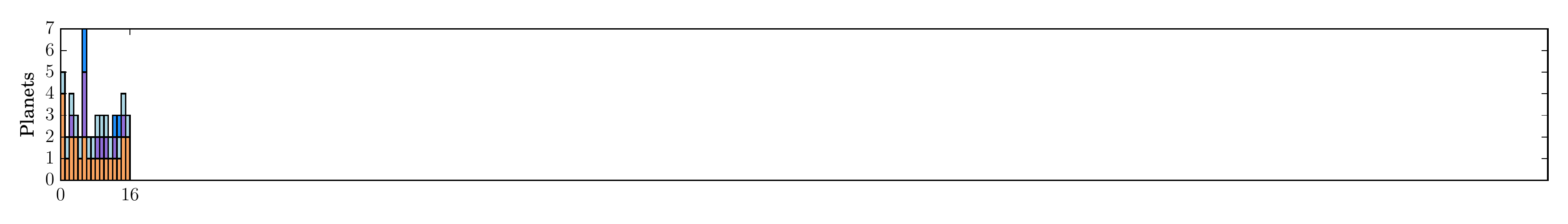}
        \caption{Terrestrial planet or Super-Earth systems with Jovians}\label{fig:multiple_exoplanet_dist_b}
	\end{subfigure}
	
	\begin{subfigure}[b]{\linewidth}
    		\centering
    		\includegraphics[width=\linewidth]{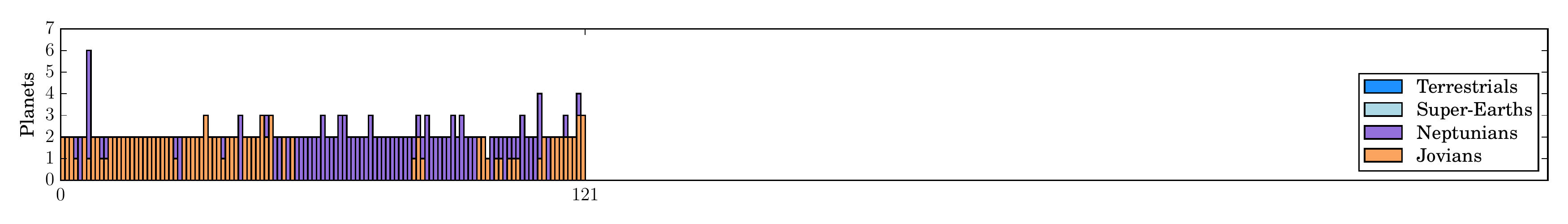}
        \caption{Non-Terrestrial planet or Super-Earth Systems}\label{fig:multiple_exoplanet_dist_c}
	\end{subfigure}
\caption{Planetary architectures of confirmed multiple planet systems. The exoplanets have been classified as per the criteria presented in Table~\ref{tab:planet_classification}. (a) The 343 multiple planet systems with Terrestrial planets or Super-Earths that do not also possess a Jovian. (b) The 16 multiple planet systems with Terrestrial planets or Super-Earths that do also possess a Jovian. (c) The 121 multiple planet systems with no Terrestrial planets or Super-Earths.}\label{fig:multiple_exoplanet_dist}
\end{figure*}

Of particular interest is an investigation into the planetary architectures of the multiple planet systems. We classify the 480 multiple systems into three broad categories based on the planet classes present in each: Non-Jovian systems, Jovian systems that coexist with smaller Terrestrial or Super-Earth planets, and Jovians and Neptunians with other giant planets. Figure~\ref{fig:multiple_exoplanet_dist_a} demonstrates that when a multiple system is found harbouring a Terrestrial  or Super-Earth planet, in the majority of cases ($343/480$, $\sim$71\%) it coexists with other Terrestrial planets, Super-Earths or Neptunians. Figure~\ref{fig:multiple_exoplanet_dist_b} shows that systems with Terrestrials or Super-Earths coexisting with a Jovian account for the small fraction of the multiple planet systems ($16/480$, $\sim$3\%), while Figure~\ref{fig:multiple_exoplanet_dist_c} shows that non-Terrestrial or Super-Earth systems account for about a quarter ($121/480$, $\sim$25\%) of the multiple planet systems. While the overall distribution of planets in multiple planet systems shows a reasonable distribution across each class (Table~\ref{tab:exoplanet_distribution}), the planet classes are not uniformly distributed in each multiple planet architecture: Terrestrial planets and Super-Earths are generally found with other Terrestrials, Super-Earths or Neptunians, whereas Jovians are generally found with other massive planets, i.e. Neptunians and/or Jovians.

Examining the entire Jovian population as they occur in both single and multiple systems, yields a total of 753 planets. We summarise our findings concerning Jovians as follows: 601 ($79\%$) Jovians are found in single planet systems, 128 ($17\%$) Jovians are found in multiple planet systems coexisting with Neptunians or other Jovians, and only 24 ($3\%$) Jovians are found in multiple planet systems coexisting with Terrestrial planets or Super-Earths. This demonstrates that for the current population of confirmed exoplanets, the majority of Jovians are either found to be in single planet systems, or to coexist with other giant Jovians or Neptunians, contrasting with our own Solar System. However, we note that this is most likely attributable to observational bias inherent in the two highest yield detection methods: the transit method and radial velocity method. The current state of the art allows for the detection of Doppler shifts to just below $1 \textrm{ m s}^{-1}$ \citep{Dumusque2012b}, making the detection of Earth-mass planets challenging \citep{Wittenmyer2011a}. As such, Jovians will completely dominate both Doppler shift signals and transit signals. The detection of Terrestrials in the HZ of Sun-like stars is made even more challenging because such planets would orbit within a few au of their host stars. The next generation of spectrographs aim to detect such planets by achieving radial velocity resolutions of around $0.1 \textrm{ m s}^{-1}$ \citep[e.g. ESPRESSO,][]{Pepe2014} and $0.01 \textrm{ m s}^{-1}$ \citep[e.g. CODEX,][]{Pasquini2010}. As the radial velocity resolution decreases, the resultant noise from stellar activity in Sun-like stars becomes significant \citep{Dumusque2010,Anglada-Escude2016}. We do not consider stellar noise in our assessment herein. The small proportion of Jovians coexisting with rocky planets and the observational biases inherent to the current state of the art provides motivation to investigate single Jovian planet systems as a subset of the existing exoplanet population which may contain smaller, Terrestrial planets in the HZ that are currently undetectable.

\cite{Giuppone2013} briefly discuss the idea of multiple planets in tightly packed configurations called compact systems. In such a system, all possible stable regions are occupied, and the system can be considered full; no additional bodies can exist on stable orbits. An excellent example of such a compact multiple planet system is the recently announced 7 planet system detected orbiting TRAPPIST-1 \citep{Gillon2017}. While single Jovian planet systems are clearly not compact, their HZs may be full, depending on the orbital parameters of the existing Jovian. It is important to determine which systems have full HZs in order to eliminate those systems as possible targets for future observations in the search for potentially habitable Earth-like planets.

In this work we aim to investigate the subset of these single Jovian planet systems that are in the Catalog of Earth-like Exoplanet Survery Targets (CELESTA) \citep{Chandler2015}. The CELESTA database calculates the HZs of nearby Sun-like stars, calculating the stellar properties needed to determine the HZs from \citet{Kopparapu2014}, and presents several possible HZ boundaries to choose from.  As a large proportion of the exoplanet population is observed around  non-Sun-like stars (e.g. M-dwarfs), the database does not contain many stars with planetary bodies. Of the 37,354 stars in CELESTA for which HZs are calculated, just 120 host confirmed exoplanets. Of these 120, just 93 are single Jovian planet systems. We cross-reference these systems from CELESTA with the Exoplanet Orbit Database \citep{Han2014a} to yield the planetary properties and orbital parameters. In this work we aim to identify which of these systems could host a $1\ \textrm{M}_\oplus$ planet in a stable orbit within the HZ. For those systems, we then determine those for which such a planet could be detected using future instruments, in order to provide a focus for future observational efforts.

\section{Method}
\label{sec:method}
We first calculate a theoretical region of chaos surrounding the existing Jovian in the  the selection of 93 CELESTA systems. To save simulation time, we remove systems that have completely stable HZs. %due to the Jovian being located sufficiently far from it. 
While these systems could host stable Terrestrial planets in their HZ, we cannot offer any further constraints on the orbits of such habitable planets. For the remaining systems, we first carry out dynamical simulations using massless test particles (TPs) spread throughout the HZ of each system to help identify regions of dynamical stability in the HZ. %For the remaining systems, we identify regions in the HZ where small planets could maintain stable orbits by carrying out dynamical simulations using massless test particles (TPs) spread throughout the HZ of each system. 
For those systems predicted to have less stable HZs, we expect significantly more interactions between TPs and the Jovian and potentially some resonant trapping. We increase the number of TPs for these systems in order to yield more robust results. Finally, we conduct a suite of simulations involving the Jovian and a $1\ \textrm{M}_\oplus$ planet to check if mutual gravitational interactions (that are absent with massless TP simulations) affect any stable regions found in the TP simulations in order to demonstrate where Terrestrial planets could be stable in those systems.

\subsection{System Selection}
\label{subsec:predicting_stable_regions}
Here we present the method used to broadly predict the overall stability of the HZ of exoplanetary systems. In cases where the Jovian is located sufficiently far from the HZ, we expect the gravitational influence of the Jovian to be negligible on the HZ and leave it completely unperturbed. Test particles within such a HZ would be capable of maintaining stable orbits and so are computationally expensive to run and provide little value, and so we want to eliminate such systems before proceeding with our numerical study. 

We consider the criterion for the onset of chaos based on the overlap of first order mean motion resonances \citep{Wisdom1980, Duncan1989}. For a planet orbiting its parent star, a region extending a distance $\delta$ around the planet will experience chaos, given by
\begin{align} 
\label{eqn:chaotic_region}
	\delta = C\mu^{2/7}a_{\textrm{planet}},
\end{align} 
where $C$ was calculated to be a constant equal to $1.57$ \citep{Duncan1989,Giuppone2013}, $\mu = M_\mathrm{planet}/M_{\star}$ is the mass ratio between the planet and its parent star, and $a_{\textrm{planet}}$ is the semi-major axis of the Jovian planet. Using this overlap criterion for the onset of chaos, \cite{Giuppone2013} present the \textit{crossing orbits criterion}, which suggests that if two planetary orbits intercept at some point, and in the absence of some kind of resonant mechanism, close encounters will occur and the system will become unstable. For a Jovian planet with an eccentric orbit, the chaotic region will extend to a distance $\delta$ exterior to the apocentre and interior to the pericentre of its orbit. Thus, the region of chaos is defined as
\begin{align} 
\label{eqn:crossing_orbit}
	a_{\textrm{planet}}(1-e) - \delta \leq \textrm{Chaotic Region} \leq a_{\textrm{planet}}(1+e) + \delta,
\end{align} 
where $e$ is the Jovian's eccentricity, and $\delta$ is defined as in equation \ref{eqn:chaotic_region}.

We use equation \ref{eqn:crossing_orbit} to calculate the region of chaos for each of the 93 single Jovian systems from the CELESTA database. We then compare the maximum and minimum semi-major axes of the chaotic region with the maximum and minimum semi-major axes of the HZ for each system, and compute the overlap between these two regions. From this, we define 3 classes of systems:

\begin{description}
  \item \textbf{Green:} if the chaotic region does not overlap the HZ,
  \item \textbf{Amber:} if the chaotic region partially overlaps the HZ,
  \item \textbf{Red:} if the chaotic region completely overlaps the HZ.
\end{description}
We predict that the green non-overlapping systems should possess entirely stable HZs, the amber partially overlapping systems should possess partially stable HZs, and the red completely overlapping systems should possess unstable HZs, except where the mutual gravitational interactions between the two bodies could stabilise specific orbits (as per the definition by \citealp{Giuppone2013}). We find that for the 93 single Jovian planet systems, 41 can be classified as green, 26 as amber, and 26 as red. 
We focus our attention on the red completely overlapping systems and amber partially overlapping systems, where the influence of the Jovian is predicted to strongly or relatively strongly influence the HZ. As the green non-overlapping systems are predicted to have stable HZs and are expected to retain the majority, if not all, of their TPs, %these systems would be computationally expensive to simulate and 
simulations would not help constrain the orbits of potentially habitable Terrestrial planets in those systems. Thus we focus on only those green systems where the Jovian is close to the HZ; that is, where the period of the Jovian, $T_{\textrm{Jovian}}$, is within one order of magnitude of the period in the HZ centre, $ T_{\textrm{HZ}}$ ($0.1\ T_{\textrm{HZ}} \leq T_{\textrm{Jovian}} \leq 10\ T_{\textrm{HZ}}$). There are 13/41 green systems that satisfy this criterion.

\subsection{Dynamical Simulations}
\label{subsec:dynamical_analysis}
We run dynamical simulations using the \textsc{swift} N-body software package \citep{Levison1994}. \textsc{swift} can integrate massive bodies that interact gravitationally, and massless test particles (TPs) that feel the gravitational forces of the massive bodies but exert no gravitational force of their own. We use the Regularised Mixed Variable Symplectic (RMVS) Method (specifically, the \texttt{rmvs3} integrator) provided in \textsc{swift} due to its advantage of being computationally faster than conventional methods \citep{Levison2000}.

We used the Runaway Greenhouse and Maximum Greenhouse scenarios presented by \cite{Kopparapu2014} for the inner and outer edges of the HZ respectively\footnote{assuming an Earth-mass planet and an Earth-like atmosphere}. The inner edge corresponds with the maximum distance from the star at which a runaway greenhouse effect would take place, causing all the surface water on the planet to evaporate. The outer edge corresponds to the maximum distance at which a cloud-free \ce{CO_2} atmosphere (with a background of \ce{N_2}) could maintain liquid water on the Terrestrial planet's surface. The Runaway Greenhouse and the Maximum Greenhouse boundaries make up the conservative HZ. The HZ boundaries have been shown to be strongly dependant on the uncertainties in stellar parameters \citep{Kane2014}. In this work, however, we take the stellar parameters given in CELESTA and the Exoplanet Orbit Database on face value. The TPs were then randomly distributed throughout the HZ, within the range of orbital parameters shown in Table~\ref{tab:tp_params}. All simulations used stellar parameters and HZ values from CELESTA \citep{Chandler2015}, and planetary properties and orbital parameters from the Exoplanet Orbit Database \citep{Han2014a}. 

The simulations were run for an integration time $T_{\textrm{sim}} = 10^7$ years, or until all the TPs were removed. The removal of a TP is defined by the ejection of the TP beyond an astrocentric distance of $250$ au. The time step for the simulations was set to $dt = 1/40$ of the smallest orbital period in the system (Jovian planet or TPs). %

\begin{table}
	\caption{The range of orbital parameters within which the test particles were randomly distributed over the HZ, and the range of orbital parameters, and number of values over each range (in equally spaced intervals) over which the $1\ \textrm{M}_\oplus$ body simulations were run.}
   	\label{tab:tp_params}
   	\centering
	\begin{tabular}{c c c c c c c c}
    		\cmidrule(r){3-4}  \cmidrule(r){5-8}  
    		& & \multicolumn{2}{c}{TPs}  & & \multicolumn{3}{c}{$1\ \textrm{M}_\oplus$ }\\
        \cmidrule(r){3-4}  \cmidrule(r){5-8}                     
        						&	& Min					& Max & 			& Min					& Max  					& $\#$ of Values	\tnote{*}\\
        \cmidrule(r){1-2} \cmidrule(r){3-4}  \cmidrule(r){5-8}  
        $a$ (AU)				&	& HZ$_{\mathrm{min}}$	& HZ$_{\mathrm{max}}$	&	& HZ$_{\mathrm{min}}$	& HZ$_{\mathrm{max}}$	& 51\\
        $e$					&	& 0.0					& 0.3 &			& 0.0					& 0.3 					& 16\\
        $i$ ($\degree$)		&	& 0.0					& 0.0 &			& 0.0					& 0.0 					& 1 \\
        $\Omega$ ($\degree$)	&	& 0.0					& 0.0 &	 		& 0.0					& 0.0 					& 1 \\
        $\omega$ ($\degree$)	&	& 0.0					& 360.0 & 		& 0.0					& 288.0 					& 5\\
        $M$ ($\degree$)		&	& 0.0					& 360.0 	&		& 0.0					& 288.0 					& 5\\
        \cmidrule(r){1-2} \cmidrule(r){3-4}  \cmidrule(r){5-8}  
   	\end{tabular}
\end{table}

\begin{table}
\caption{A description and size of the sets of simulations run as part of our simulation suite.}
    \label{tab:sim_sets}
    \centering	
    \begin{tabular}{c p{6.5cm} c}
        \toprule
        Set		&  \centering {Description} &\\
        \toprule
        	\rom{1}	& A set of 13 simulations with 1000 TPs in the HZ for all green non-overlapping systems where the orbital period of the Jovian was within one order of magnitude of the period in the centre of the HZ ($0.1\ T_{\textrm{HZ}} \leq T_{\textrm{Jovian}} \leq 10\ T_{\textrm{HZ}}$).&\\
        	\midrule
        	\rom{2}	& A set of 26 simulations with 5000 TPs in the HZ for all amber partially overlapping systems.&\\
        	\midrule
        	\rom{3}	& A set of 26 simulations with 10,000 TPs in the HZ for all red completely overlapping systems.&\\
        	\midrule
        	\rom{4}	& A set of 20,400 simulations with a $1\ \textrm{M}_\oplus$ planet for the 26 red completely overlapping systems (530,400 simulations in total). For each system,  20,400 simulations were run, sweeping a $1\ \textrm{M}_\oplus$ planet over the orbital parameter space as outlined in Table~\ref{tab:tp_params}.&\\
        	\midrule
        	\rom{5}	& A set of 20,400 $1\ \textrm{M}_\oplus$ planet simulations for those red systems found to be stable in a narrow region of resonant stability ($15/26$ systems) for a simulation time $T_{\textrm{sim}} = 10^8$ years.\\
        \bottomrule
    \end{tabular}
\end{table}

Table~\ref{tab:sim_sets} describes the sets of simulations that were carried out. Set \rom{1} tests the sub-set of the green non-overlapping systems that have their Jovians nearest to their respective HZs. Set \rom{2} tests the amber partially overlapping systems with 5000 TPs, and set \rom{3} tests the red completely overlapping systems with 10,000 TPs. Increasingly more TPs were used for those systems with predictably more interacting HZs to achieve higher resolution maps when analysing the results.

Simulation set \rom{4} comprises a suite of simulations for each red completely overlapping system with a $1\ \textrm{M}_\oplus$ planet in the HZ, along with the system's Jovian. Assuming co-planar planets, these simulations explored the semi-major axis ($a$), eccentricity ($e$), argument of periastron ($\omega$) and mean anomoly ($M$) parameter space of the $1\ \textrm{M}_\oplus$ planet. Table~\ref{tab:tp_params} shows the range of orbital parameters and the number of equally spaced intervals within each range. In total a suite of 20,400 simulations were carried out for each system, with each simulation representing a unique set of planetary orbital parameters. As there are 5 values explored for both $\omega$ and $M$, this means there are 25 simulations for a given pair of ($a$, $e$) values. The $1\ \textrm{M}_\oplus$ simulations were ran for $T_{\rm sim} = 10^7$ years, or until one of the planets were removed or was involved in a collision. As all the Jovian planets in these red completely overlapping sample were located in the vicinity of the HZ (which was located well within 10 au), a planet removal was defined following \cite{Robertson2012}: if either planet exceeded an astrocentric distance of 10 au. A collision was defined as occurring when the planets approached within 1 Hill radii of each other. The time step for these $1\ \textrm{M}_\oplus$ simulations was set to $1/20$ of the smallest orbital period of the Jovian and $1\ \textrm{M}_\oplus$ planet. Simulation set \rom{5} repeats these $1\ \textrm{M}_\oplus$ body simulations for red systems which hosted some stable regions for an extended integration time of $T_{\textrm{sim}} = 10^8$ years.

\subsection{Simulation Analysis}
\label{subsec:defining_stable_regions}
The results of the simulations were interpreted using stability maps and resonant angle plots. The stability maps are plotted over the semi-major axis--eccentricity ($a$,$e$) parameter space. This 2-dimensional map presents the lifetimes of bodies as a function of their initial semi-major axis ($x$-axis) and eccentricity ($y$-axis) values.

\begin{figure}
	\begin{subfigure}{\linewidth}
    		\centering
    		\includegraphics[width=\linewidth]{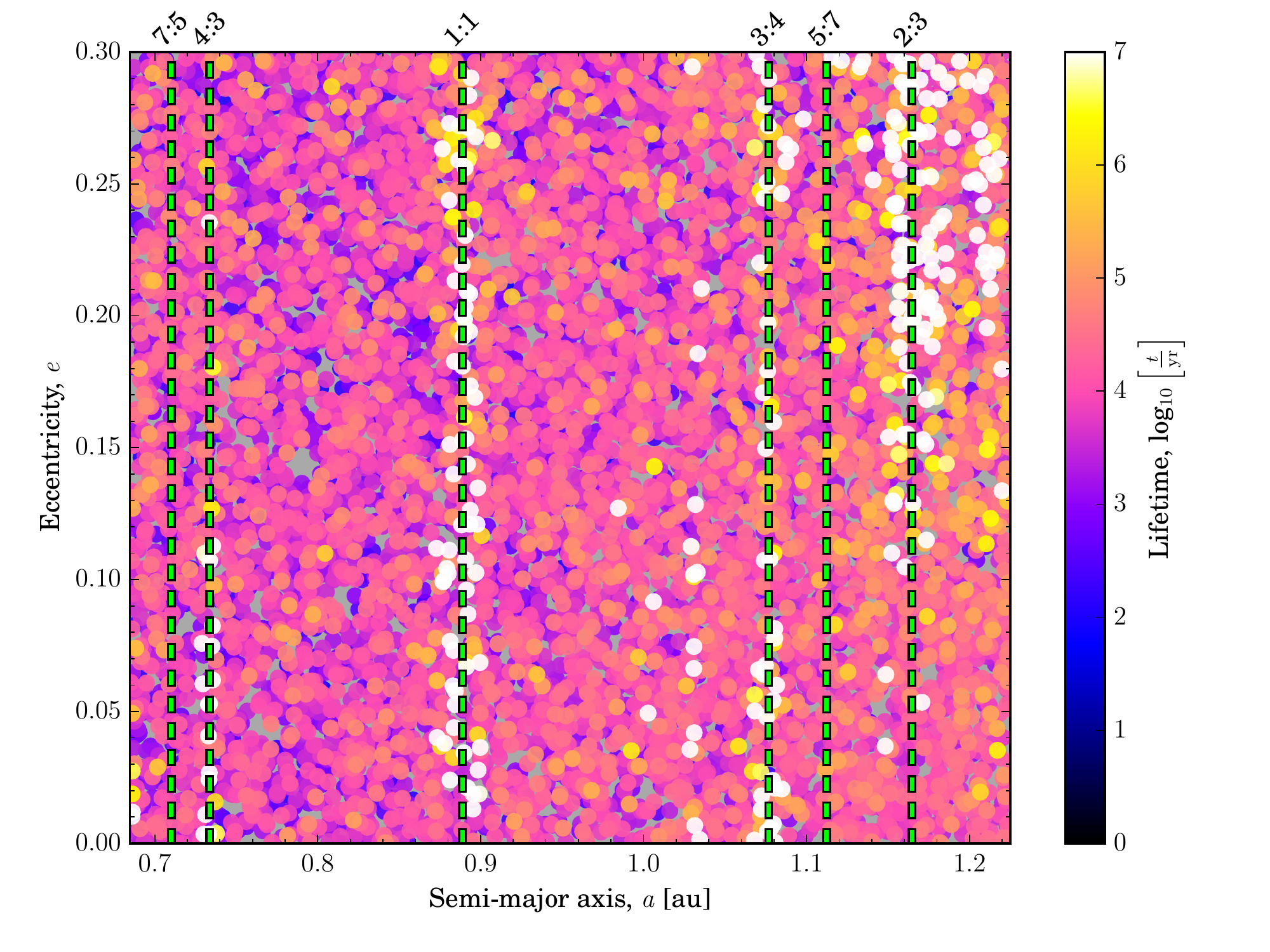}
        \caption{10,000 TP}\label{fig:red_example_a}
	\end{subfigure}
	
    \begin{subfigure}{\linewidth}
	\centering
		\includegraphics[width=\linewidth]{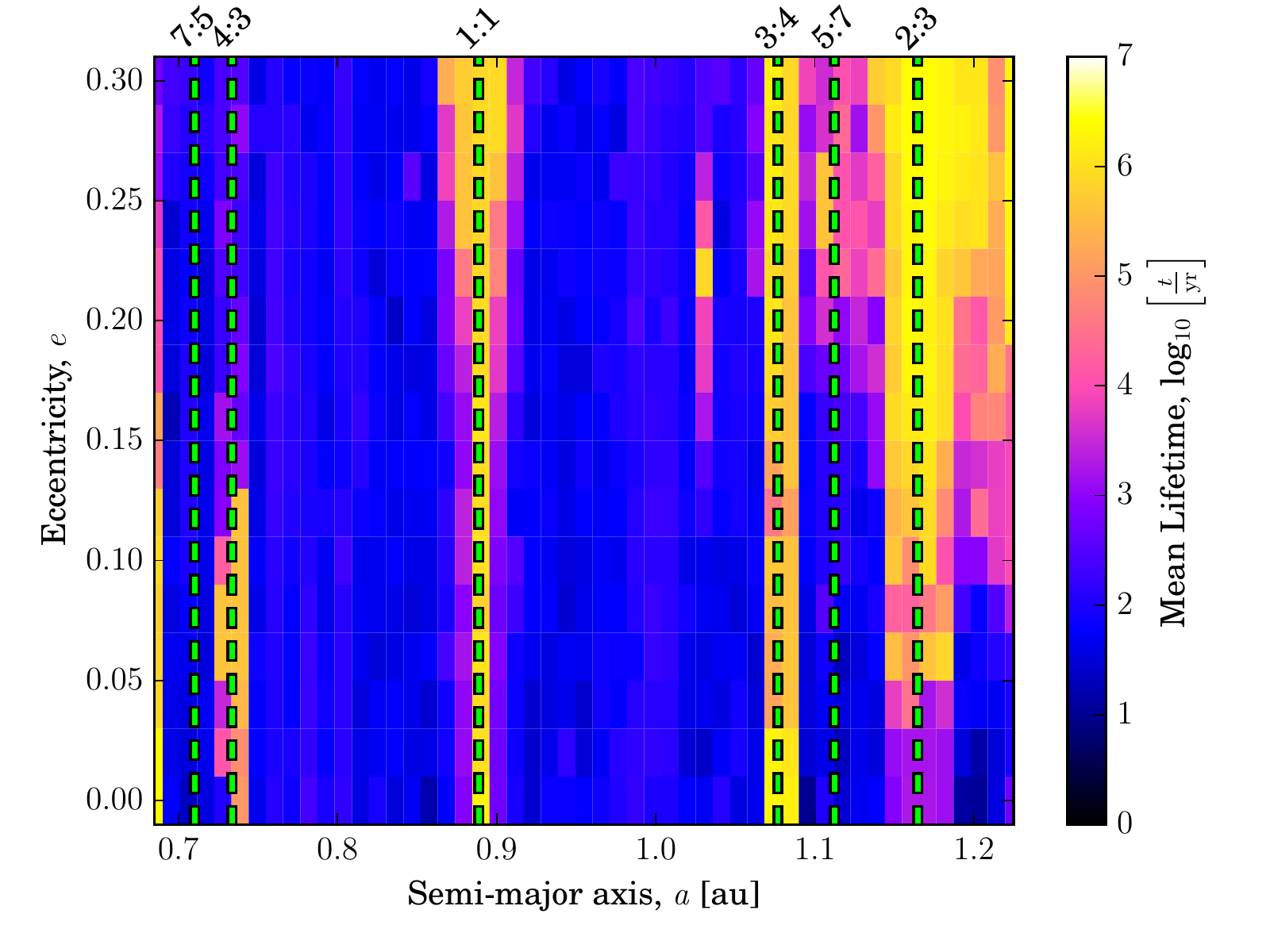}
		\caption{$1\ \textrm{M}_\oplus$}\label{fig:red_example_b}
	\end{subfigure}
\caption{A comparison between the stability maps for the simulations of (a) 10,000 TP in the HZ, and (b) the 20,400 $1\ \textrm{M}_\oplus$ simulations, of the red completely overlapping system HD 137388. We mark the location of several first and second order MMRs with green dashed lines.}\label{fig:red_example}
\end{figure}

For the Earth mass planet simulations (sets \rom{4} and \rom{5}), each simulation had the $1\ \textrm{M}_\oplus$ planet at a specified initial ($a$, $e$). As mentioned above, at each ($a$, $e$) position there are 25 simulations exploring the ($\omega$, $M$) parameter space. As such, the maps combine the results of the 25 simulations over the ($\omega$, $M$) parameter space by plotting the mean lifetime of all bodies with those ($a$, $e$) values \citep[similar to previous work by][]{Wittenmyer2012,Robertson2012}.  Figure~\ref{fig:red_example} shows a comparison of the two types of stability map: the lifetime of randomly distributed TPs across the HZ (Fig.~\ref{fig:red_example_a}), and the average lifetime of a $1\ \textrm{M}_\oplus$ body being swept through the orbital parameter space (Fig.~\ref{fig:red_example_b}).

\begin{figure}
	\centering
	\includegraphics[width=\linewidth]{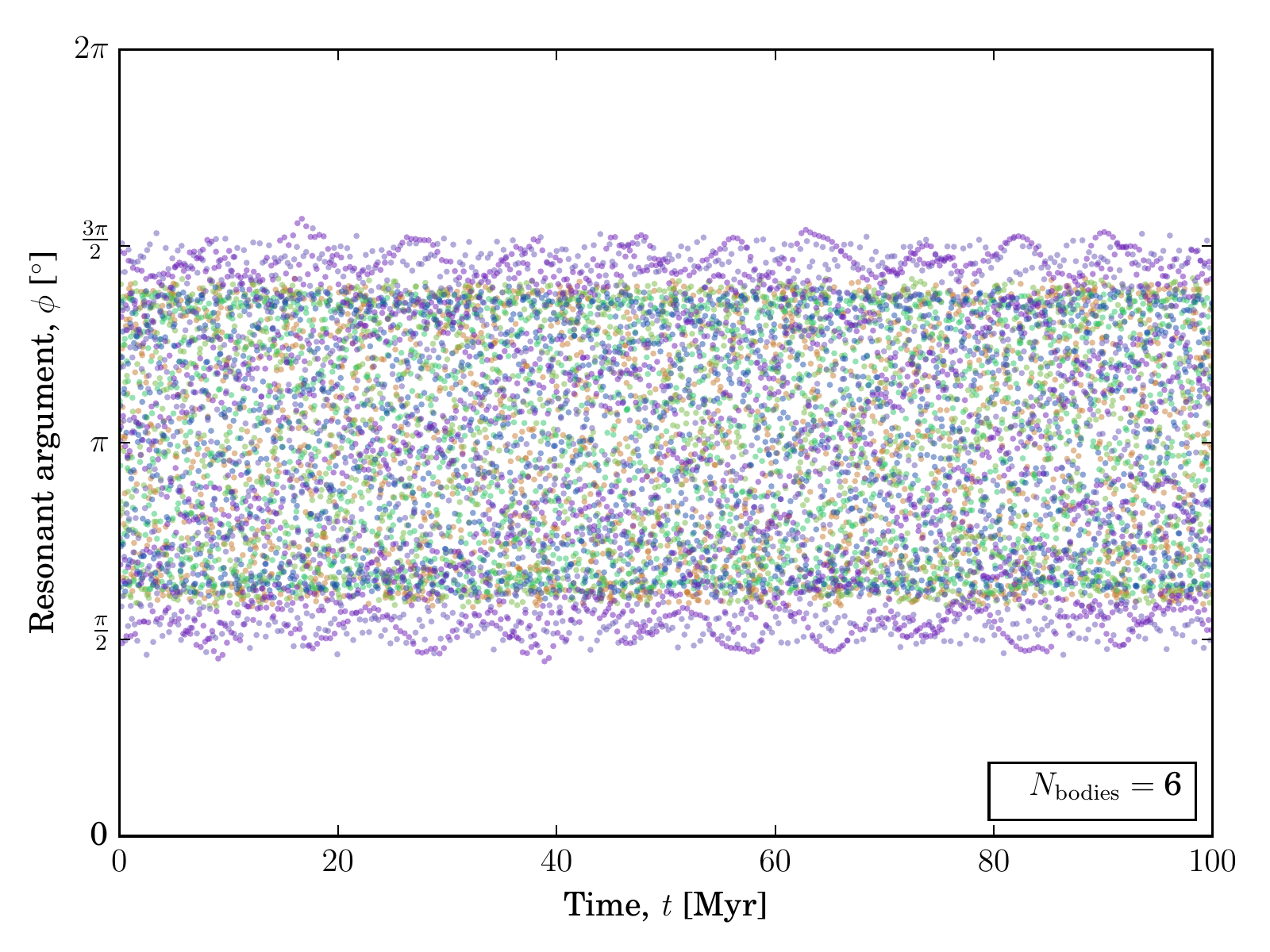}
	\caption{The librating resonant angle $\phi=4\lambda'-3\lambda-\omega'$ versus time for the stable bodies (6) of the 4:3 MMR with the Jovian in the red completely overlapping system HD 137388. Note that each body is run in its own simulation, with the resonant angle from all simulations stacked.}\label{fig:res_example}
\end{figure}

Figure~\ref{fig:res_example} shows the time evolution of the resonant angle for all $1\ \textrm{M}_\oplus$ bodies that were trapped in 4:3 resonance with the Jovian planet in the HD 137388 system from our simulations. Such plots reveal whether potentially resonant $1\ \textrm{M}_\oplus$ bodies librate, and can therefore be considered to be trapped in mean-motion resonance.


\section{Results and Discussion}
\label{sec:results}
\begin{figure}
	\centering
	\includegraphics[width=\linewidth]{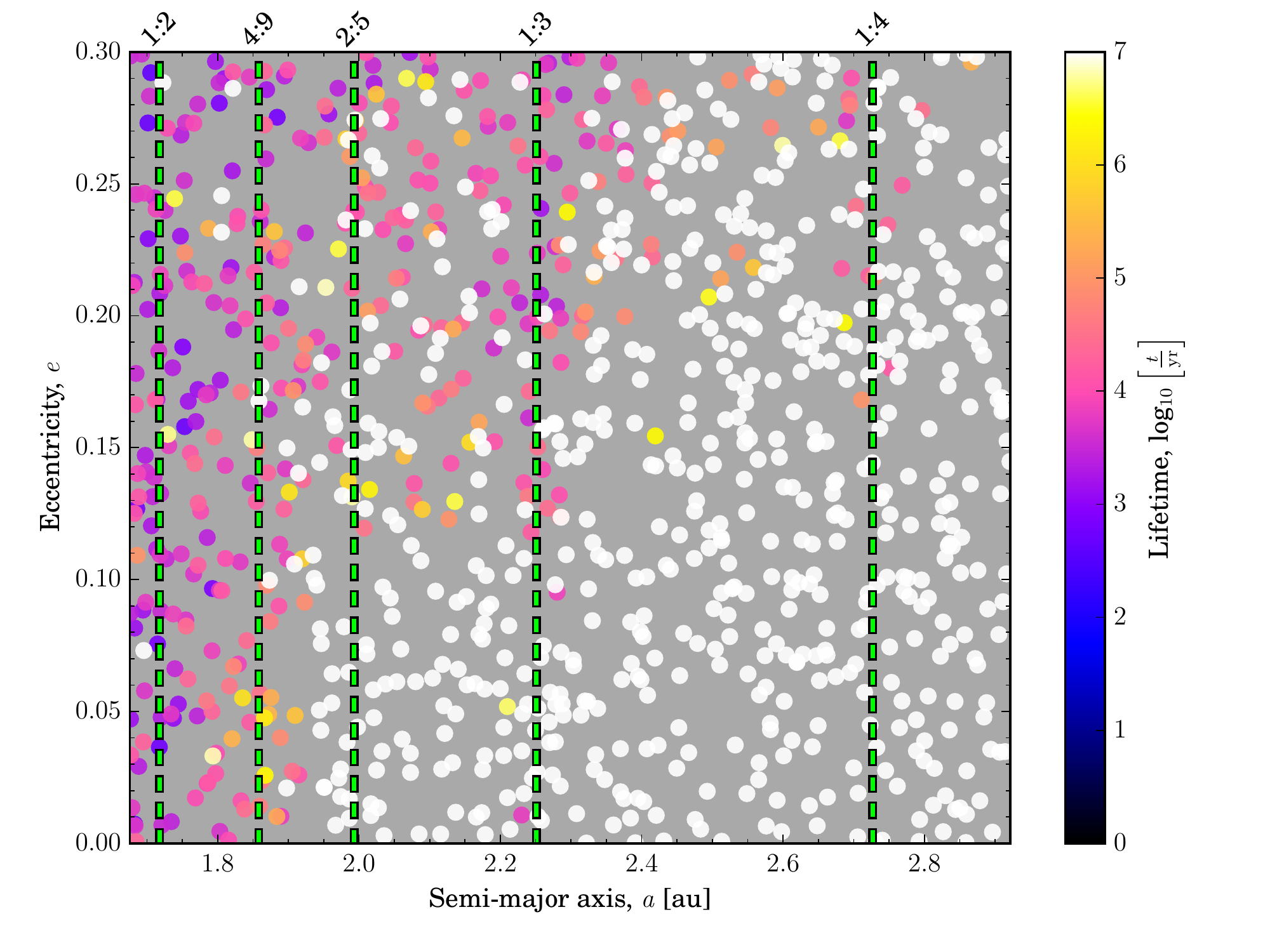}
	\caption{The stability map of the green non-overlapping system HD 67087 with 1000 TPs in the HZ. The Jovian planet is located interior to the HZ. We mark the location of several first and second order MMRs with green dashed lines.}\label{fig:green_map}
\end{figure}

For the green non-overlapping systems where the orbital period of the Jovian was within one order of magnitude of the period in the centre of the HZ which we simulated in set \rom{1}, %($0.1\ T_{\textrm{HZ}} \leq T_{\textrm{Jovian}} \leq 10\ T_{\textrm{HZ}}$), 
we found that some TPs in the HZ were still disrupted by the presence of such a Jovian. An example system is shown in Figure~\ref{fig:green_map}. Despite this, our results demonstrated that the majority of the TPs remain in stable orbits in the HZ. As a result of their stability, these systems were computationally expensive to simulate, since typically they retain the majority, if not all, of their TPs. Due to the presence of these large, unperturbed regions of the HZ within which TPs are dynamically stable, we conclude that it is dynamically possible for a Terrestrial planet to be hidden in the HZ of green  non-overlapping systems for which the chaotic region does not overlap the HZ. Given we cannot further constrain the location of these potentially habitable Terrestrials, the green systems were not tested further.

\begin{figure}
	\centering
	\includegraphics[width=\linewidth]{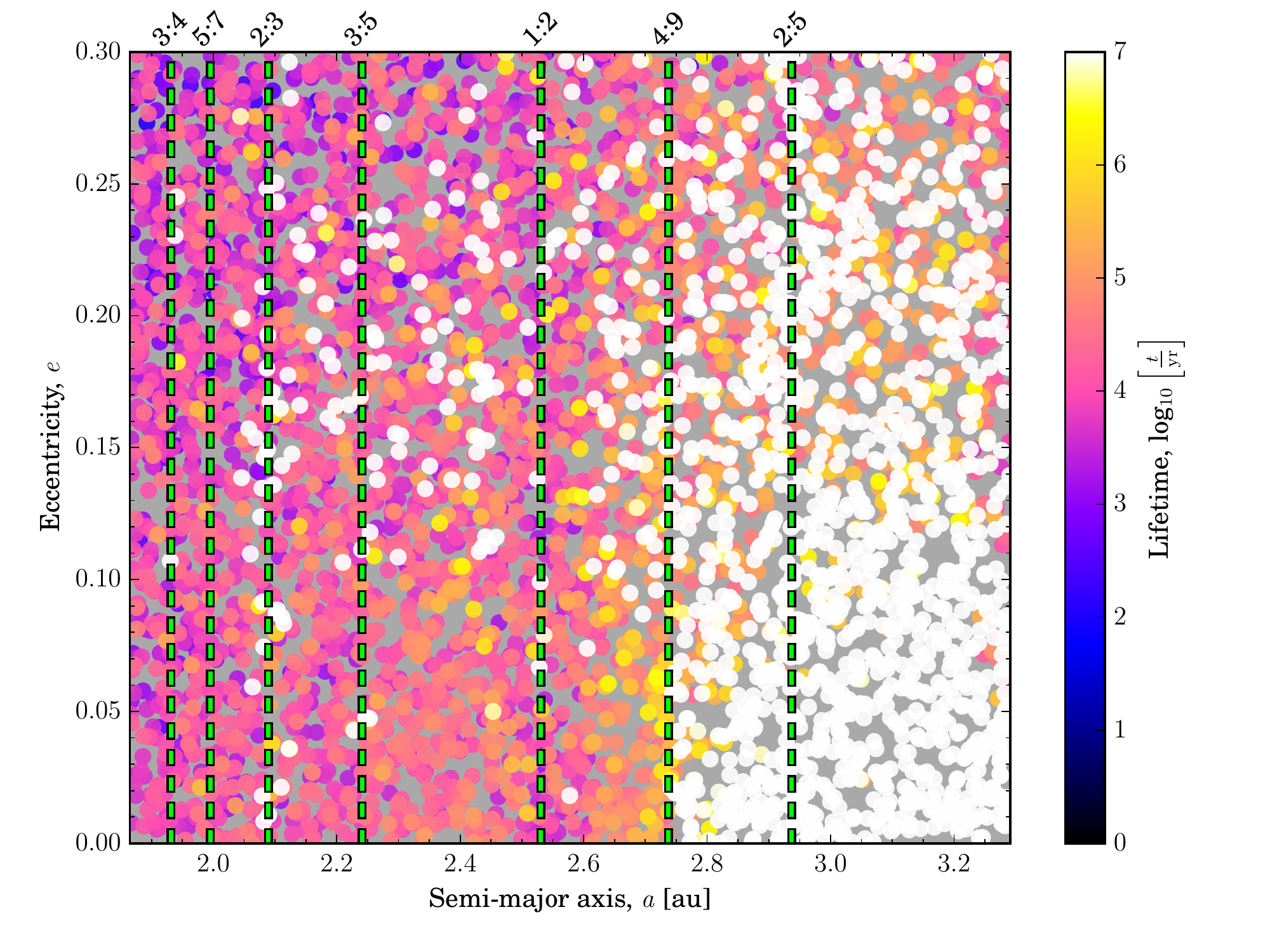}
	\caption{The stability map of the the amber partially overlapping system HD 48265 with 5000 TPs in the HZ. The Jovian planet is located interior to the HZ. We mark the location of several first and second order MMRs with green dashed lines.}\label{fig:amber_map}
\end{figure}

Based on the classification and selection scheme outlined in section \ref{subsec:predicting_stable_regions}, the results from the amber partially overlapping systems (set \rom{2}) behave as expected. We can see in Figure~\ref{fig:amber_map} that there is a gradient of stability across the HZ, moving from more stable regions farther from the Jovian, to more unstable in regions nearer to the Jovian. Similar to the green non-overlapping systems, the presence of large, unperturbed regions of the HZ where TPs are dynamically stable in the amber partially overlapping systems suggest that it is dynamically possible for a Terrestrial planet to be hidden in the HZ of these systems. Our simulations cannot further constrain these locations, so no further investigation of these systems is conducted.

\begin{figure*}
\centering
	\begin{subfigure}{0.31\textwidth}
  		\centering
  		\includegraphics[width=\textwidth]{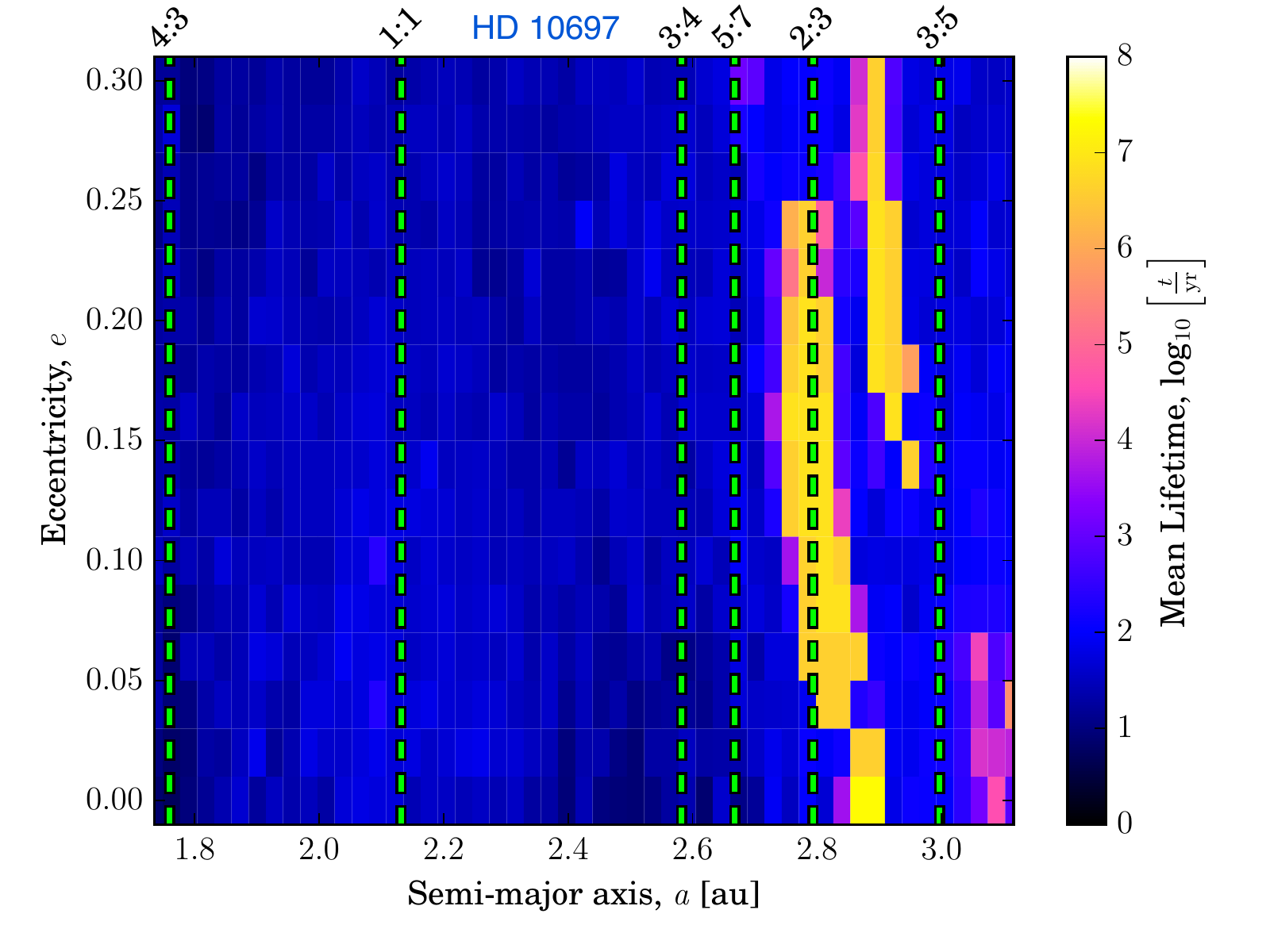}
  		\caption{HD 10697}
	\end{subfigure}
	\begin{subfigure}{0.31\textwidth}
  		\centering
  		\includegraphics[width=\textwidth]{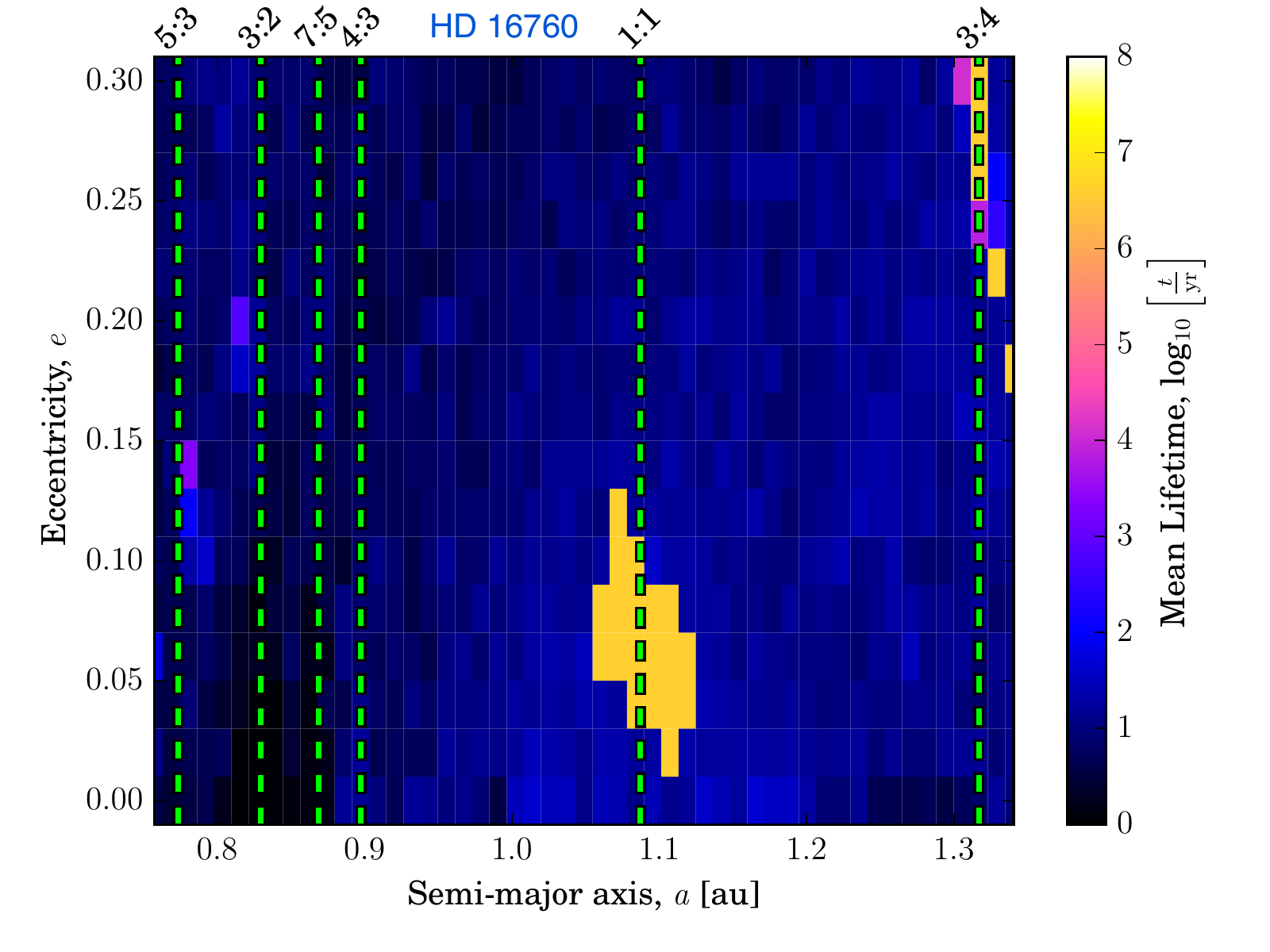}
  		\caption{HD 16760}
	\end{subfigure}
	\begin{subfigure}{0.31\textwidth}
  		\centering
  		\includegraphics[width=\textwidth]{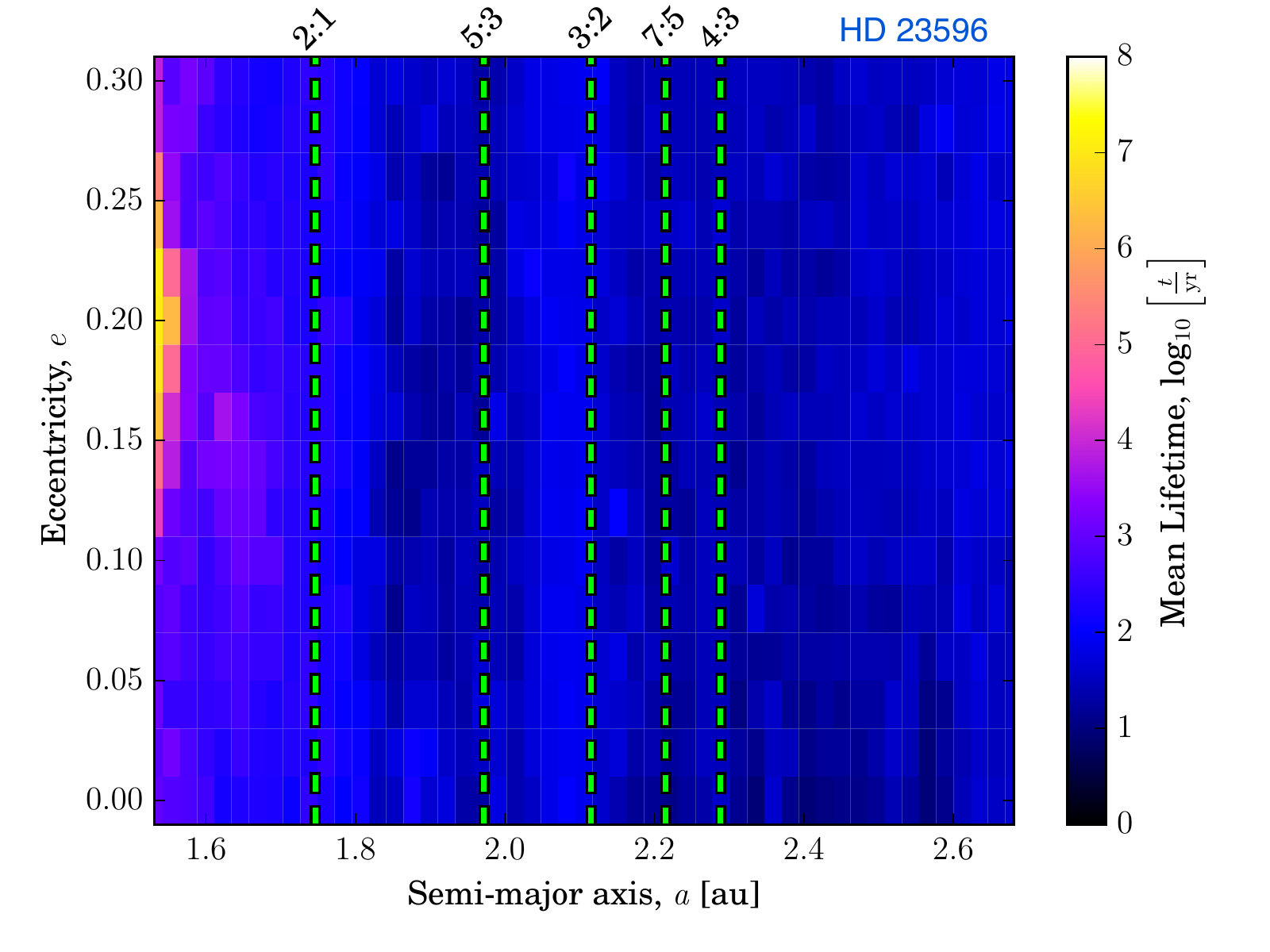}
  		\caption{HD 23596}
	\end{subfigure}
	
	\begin{subfigure}{0.31\textwidth}
  		\centering
  		\includegraphics[width=\textwidth]{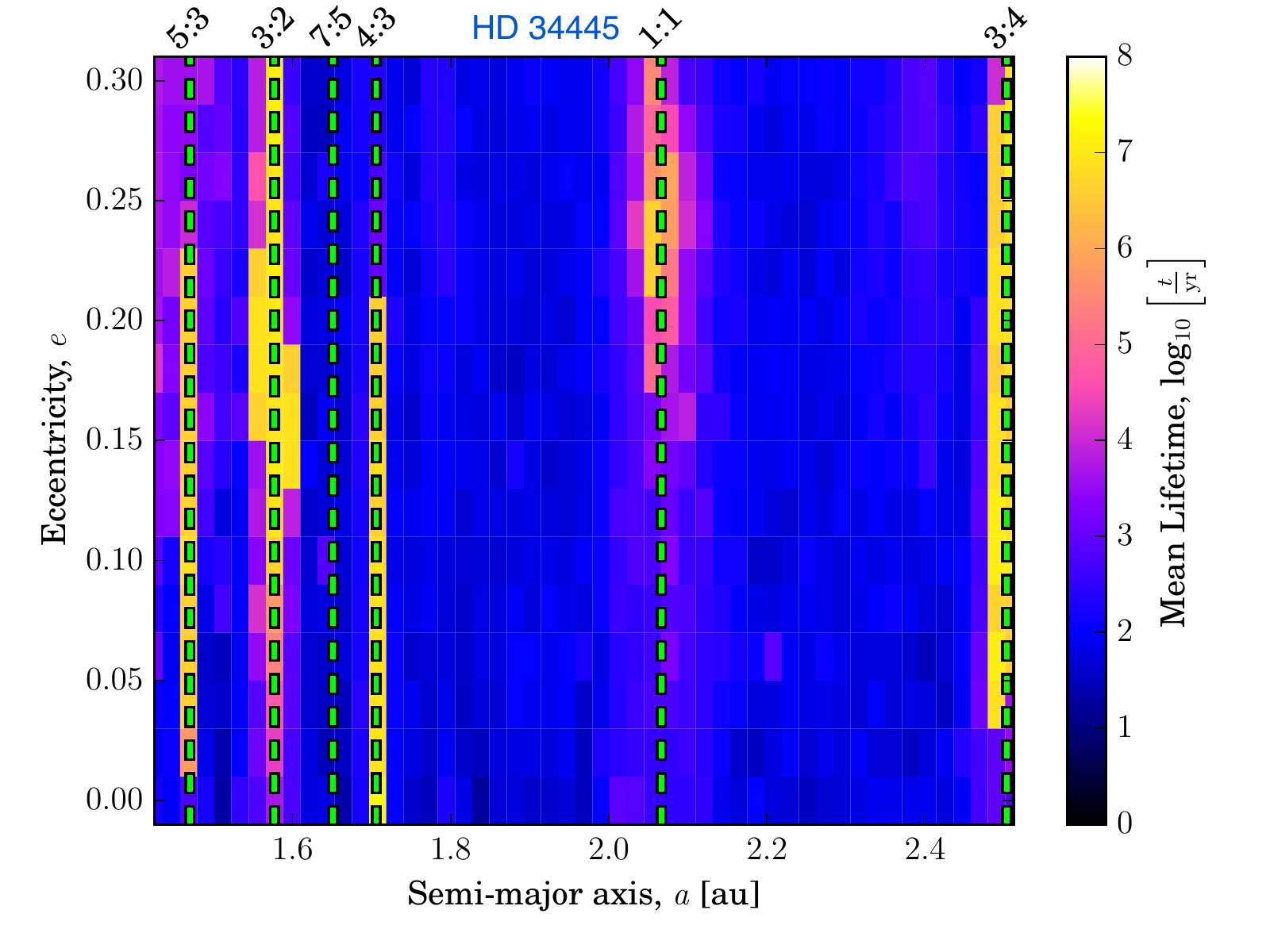}
  		\caption{HD 34445}
	\end{subfigure}
	\begin{subfigure}{0.31\textwidth}
  		\centering
  		\includegraphics[width=\textwidth]{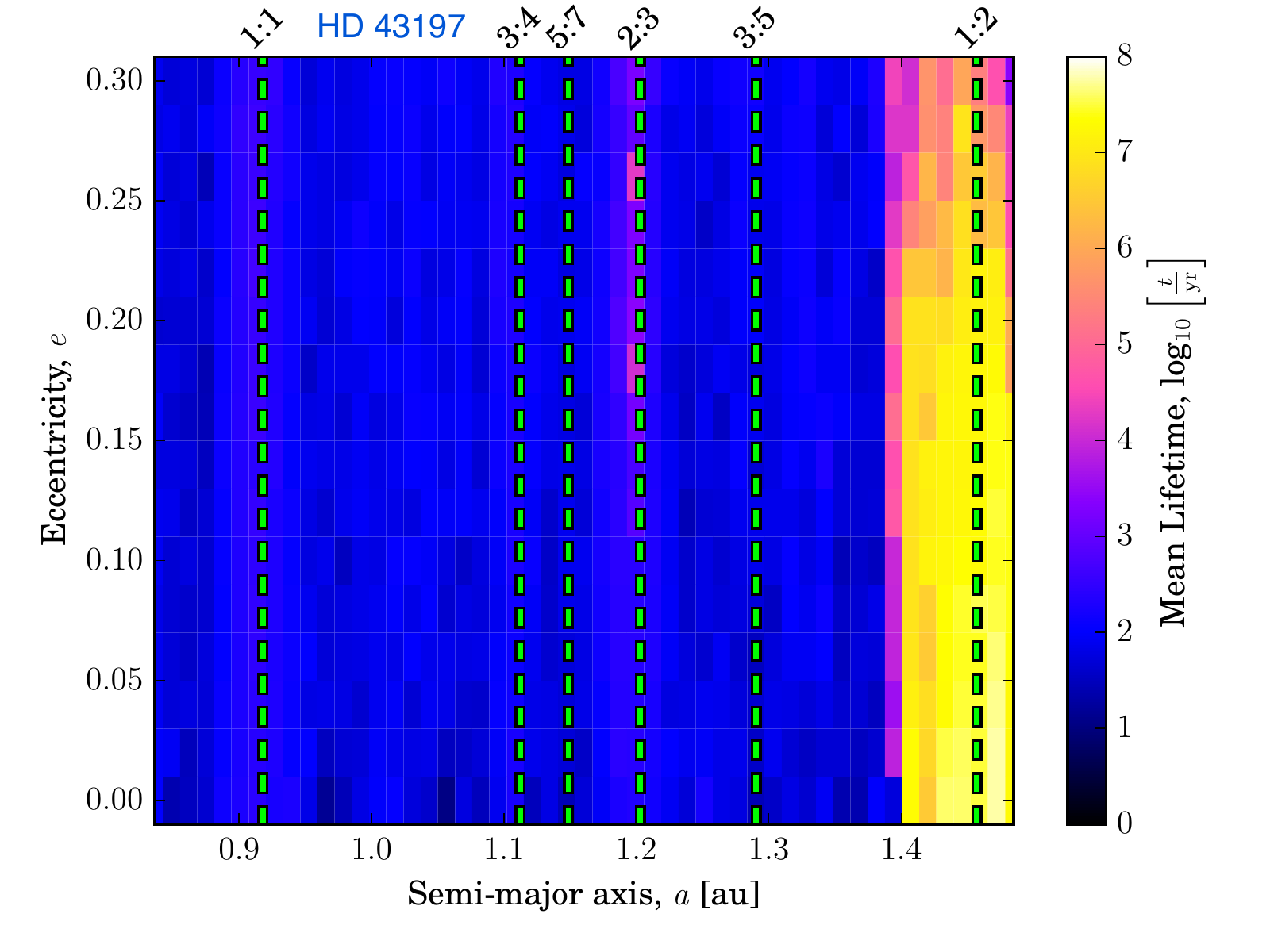}
  		\caption{HD 43197}
	\end{subfigure}
	\begin{subfigure}{0.31\textwidth}
  		\centering
  		\includegraphics[width=\textwidth]{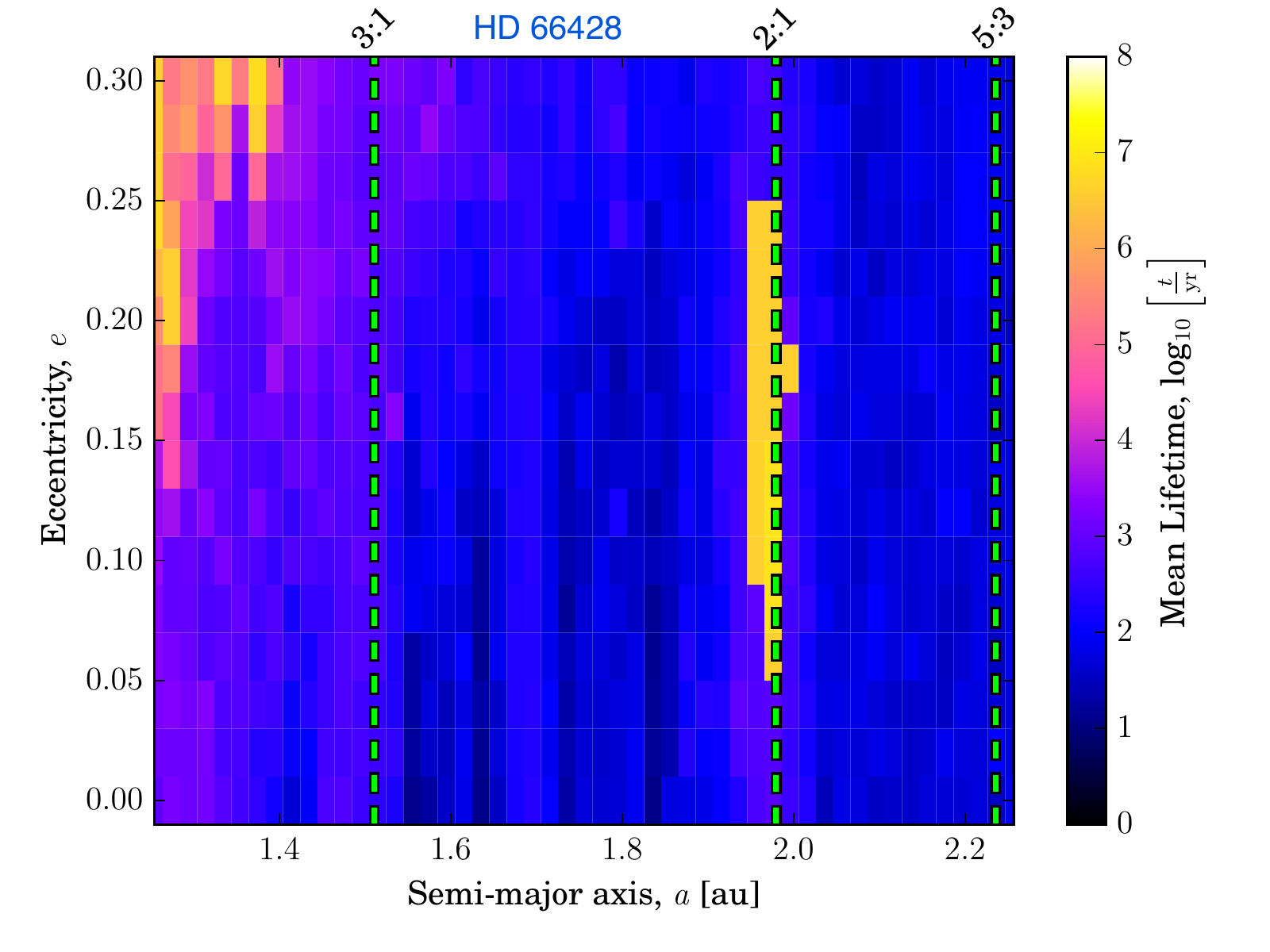}
  		\caption{HD 66428}
	\end{subfigure}
	
	\begin{subfigure}{0.31\textwidth}
  		\centering
  		\includegraphics[width=\textwidth]{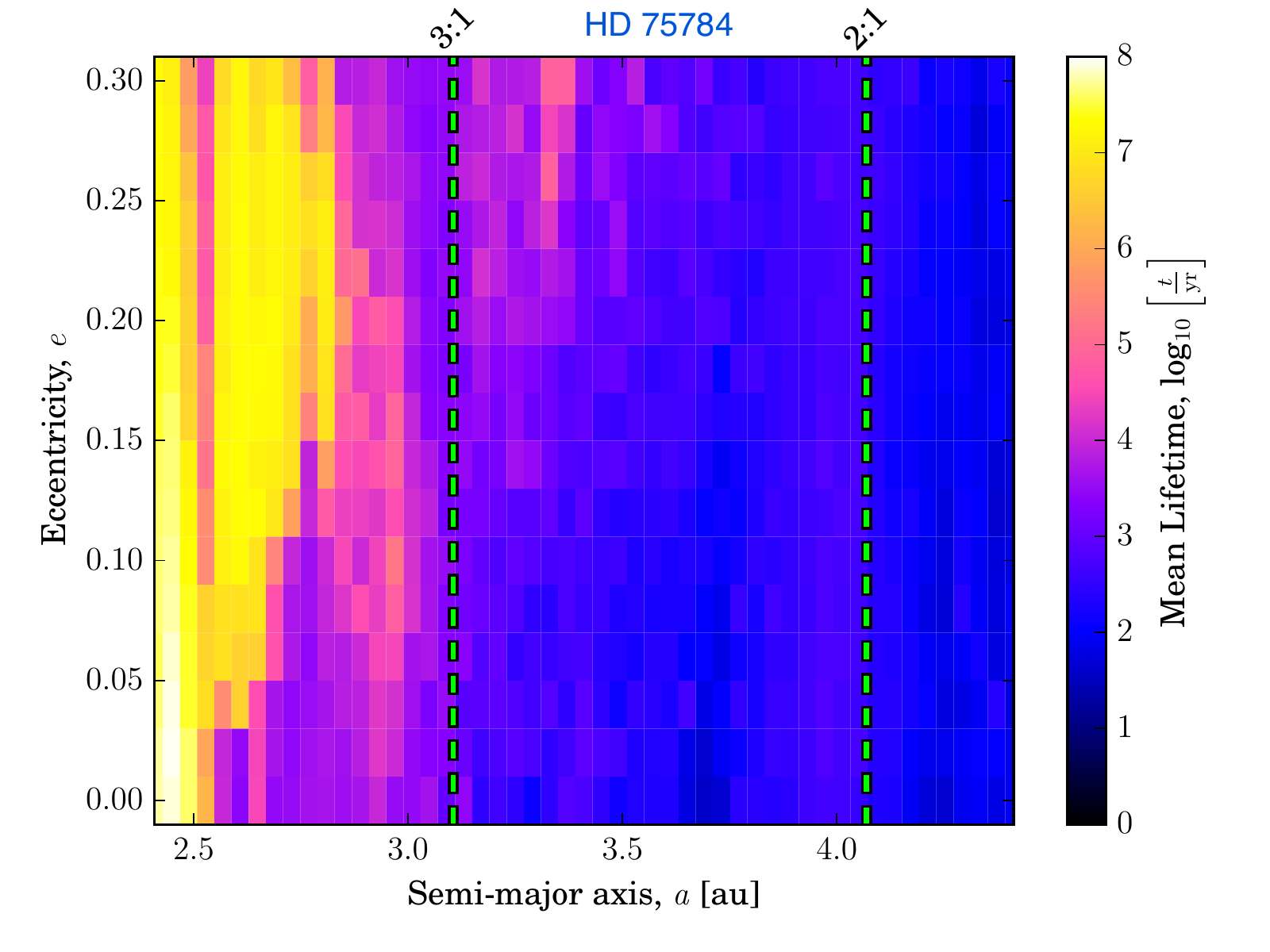}
  		\caption{HD 75784}
	\end{subfigure}
	\begin{subfigure}{0.31\textwidth}
  		\centering
  		\includegraphics[width=\textwidth]{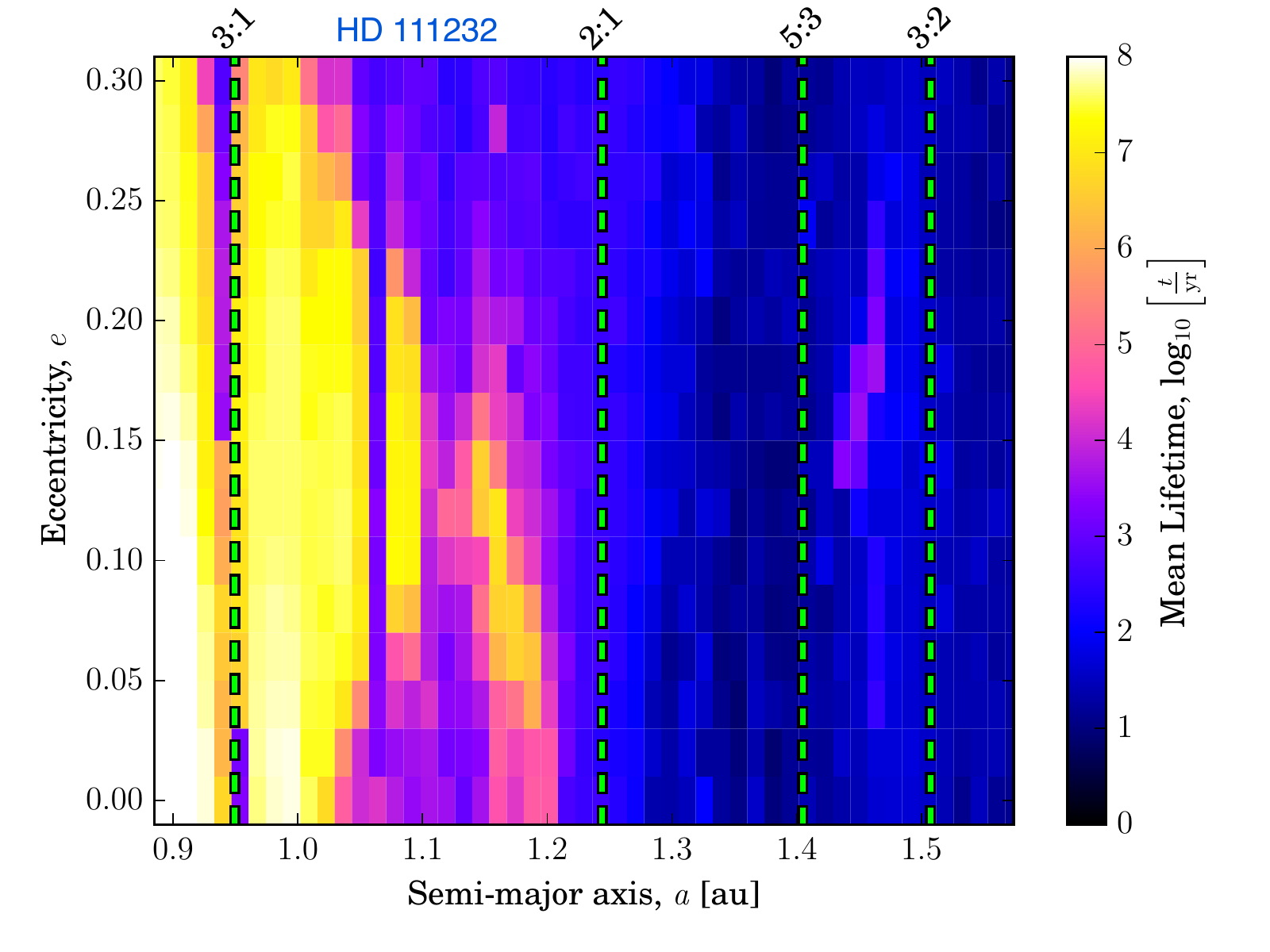}
  		\caption{HD 111232}
	\end{subfigure}
	\begin{subfigure}{0.31\textwidth}
  		\centering
  		\includegraphics[width=\textwidth]{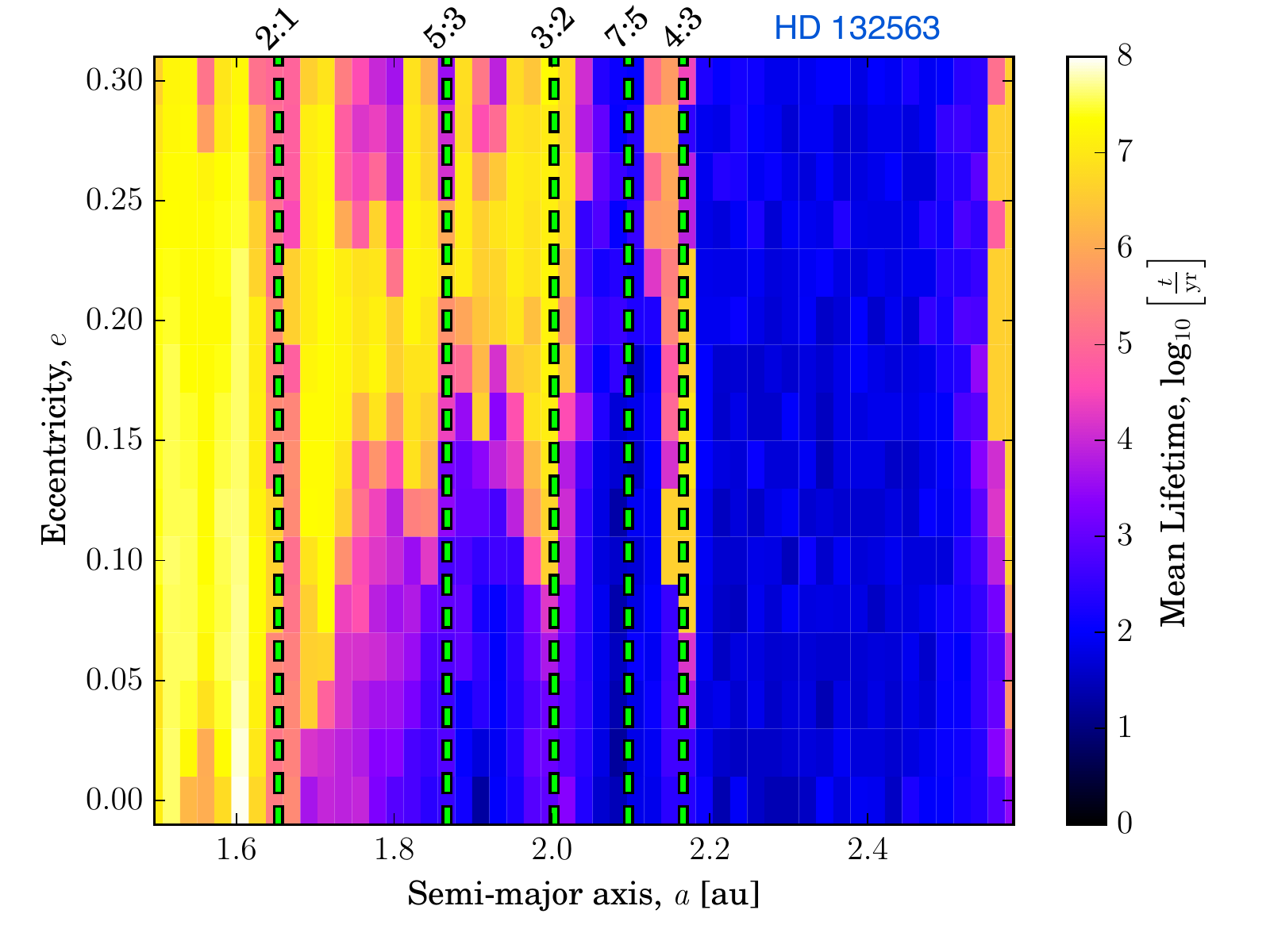}
  		\caption{HD 132563 B}
	\end{subfigure}
	
	\begin{subfigure}{0.31\textwidth}
  		\centering
  		\includegraphics[width=\textwidth]{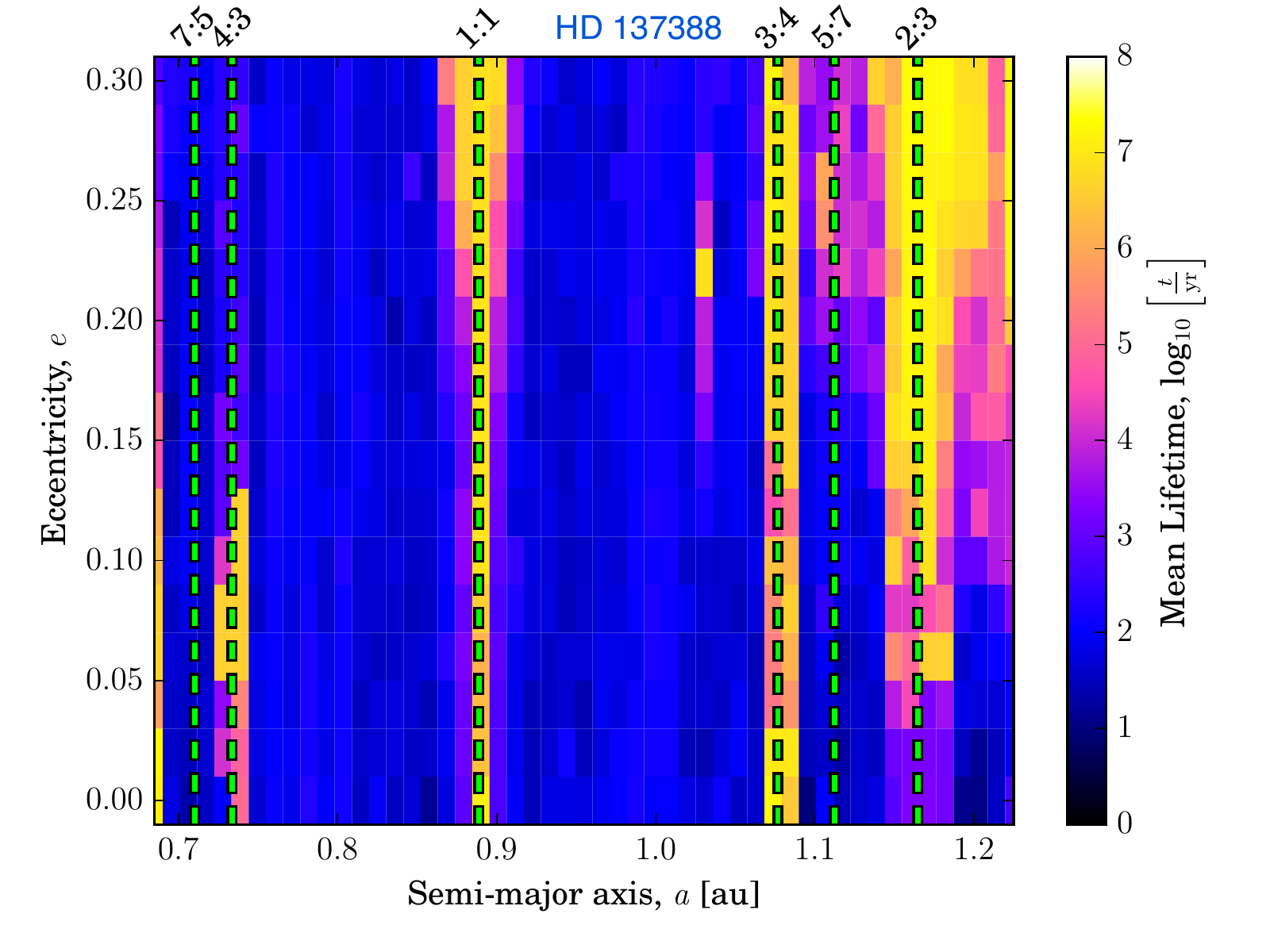}
  		\caption{HD 137388}
	\end{subfigure}
	\begin{subfigure}{0.31\textwidth}
  		\centering
  		\includegraphics[width=\textwidth]{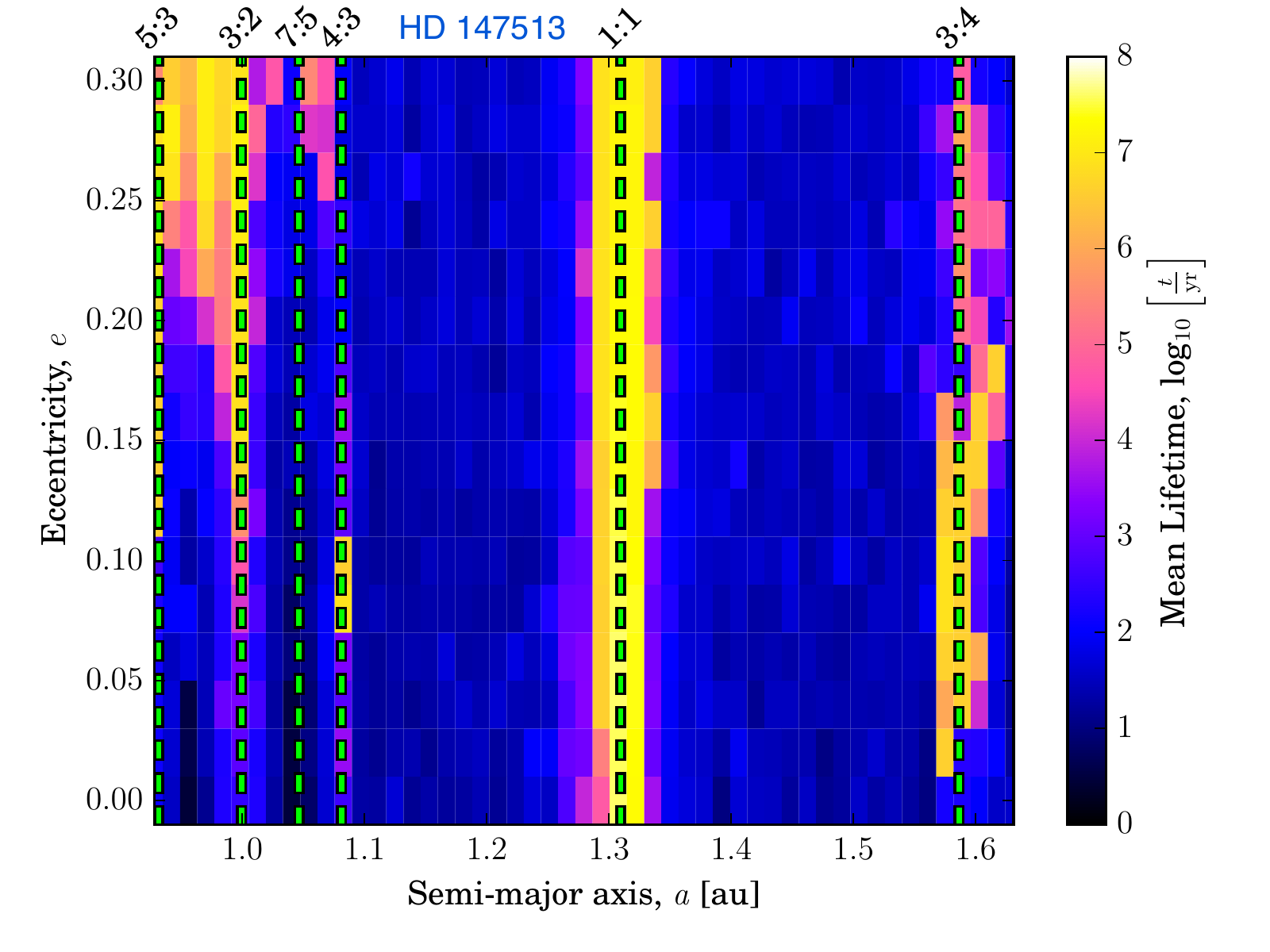}
  		\caption{HD 147513}
	\end{subfigure}
	\begin{subfigure}{0.31\textwidth}
  		\centering
  		\includegraphics[width=\textwidth]{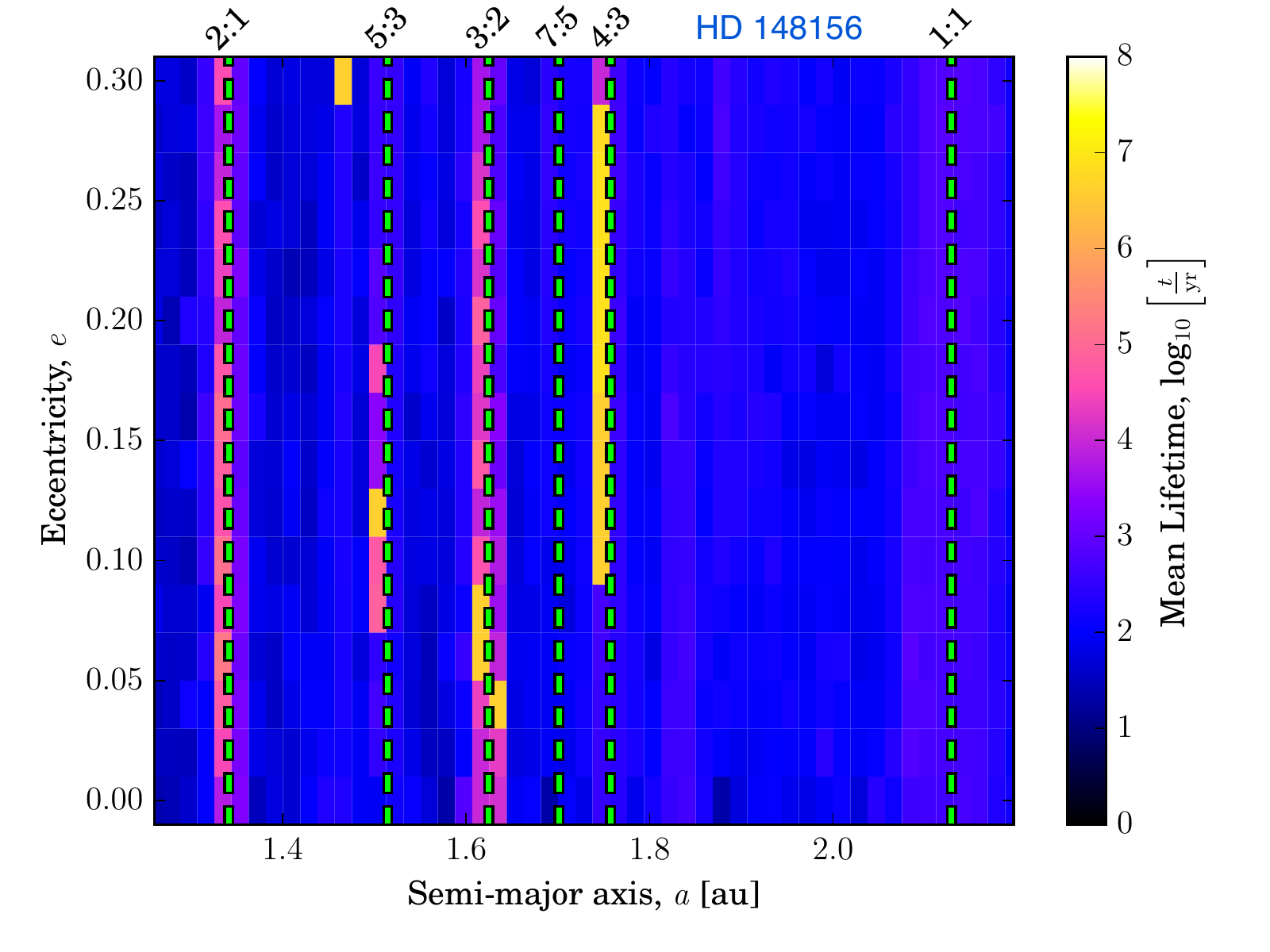}
  		\caption{HD 148156}
	\end{subfigure}
	
	\begin{subfigure}{0.31\textwidth}
  		\centering
  		\includegraphics[width=\textwidth]{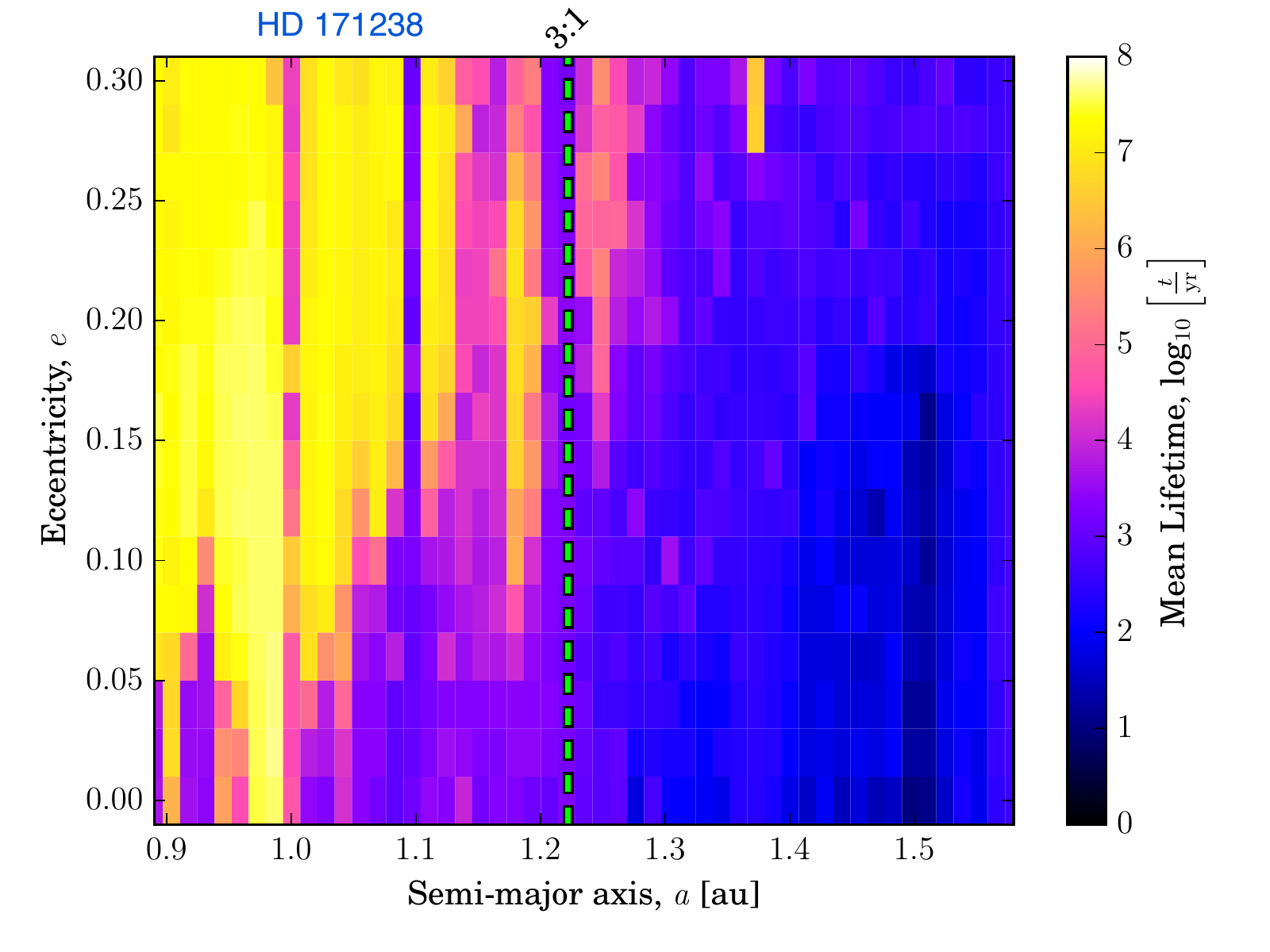}
  		\caption{HD 171238}
	\end{subfigure}
	\begin{subfigure}{0.31\textwidth}
  		\centering
  		\includegraphics[width=\textwidth]{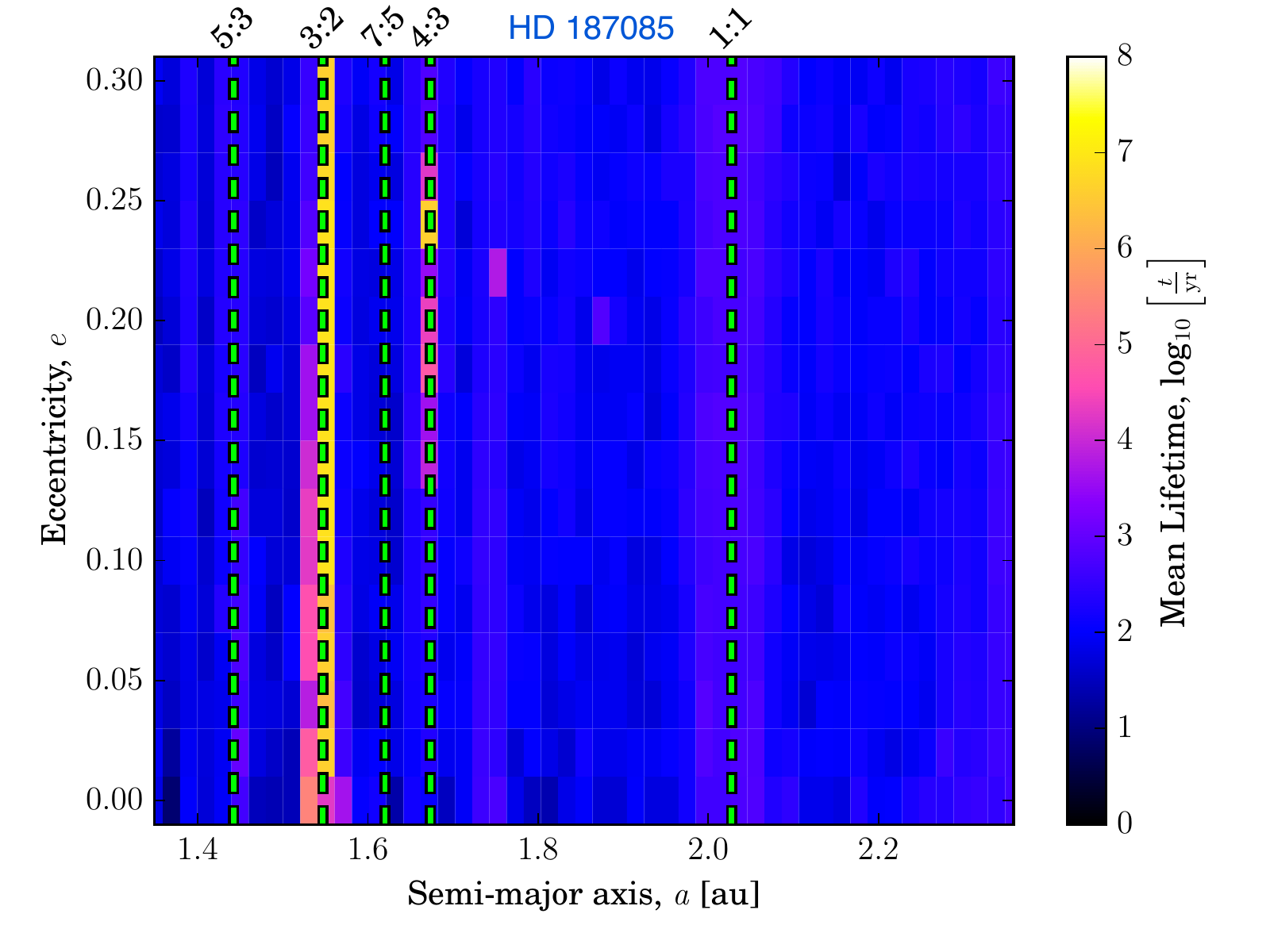}
  		\caption{HD 187085}
	\end{subfigure}
	\begin{subfigure}{0.31\textwidth}
  		\centering
  		\includegraphics[width=\textwidth]{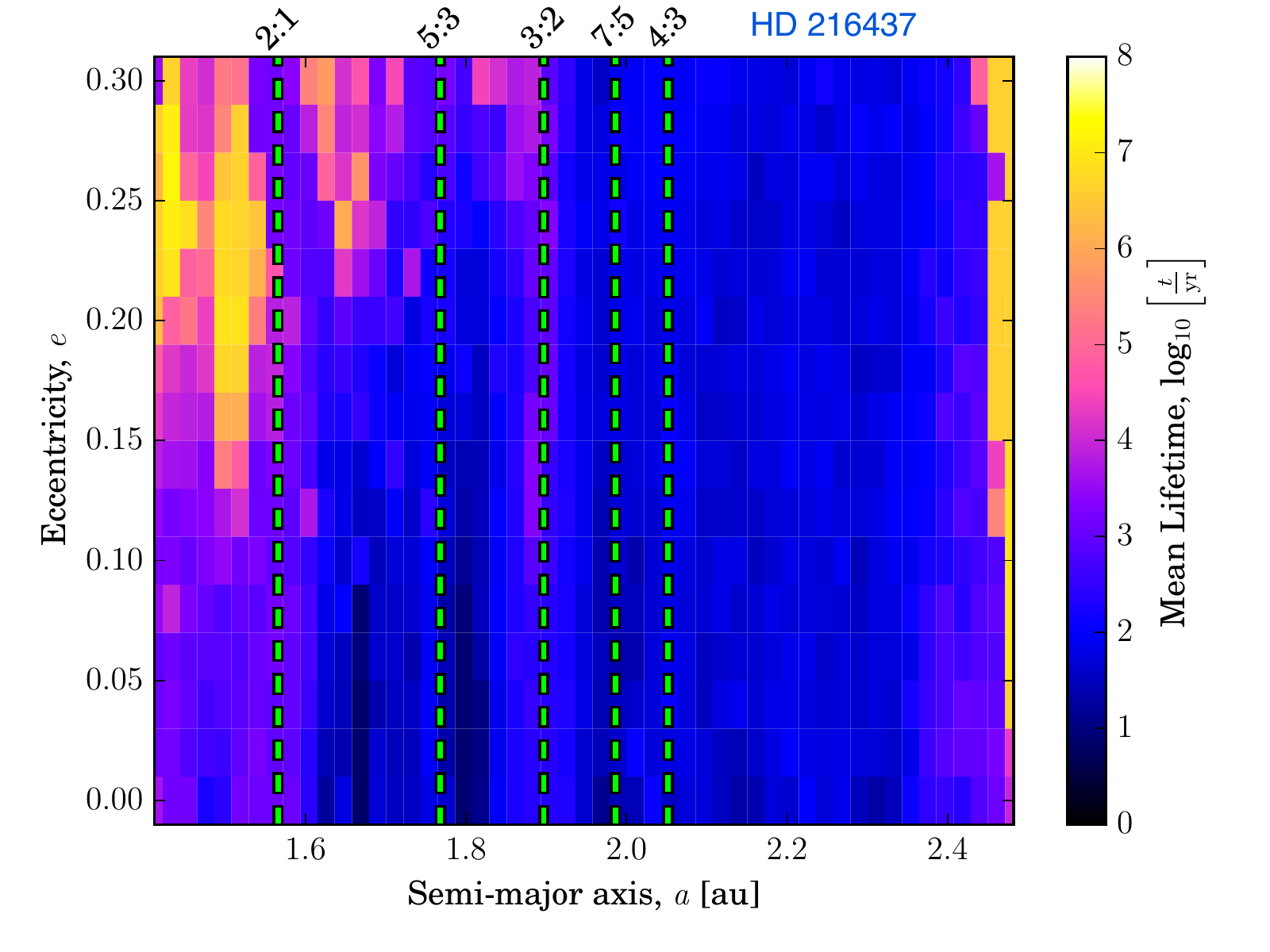}
  		\caption{HD 216437}
	\end{subfigure}
	\caption{The stability maps for all the red systems with stable MMRs. These are the 15 systems re-classifed as blue resonant systems.}
	\label{fig:red_maps}
\end{figure*}

More than half of the red systems were found to contain regions of stability, some of which were aligned with the MMRs of the Jovian. As the HZs of these systems were significantly influenced by the presence of the Jovian, it would be reasonable to consider whether mutual interactions with the massive planet affected the stability. We continued this investigation with additional simulations in which we replaced the massless TPs with a $1\ \textrm{M}_\oplus$ planet. Set \rom{4} examined all 26 of the red completely overlapping systems, and identified 15 systems for which $1\ \textrm{M}_\oplus$ planets might prove stable at some location within the HZ. For Set \rom{5}, we took this subset of 15 stable systems and performed significantly longer simulations of duration $T_{\textrm{sim}} = 10^8$ years. The results showed that all 15 of these systems were found to be capable of hosting a $1\ \textrm{M}_\oplus$ planet on a stable orbit within their HZ, and were then reclassified as blue resonant systems.\footnote{Note that this label is semantic, as it was found that some of the stable bodies do not appear to be in resonant configurations (stable bodies that are not in an MMR do not show up in the libration plots shown in Figure~\ref{fig:res_plots}).} Figure~\ref{fig:red_maps} shows the stability maps of all 15 of these blue resonant systems from set \rom{5}, while Table~\ref{tab:landscape} shows the system properties of the 11 remaining red completely overlapping systems and the 15 reclassified blue resonant systems.

\begin{table*}
	\setlength{\tabcolsep}{2pt}
  	\caption{The system properties and orbital parameters of the blue resonant (upper) and red completely overlapping (lower) systems. All Jovians were detected via the radial velocity method.}
  	\label{tab:landscape}
  	\centering
  	\begin{threeparttable}
  	\centering
  	\makebox[0.935\linewidth]{
  	\centering
	\begin{tabular*}{0.975\textwidth}{l l l l l l l l l l l l l}
  		\cmidrule(r){3-6}\cmidrule(r){8-13}
		& & \multicolumn{4}{c}{Star} & & \multicolumn{6}{c}{Jovian\tnote{$\dagger$}} \\
    		\cmidrule(r){3-6}\cmidrule(r){8-13}
     				&&Type& Mass 			& HZ$_{\textrm{inner}}$ 	& HZ$_{\textrm{outer}}$ 	&& $M \sin{i}$ 				& $a$ 	& $e$  	& $\omega$ 		& Instrument		& Detection Reference\tnote{*}\\
     				&& 		& (M$_{\odot}$)	& (AU)					& (AU)					&& (M$_{\textrm{Jupiter}}$)	& (AU)	&		& ($\degree$)	& 						& \\
    		\cmidrule(r){1-1}\cmidrule(r){3-6}\cmidrule(r){8-13}
    		HD 10697		&& G5 IV	& 0.847			& 1.735					& 3.116				&& 6.23505					& 2.13177	& 0.099	& 111.2			& HIRES\tnote{a}& \cite{American2000}\\
    		HD 16760 	&& G5 V	& 0.991			& 0.7573					& 1.3407				&& 13.2921					& 1.08727	& 0.067	& 232			& HIRES\tnote{a}, HDS\tnote{b}	& \cite{Sato2009}\\
    		HD 23596 	&& F8	& 1.19			& 1.5307					& 2.681				&& 7.74272					& 2.77219	& 0.266	& 272.6			& ELODIE\tnote{c}	& \cite{Perrier2003a}\\
    		HD 34445 	&& G0 V	& 1.06			& 1.425					& 2.5121				&& 0.790506					& 2.06642	& 0.27	& 104			& HIRES\tnote{a}	& \cite{W.Howard2010a}\\
    		HD 43197 	&& G8 V	& 0.945			& 0.8359					& 1.485				&& 0.596868					& 0.918027	& 0.83	& 251			& HARPS	\tnote{d} & \cite{Naef2010}\\
    		HD 66428 	&& G5	& 0.83			& 1.254					& 2.257				&& 2.74962					& 3.14259	& 0.465	& 152.9			& HIRES\tnote{a}	& \cite{Butler2006}\\
    		HD 75784 	&& K3 IV	& 0.719			& 2.408					& 4.4125				&& 5.6						& 6.45931	& 0.36	& 301			& HIRES\tnote{a}	& \cite{From2015}\\
		HD 111232  	&& G5 V	& 0.933			& 0.8849					& 1.574				&& 6.84182					& 1.97489	& 0.2	& 98				& CORALIE\tnote{e}	& \cite{Mayor2004}\\
		HD 132563 B 	&& -		& 1.53			& 1.496					& 2.585				&& 1.492470					& 2.62431	& 0.22	& 158			& SARG\tnote{f}	& \cite{Desidera2011}\\
		HD 136118 	&& F9 V	& 1.09			& 1.67					& 2.939				&& 11.6809					& 2.33328	& 0.338	& 319.9			& Hamilton\tnote{g}	& \cite{Fischer2002}\\
		HD 137388 	&& K0/K1 V	& 0.8819		& 0.6848					& 1.225				&& 0.227816					& 0.88883	& 0.36	& 86				& HARPS	\tnote{d}	& \cite{Dumusque2011} \\
		HD 147513 	&& G3/G5 V 	& 1.109		& 0.9279					& 1.6312				&& 1.179650					& 1.30958	& 0.26	& 282			& CORALIE\tnote{e}	& \cite{Mayor2004}\\
		HD 148156 	&& F8 V	& 1.324			& 1.26					& 2.197				&& 0.847612					& 2.12913	& 0.52	& 35				& HARPS	\tnote{d} & \cite{Naef2010}\\
		HD 171238 	&& K0 V	& 0.955			& 0.89014				& 1.58				&& 2.60901					& 2.54268	& 0.4	& 47				& CORALIE\tnote{e}	& \cite{Segransan2009}\\
		HD 187085 	&& G0 V	& 1.24			& 1.349					& 2.359				&& 0.803694					& 2.02754	& 0.47	& 94				& UCLES\tnote{h}	& \cite{Jones2006}\\
		\cmidrule(r){1-1}\cmidrule(r){3-6}\cmidrule(r){8-13}
		HD 216437 	&& G2/G3 IV	& 1.102		& 1.4115					& 2.4825				&& 2.16817					& 2.48556	& 0.319	& 67.7			& UCLES\tnote{h}	& \cite{Jones2002a}\\
		HD 131664 	&& G3 V	& 1.122			& 1.165					& 2.047				&& 18.3282					& 3.17098	& 0.638	& 149.7			& HARPS\tnote{d}	& \cite{Mayor2009}\\
		HD 132406 	&& G0 V	& 0.848			& 1.34					& 2.406				&& 5.60495					& 1.98227	& 0.34	& 214			& ELODIE\tnote{c}	& \cite{DaSilva2007}\\
		HD 141937 	&& G2/G3 V & 1.13		& 0.9993					& 1.755				&& 9.4752					& 1.50087	& 0.41	& 187.72			& CORALIE\tnote{e}	& \cite{The2002}\\
		HD 16175 	&& G0	& 1.15			& 1.749					& 3.069				&& 4.37946					& 2.1185		& 0.6	& 222			& Hamilton\tnote{g}& \cite{Peek2009} \\
		HD 190228 	&& G5 IV	& 0.962			& 2.014					& 3.574				&& 5.94193					& 2.60478	& 0.531	& 101.2			& ELODIE\tnote{c}	& \cite{Perrier2003a}\\
		HD 2039 		&& G2/G3 IV/V	& 1.16	& 1.399					& 2.453				&& 5.92499					& 2.19755	& 0.715	& 344.1			& UCLES\tnote{h}	& \cite{Tinney2003} \\
		HD 22781 	&& K0 V	& 0.83			& 0.58511				& 1.053				&& 13.8403					& 1.16847	& 0.8191& 315.92			& SOPHIE\tnote{i}	& \cite{Diaz2015} \\
		HD 45350 	&& G5 V	& 0.999			& 1.176					& 2.081				&& 1.83614					& 1.94413	& 0.778	& 343.4			& HIRES\tnote{a}& \cite{Marcy2005}\\
		HD 50554 	&& F8 V	& 1.18			& 1.103					& 1.9319				&& 4.39876					& 2.26097	& 0.444	& 7.4			& HIRES\tnote{a}, Hamilton\tnote{g} 	& \cite{Fischer2002}\\
		HD 86264 	&& F7 V	& 1.4			& 1.866					& 3.245				&& 6.62738					& 2.84117	& 0.7	& 306			& Hamilton\tnote{g}& \cite{Fischer2009}\\
    		\cmidrule(r){1-1}\cmidrule(r){3-6}\cmidrule(r){8-13}
  	\end{tabular*}}
  	\begin{tablenotes}[flushleft]                                                                                                                                                                                                                                                                                                                                                                                                                                                                                                                                                                                                                                                                                                                                                                      		\item[$\dagger$] $I$, $\Omega$ and $M$ were 0.0\degree\ for all systems simulated.
  		\item[*] Detection reference from the Exoplanet Orbit Database \citep{Han2014a}.
  		\item[a] HIRES (High Resolution Echelle Spectrometer) at Keck Observatory.
  		\item[b] HDS (High Dispersion Spectrograph) at the Subaru Telescope.
  		\item[c] ELODIE echelle spectrograph at the Haute-Provence Observatory.
  		\item[d] HARPS (High Accuracy Radial velocity Planet Searcher) speectrograph at La Silla Observatory.
  		\item[e] CORALIE echelle spectrograph at La Silla Observatory.
  		\item[f] SARG high resolution spectrograph at the Telescopio Nazionale Galileo (TNG)
  		\item[g] Hamilton echelle spectrograph at Lick Observatory.
  		\item[h] UCLES (University College London Echelle Spectrograph) at the Anglo-Australian Telescope.
  		\item[i] SOPHIE echelle spectrograph at the Haute-Provence Observatory.
  	\end{tablenotes}
  	\end{threeparttable}
\end{table*}

\begin{figure*}
	\centering
	\includegraphics[width=\linewidth]{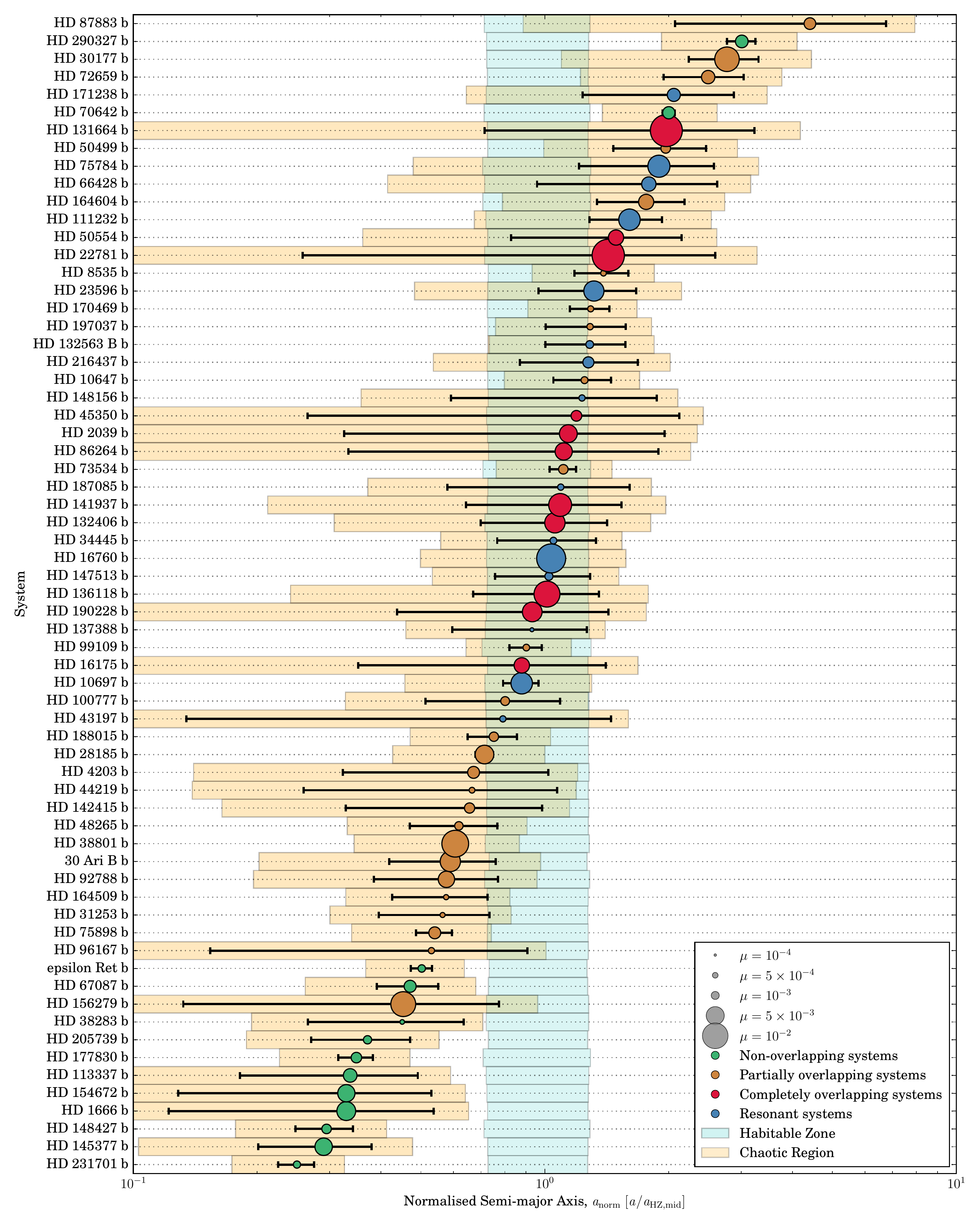}
	\caption{The planetary architectures of all the simulated single Jovian planet systems (65/93). The aqua shaded region indicates the HZ for each system as per the equations presented by \protect\cite{Kopparapu2014}, while the orange region indicates the chaotic region as per the equations presented by \protect\cite{Giuppone2013}. The size of each planet represents the mass ratio, $\mu = M_\mathrm{planet}/M_{\star}$, of the system. The error bars indicate the apsides of the orbit of the Jovian. The colour represents the system class, with the blue class representing those red systems that are found to have stable MMR zones.} 
	\label{fig:stable_arch}
\end{figure*}

We next consider the architectures of all 65 systems we simulated. Figure~\ref{fig:stable_arch} plots normalised semi-major axis ($a/a_\mathrm{HZ,mid}$) along the $x$-axis and all the systems along the $y$-axis in increasing order of their normalised semi-major axis. The normalised semi-major axis indicates where an object %(test particle or massive body) 
is located relative to the centre of the HZ ($a_\mathrm{HZ,mid}$), and can also be used to indicate the locations of the inner and outer boundaries of the HZ and chaotic region relative to $a_\mathrm{HZ,mid}$. The advantage of the normalised semi-major axis is that it allows for a clearer comparison across systems. We plot the normalised semi-major axis for the position of the Jovian with error bars that represent the periastron and apastron of its orbit (so systems with larger error bars indicate a higher eccentricity), the HZ of each system in aqua (the inner and outer edges), and the chaotic region of each Jovian in orange. Each Jovian is then plotted with a size corresponding to its mass ratio, $\mu$, and a colour corresponding to its overlapping classification (green, amber, red, or blue).

Figure \ref{fig:stable_arch} highlights some interesting architectural characteristics. All systems with Jovian planets interior to the HZ exhibit at least partial or complete regions of stability, i.e. they are either amber or green. For Jovians significantly interior to the HZ, such as hot Jupiters on orbits with radii of $\sim$0.05 au, this seems intuitive. However, it highlights a potential asymmetry on either side of the HZ. We also find that a number of the red systems with a Jovian exterior to the HZ can host  stable regions, i.e. some become blue systems. While it might be thought that a Jovian interior to the HZ would pose challenges in regards to planetary formation, several studies have suggested that there may still be sufficient material available for Terrestrial planet formation in the HZ after the inward migration of a Jovian to the inner regions of a planetary system \citep{Mandell2003a,Fogg2005a,Raymond2006,Mandell2007}. However, %despite the fact that terrestrial planet formation does not theoretically appear to be suppressed, 
observational evidence has not yet inferred the presence of nearby companions to hot Jupiters \citep{Steffen2012a}. In contrast, warm Jupiters and hot Neptunes have been found to coexist with nearby companions \citep{Huang2016}. \cite{Steffen2012a} highlights that while this may indicate the companions do not exist, there is still the possibility that they are too small to be detected or are being missed (e.g. because they have very large transit timing variations and are missed by the transit search algorithm).

\begin{figure*}
	\centering
	\includegraphics[width=\linewidth]{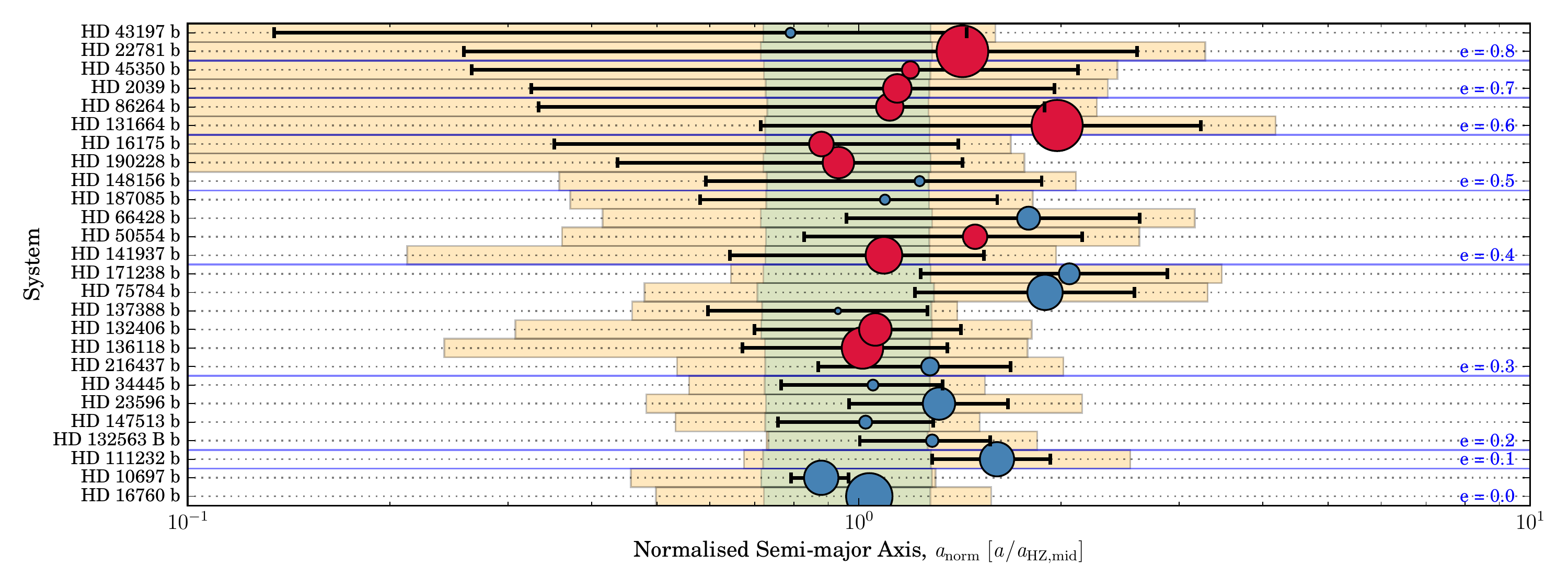}
	\caption{The planetary architectures of all the red completely overlapping and blue resonant systems (26/93) ordered vertically by eccentricity. The aqua shaded region indicates the HZ for each system as per the equations presented by \protect\cite{Kopparapu2014}, while the orange region indicates the chaotic region as per the equations presented by \protect\cite{Giuppone2013}. The size of each point represents the mass ratio, $\mu = M_\mathrm{planet}/M_{\star}$, of the system. The error bars indicate the apsides of the orbit of the Jovian. The colour represents the system class, with the blue class representing those red systems that were found to have stable MMR zones. The solid blue lines mark threshold eccentricity values.}
	\label{fig:ecc_arch}
\end{figure*}

In Figure~\ref{fig:ecc_arch} we show the system architectures in order of increasing eccentricity for all the red completely overlapping and blue resonant systems. It should be noted that the Jovian's eccentricity is determined from the best fit of the observed data and is often overestimated in radial velocity studies. A similar signature could result from a multiple planet system with lower eccentricities \citep{Anglada-Escud2010a,Anglada-Escud2010b,Wittenmyer2013}. This highlights more clearly the influence of a Jovian's eccentricity on its ability to coexist with Earth mass planets in stable MMRs. With an eccentricity greater than $\sim 0.4$, a Jovian is much less likely to host a stable MMR that could be occupied by an Earth mass planet. Those that could coexist with Earth mass planets in stable orbits in the HZ possess very low $\mu$ ratios. This result highlights that systems with a Jovian with $e \gtrsim 0.4$ near the HZ seem unlikely to be able to host a rocky planet within the HZ. This conclusion is based on the architecture of the system as it is today, and does not take into account the dynamical evolution of the system to this point. However, other studies on the dynamical evolution of multiple Jovian and massive body systems independently draw a similar conclusion \citep[e.g.][]{Carrera2016,Matsumura2016}, suggesting that massive bodies with $e \gtrsim$0.4 result from planetary scattering and that a rocky planet is unlikely to survive in the HZ of such systems.

\subsection{Searching for Exo-Earths in Single Jupiter Systems}
\begin{figure}
	\centering
	\includegraphics[width=\linewidth]{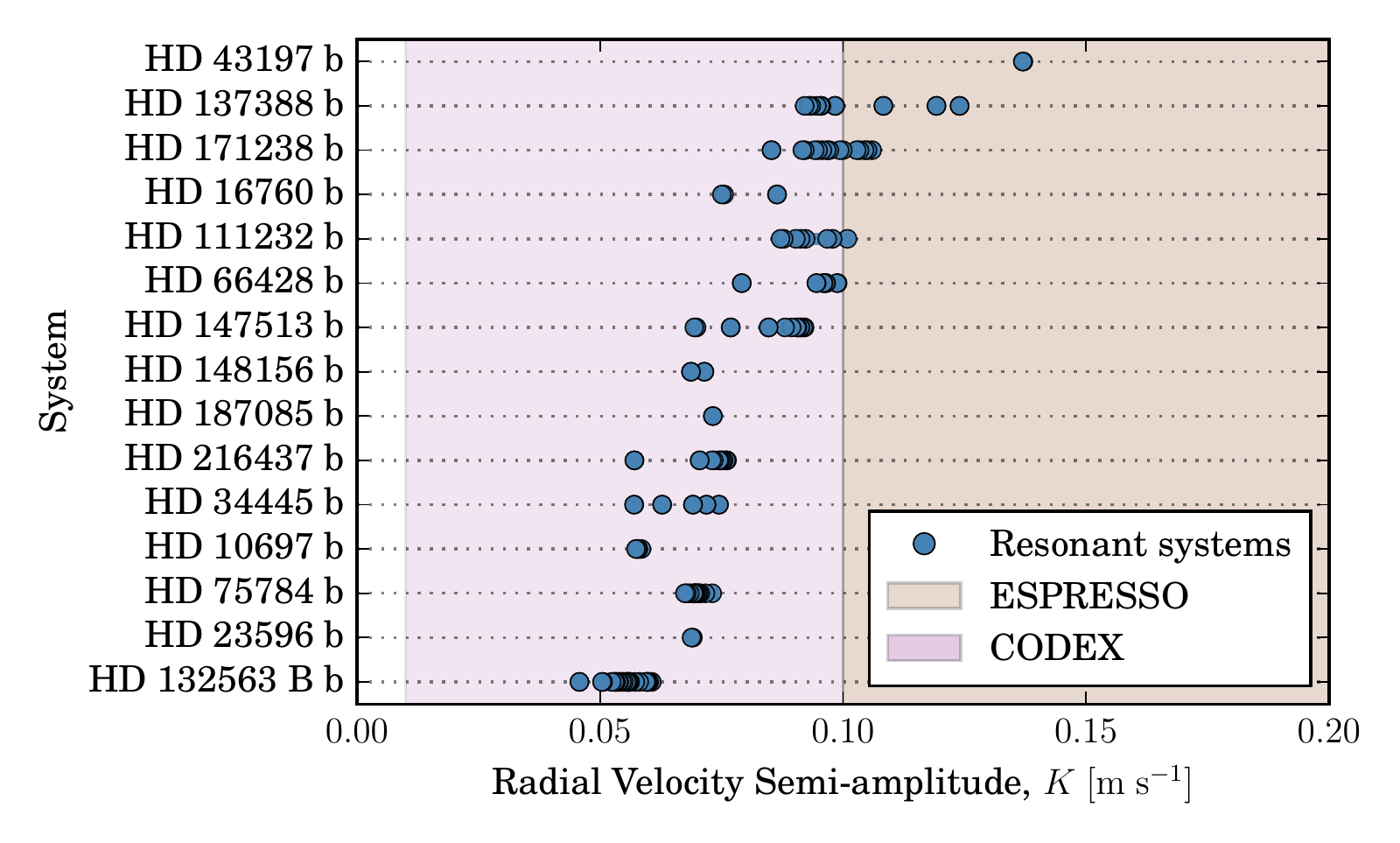}
	\caption{The semi-amplitude of Doppler wobble induced on all fifteen $\textrm{M}_\oplus$ simulated systems that were found to be capable of hosting a $1\ \textrm{M}_\oplus$ in its HZ. At stable semi-major axes positions, the semi-amplitude of the induced Doppler wobble was calculated with equation \ref{eqn:doppler_shift}. The systems are ordered by strength of the radial veloctiy semi-amplitude. The brown shaded and pink regions indicate the detection limits of the future instruments ESPRESSO ($0.1 \textrm{ m s}^{-1}$) and CODEX ($0.01 \textrm{ m s}^{-1}$) respectively.}
	\label{fig:doppler_shift_pl}
\end{figure}

\begin{figure}
	\centering
	\includegraphics[width=\linewidth]{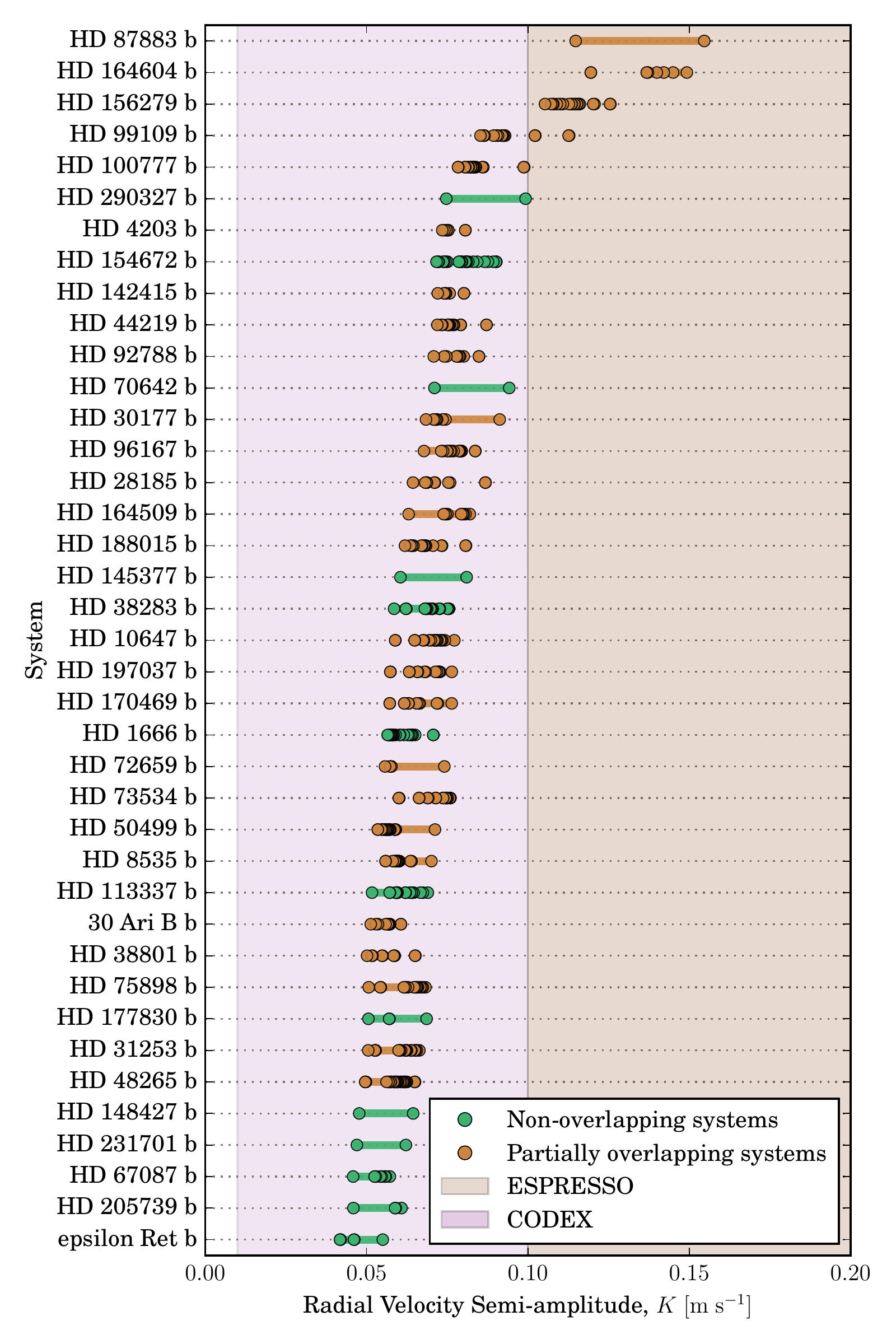}
	\caption{As per Fig.~\ref{fig:doppler_shift_pl}, but for the thirty-nine test particle simulated systems.}
	\label{fig:doppler_shift_tp}
\end{figure}

\begin{figure}
	\centering
	\includegraphics[width=\linewidth]{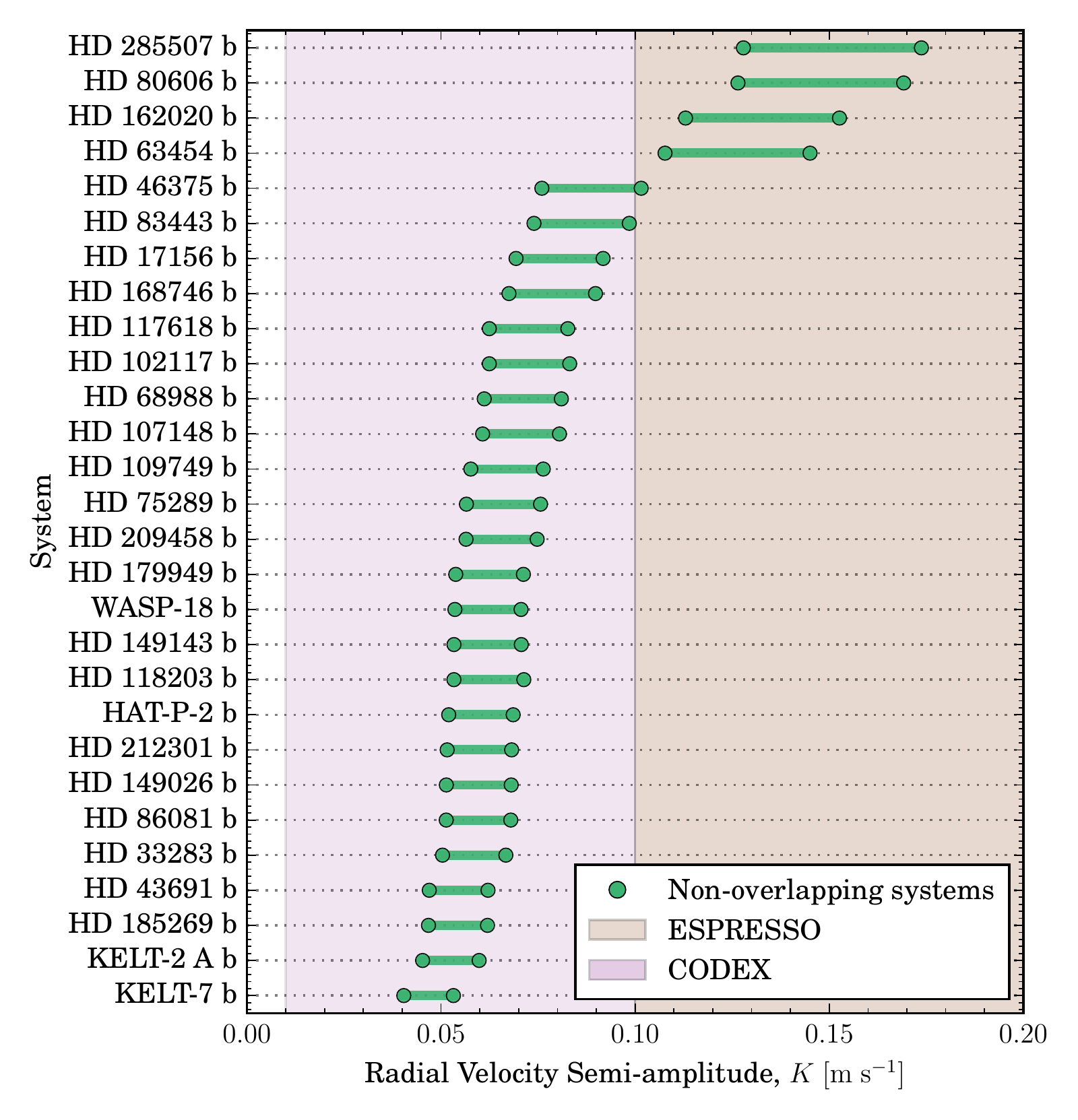}
	\caption{As per Fig.~\ref{fig:doppler_shift_pl}, but for the twenty-eight green systems that we did not simulate as discussed in Section \ref{subsec:predicting_stable_regions}.}
	\label{fig:doppler_shift_theory}
\end{figure}

Our dynamical study of 65 single Jovian systems has revealed a range of semi-major axes in the HZ of the systems that could host stable orbits.
If a $1\ \textrm{M}_\oplus$ planet were to exist in these regions, would it be detectable with current or future instruments? We can determine the magnitude of the Doppler wobble a $1\ \textrm{M}_\oplus$ planet located at these stable semi-major axes would induce on its host star. The semi-amplitude of the observable Doppler shift is given by
\begin{align} 
\label{eqn:doppler_shift}
	K &= \left( \frac{2\pi G}{T_\oplus} \right)^{\frac{1}{3}} \frac{M_\oplus\sin{I}}{\left( M_\star + M_\oplus \right)^{\frac{2}{3}}} \frac{1}{\sqrt{1-e_\oplus^2}} \,
\end{align} 
where $G$ is the gravitational constant, $M_\star$ is the mass of the host star, $I$ is the inclination of the planet's orbit (with respect to our line of sight) and $T_\oplus$, $e_\oplus$ and $M_\oplus$ are the period, mass and eccentricity of the $1\ \textrm{M}_\oplus$ planet respectively. Performing this calculation for all systems found with stable regions in the HZ provides a guide to which systems would be good targets for future observational follow-up.  
Figures \ref{fig:doppler_shift_pl} and \ref{fig:doppler_shift_tp} show the radial velocity sensitivity required to detect a $1\ \textrm{M}_\oplus$ planet at the corresponding stable semi-major axes of all of the $1\ \textrm{M}_\oplus$ simulated blue systems and the TP simulated green and amber systems respectively. Figure \ref{fig:doppler_shift_theory} similarly shows the radial velocity sensitivity required to detect a $1\ \textrm{M}_\oplus$ in the HZ of those green systems we did not simulate because they are predicted to have completely stable HZs due to the Jovian being located sufficiently far from the HZ as discussed in Section \ref{subsec:defining_stable_regions}.

Current instruments cannot resolve Doppler shifts much smaller than $1 \textrm{ m s}^{-1}$ \citep{Dumusque2012b,Swift2015}, and so $1\ \textrm{M}_\oplus$ planets in the stable regions of the HZ of these systems are currently undetectable. The sensitivities of the future instruments, such as ESPRESSO (\textbf{E}chelle \textbf{SP}ectrograph for \textbf{R}ocky \textbf{E}xoplanet- and \textbf{S}table \textbf{S}pectroscopic \textbf{O}bservations) for the Very Large Telescope and CODEX (\textbf{CO}smic \textbf{D}ynamics and \textbf{EX}o-earth experiment) for the European Extremeley Large Telescope, aim to resolve Doppler shifts to as low as $0.1 \textrm{ m s}^{-1}$ \citep{Pepe2014} and $0.01 \textrm{ m s}^{-1}$ \citep{Pasquini2010} respectively. As mentioned earlier, the resultant noise from stellar activity in Sun-like stars is not considered in our assessment but will need to be taken into account at such low detection limits to avoid false positives \citep{Robertson2014}.

We overlay the radial velocity sensitivities of ESPRESSO and CODEX on Figures \ref{fig:doppler_shift_pl}, \ref{fig:doppler_shift_tp} and \ref{fig:doppler_shift_theory} to demonstrate the detection limit of both of these future instruments (the coffee region representing ESPRESSO and the purple region representing CODEX). Those systems which have points or spans only in the coffee coloured region of the plots indicate that, if a stable Earth-mass planet exists in the HZ, it will be detectable by ESPRESSO. The detectability of systems for which points or spans straddle both the coffee and purple coloured zones of the plots is uncertain, since we cannot further constrain the location of a stable Earth-mass planet in the HZ of these systems. The remaining systems (those that reside only in the purple zone of the plots) will require CODEX to detect any potential Earth-mass planets.

We identify 8 systems for which a stable $1\ \textrm{M}_\oplus$ planet in the HZ is completely within ESPRESSOs detection limit, (i.e., those systems in Figures \ref{fig:doppler_shift_pl}, \ref{fig:doppler_shift_tp} and \ref{fig:doppler_shift_theory} for which the points or spans are only in the coffee region of the plots) suggesting they would be good candidates for future observational follow up. These include one system identified via the $1\ \textrm{M}_\oplus$ planet simulations (HD 43197; Fig.~\ref{fig:doppler_shift_pl}), 3 via the test particle simulations (HD 87883, HD 164604 and HD 156279; Fig.~\ref{fig:doppler_shift_tp}) and 4 via the crossing orbits criterion presented by \cite{Giuppone2013} (HD 285507, HD 80606, HD 162020 and HD 63454; Fig.~\ref{fig:doppler_shift_theory}). These systems should be a priority for ESPRESSO. We also identify 5 additional systems for which the points or spans straddle both the coffee and purple coloured zones. These systems should be a second priority in ESPRESSO target lists. These systems include three identified via the $1\ \textrm{M}_\oplus$ planet simulations (HD 137388, HD 171238 and HD 111232; Fig.~\ref{fig:doppler_shift_pl}), one from test particle simulations (HD 99109; Fig.~\ref{fig:doppler_shift_tp}) and one from the crossing orbits criterion (HD 46375; Fig.~\ref{fig:doppler_shift_theory}). It should be emphasised that the induced Doppler shifts are all for $1\ \textrm{M}_\oplus$ planets and so more massive planets could still be found within ESPRESSO's detection limits. CODEX will reach a low enough detection limit that all of the $1\ \textrm{M}_\oplus$ planets in the stable regions of each system's HZ would be detectable, if they exist.

\section{Conclusions}
\label{sec:summary}
We have taken a systematic approach to investigate all 93 single Jovian planet systems in the CELESTA database in order to identify promising candidates for future observational follow up and to better identify the properties of planetary architecture in those systems that could harbour an Earth-mass planet in their HZ. As a 3-body system, the dynamics of star-Jovian-Terrestrial systems are unsolvable analytically, and so it is difficult to predict in which systems Jovian and Terrestrial planets can coexist. We first use an analytic classification scheme to remove systems with completely stable HZs for which we are unable to further constrain the location of stable orbits from numerical studies. We then use N-body simulations of the evolution of massless test particles to identify regions within the HZ which could host dynamically stable orbits, and follow these with a suite of simulations with a $1\ \textrm{M}_\oplus$ body, to make prediction of which systems could harbour a Terrestrial planet in their HZs. Our key findings include:

\begin{itemize}

\item For the 67 systems in which the chaotic region of the Jovian does not overlap -- or only partially overlaps -- the HZ, there are large regions of stability in which test particles can maintain stable orbits in the HZ, and so we predict that a $1\ \textrm{M}_\oplus$ body could also do so in these systems.
\item For the 26 systems in which the chaotic region of the Jovian completely overlaps the HZ, numerical simulations show that a $1\ \textrm{M}_\oplus$ body can still maintain stable orbits in the HZ of some of these systems (15/26 systems; see Table~\ref{tab:landscape}), often as a result of the body being trapped in MMRs with the Jovian.
\item Of all the single Jovian planet systems we investigate, only 11/93 ($\sim12\%$) were incapable of hosting a small body in a stable orbit within the HZ.
\item We find that Jovians with $e \gtrsim$ 0.4 seem  unlikely to coexist with Terrestrial planets in the system's HZ. %This constraint is compounded by the fact that the dynamical evolution of a system to produce such a planetary architecture is likely the result of dynamical instabilities that would have resulted in the collision or ejection of other planets in the HZ \citep{Carrera2016}.
Systems containing Jovians with such high eccentricities are thought to be the result of dynamical instabilities that would have resulted in the collision or ejection of other planets in the HZ \citep{Carrera2016}. Given that the fitting of radial velocity data can overestimate the eccentricity of observed single Jovian planets \citep{Anglada-Escud2010a,Anglada-Escud2010b,Wittenmyer2013}, this points to the need for ongoing follow up work to better constrain the orbits of such systems.
\item Interior Jovians do not overlap as strongly with the HZ, while exterior Jovians tend to overlap with more of the HZ. However, interior Jovians raise potential problems in the formation and migration of the Jovian to such a position, and this may pose problems for Terrestrial planet formation in the HZ after such migration \citep{Armitage2003}. However, studies have shown that there can still be sufficient material for Terrestrial planet formation \citep{Mandell2003a,Fogg2005a,Raymond2006,Mandell2007}. Conversely, exterior Jovians do not pose the same formation and migration problems and have demonstrably stable MMRs that can host a $1\ \textrm{M}_\oplus$ in the HZ.
\item We identify 8 systems for which stable $1\ \textrm{M}_\oplus$ planets in the HZ are dynamically stable and could be detected with the future ESPRESSO spectrograph, if they exist: HD 43197, HD 87883, HD 164604, HD 156279, HD 285507, HD 80606, HD 162020 and HD 63454. We also identify 5 additional systems that can support $1\ \textrm{M}_\oplus$ planets in the HZ and may be detectable, but that also have stable regions within the HZ outside of the detection limit of ESPRESSO: HD 137388, HD 171238, HD 111232, HD 99109 and HD 46375.

\end{itemize}

\section*{Acknowledgements}

We wish to thank the anonymous referee for helpful comments and suggestions. MTA was supported by an Australian Postgraduate Award (APA). ET was supported by a Swinburne University Postgraduate Research Award (SUPRA). This work was performed on the gSTAR national facility at Swinburne University of Technology. gSTAR is funded by Swinburne and the Australian Government’s Education Investment Fund. This research has made use of the Exoplanet Orbit Database, the Exoplanet Data Explorer at exoplanets.org and the NASA Exoplanet Archive, which is operated by the California Institute of Technology, under contract with the National Aeronautics and Space Administration under the Exoplanet Exploration Program.

%%%%%%%%%%%%%%%%%%%%%%%%%%%%%%%%%%%%%%%%%%%%%%%%%%



\bibliographystyle{mnras}
\bibliography{project_1_final.bib} % if your bibtex file is called example.bib

\begin{thebibliography}{}
\makeatletter
\relax
\def\mn@urlcharsother{\let\do\@makeother \do\$\do\&\do\#\do\^\do\_\do\%\do\~}
\def\mn@doi{\begingroup\mn@urlcharsother \@ifnextchar [ {\mn@doi@}
  {\mn@doi@[]}}
\def\mn@doi@[#1]#2{\def\@tempa{#1}\ifx\@tempa\@empty \href
  {http://dx.doi.org/#2} {doi:#2}\else \href {http://dx.doi.org/#2} {#1}\fi
  \endgroup}
\def\mn@eprint#1#2{\mn@eprint@#1:#2::\@nil}
\def\mn@eprint@arXiv#1{\href {http://arxiv.org/abs/#1} {{\tt arXiv:#1}}}
\def\mn@eprint@dblp#1{\href {http://dblp.uni-trier.de/rec/bibtex/#1.xml}
  {dblp:#1}}
\def\mn@eprint@#1:#2:#3:#4\@nil{\def\@tempa {#1}\def\@tempb {#2}\def\@tempc
  {#3}\ifx \@tempc \@empty \let \@tempc \@tempb \let \@tempb \@tempa \fi \ifx
  \@tempb \@empty \def\@tempb {arXiv}\fi \@ifundefined
  {mn@eprint@\@tempb}{\@tempb:\@tempc}{\expandafter \expandafter \csname
  mn@eprint@\@tempb\endcsname \expandafter{\@tempc}}}

\bibitem[\protect\citeauthoryear{{Agnor} \& {Lin}}{{Agnor} \&
  {Lin}}{2012}]{Agnor2012}
{Agnor} C.~B.,  {Lin} D.~N.~C.,  2012, \mn@doi [\apj]
  {10.1088/0004-637X/745/2/143}, \href
  {http://adsabs.harvard.edu/abs/2012ApJ...745..143A} {745, 143}

\bibitem[\protect\citeauthoryear{{Anglada-Escud{\'e}} \&
  {Dawson}}{{Anglada-Escud{\'e}} \& {Dawson}}{2010}]{Anglada-Escud2010a}
{Anglada-Escud{\'e}} G.,  {Dawson} R.~I.,  2010, preprint, \href
  {http://adsabs.harvard.edu/abs/2010arXiv1011.0186A} {} (\mn@eprint {arXiv}
  {1011.0186})

\bibitem[\protect\citeauthoryear{{Anglada-Escud{\'e}}, {L{\'o}pez-Morales}  \&
  {Chambers}}{{Anglada-Escud{\'e}} et~al.}{2010}]{Anglada-Escud2010b}
{Anglada-Escud{\'e}} G.,  {L{\'o}pez-Morales} M.,   {Chambers} J.~E.,  2010,
  \mn@doi [\apj] {10.1088/0004-637X/709/1/168}, \href
  {http://adsabs.harvard.edu/abs/2010ApJ...709..168A} {709, 168}

\bibitem[\protect\citeauthoryear{{Anglada-Escud{\'e}}
  et~al.,}{{Anglada-Escud{\'e}} et~al.}{2016}]{Anglada-Escude2016}
{Anglada-Escud{\'e}} G.,  et~al., 2016, \mn@doi [\nat] {10.1038/nature19106},
  \href {http://adsabs.harvard.edu/abs/2016Natur.536..437A} {536, 437}

\bibitem[\protect\citeauthoryear{{Armitage}}{{Armitage}}{2003}]{Armitage2003}
{Armitage} P.~J.,  2003, \mn@doi [\apjl] {10.1086/346198}, \href
  {http://adsabs.harvard.edu/abs/2003ApJ...582L..47A} {582, L47}

\bibitem[\protect\citeauthoryear{{Bond}, {Lauretta}  \& {O'Brien}}{{Bond}
  et~al.}{2010}]{Carter-Bond2010}
{Bond} J.~C.,  {Lauretta} D.~S.,   {O'Brien} D.~P.,  2010, \mn@doi [\icarus]
  {10.1016/j.icarus.2009.07.037}, \href
  {http://adsabs.harvard.edu/abs/2010Icar..205..321B} {205, 321}

\bibitem[\protect\citeauthoryear{{Borucki} et~al.,}{{Borucki}
  et~al.}{2013}]{Borucki2012}
{Borucki} W.~J.,  et~al., 2013, \mn@doi [Science] {10.1126/science.1234702},
  \href {http://adsabs.harvard.edu/abs/2013Sci...340..587B} {340, 587}

\bibitem[\protect\citeauthoryear{{Brasser}, {Matsumura}, {Ida}, {Mojzsis}  \&
  {Werner}}{{Brasser} et~al.}{2016}]{Brasser2016}
{Brasser} R.,  {Matsumura} S.,  {Ida} S.,  {Mojzsis} S.~J.,   {Werner} S.~C.,
  2016, \mn@doi [\apj] {10.3847/0004-637X/821/2/75}, \href
  {http://adsabs.harvard.edu/abs/2016ApJ...821...75B} {821, 75}

\bibitem[\protect\citeauthoryear{{Butler} et~al.,}{{Butler}
  et~al.}{2006}]{Butler2006}
{Butler} R.~P.,  et~al., 2006, \mn@doi [\apj] {10.1086/504701}, \href
  {http://adsabs.harvard.edu/abs/2006ApJ...646..505B} {646, 505}

\bibitem[\protect\citeauthoryear{{Carrera}, {Davies}  \& {Johansen}}{{Carrera}
  et~al.}{2016}]{Carrera2016}
{Carrera} D.,  {Davies} M.~B.,   {Johansen} A.,  2016, \mn@doi [\mnras]
  {10.1093/mnras/stw2218}, \href
  {http://adsabs.harvard.edu/abs/2016MNRAS.tmp.1330C} {}

\bibitem[\protect\citeauthoryear{{Carter-Bond}, {O'Brien}  \&
  {Raymond}}{{Carter-Bond} et~al.}{2012a}]{Carter-Bondb2012}
{Carter-Bond} J.~C.,  {O'Brien} D.~P.,   {Raymond} S.~N.,  2012a, Meteoritics
  and Planetary Science Supplement, \href
  {http://adsabs.harvard.edu/abs/2012M%26PSA..75.5009C} {75, 5009}

\bibitem[\protect\citeauthoryear{{Carter-Bond}, {O'Brien}  \&
  {Raymond}}{{Carter-Bond} et~al.}{2012b}]{Carter-Bonda2012}
{Carter-Bond} J.~C.,  {O'Brien} D.~P.,   {Raymond} S.~N.,  2012b, \mn@doi
  [\apj] {10.1088/0004-637X/760/1/44}, \href
  {http://adsabs.harvard.edu/abs/2012ApJ...760...44C} {760, 44}

\bibitem[\protect\citeauthoryear{{Chandler}, {McDonald}  \& {Kane}}{{Chandler}
  et~al.}{2016}]{Chandler2015}
{Chandler} C.~O.,  {McDonald} I.,   {Kane} S.~R.,  2016, \mn@doi [\aj]
  {10.3847/0004-6256/151/3/59}, \href
  {http://adsabs.harvard.edu/abs/2016AJ....151...59C} {151, 59}

\bibitem[\protect\citeauthoryear{{Charbonneau}, {Brown}, {Latham}  \&
  {Mayor}}{{Charbonneau} et~al.}{2000}]{Charbonneau2000}
{Charbonneau} D.,  {Brown} T.~M.,  {Latham} D.~W.,   {Mayor} M.,  2000, \mn@doi
  [\apjl] {10.1086/312457}, \href
  {http://adsabs.harvard.edu/abs/2000ApJ...529L..45C} {529, L45}

\bibitem[\protect\citeauthoryear{{Deienno}, {Gomes}, {Walsh}, {Morbidelli}  \&
  {Nesvorn{\'y}}}{{Deienno} et~al.}{2016}]{Deienno2016}
{Deienno} R.,  {Gomes} R.~S.,  {Walsh} K.~J.,  {Morbidelli} A.,
  {Nesvorn{\'y}} D.,  2016, \mn@doi [\icarus] {10.1016/j.icarus.2016.02.043},
  \href {http://adsabs.harvard.edu/abs/2016Icar..272..114D} {272, 114}

\bibitem[\protect\citeauthoryear{{Desidera} et~al.,}{{Desidera}
  et~al.}{2011}]{Desidera2011}
{Desidera} S.,  et~al., 2011, \mn@doi [\aap] {10.1051/0004-6361/201117191},
  \href {http://adsabs.harvard.edu/abs/2011A%26A...533A..90D} {533, A90}

\bibitem[\protect\citeauthoryear{{D{\'{\i}}az} et~al.,}{{D{\'{\i}}az}
  et~al.}{2016}]{Diaz2015}
{D{\'{\i}}az} R.~F.,  et~al., 2016, \mn@doi [\aap]
  {10.1051/0004-6361/201526729}, \href
  {http://adsabs.harvard.edu/abs/2016A%26A...585A.134D} {585, A134}

\bibitem[\protect\citeauthoryear{{Dumusque}, {Udry}, {Lovis}, {Santos}  \&
  {Monteiro}}{{Dumusque} et~al.}{2011a}]{Dumusque2010}
{Dumusque} X.,  {Udry} S.,  {Lovis} C.,  {Santos} N.~C.,   {Monteiro}
  M.~J.~P.~F.~G.,  2011a, \mn@doi [\aap] {10.1051/0004-6361/201014097}, \href
  {http://adsabs.harvard.edu/abs/2011A%26A...525A.140D} {525, A140}

\bibitem[\protect\citeauthoryear{{Dumusque} et~al.,}{{Dumusque}
  et~al.}{2011b}]{Dumusque2011}
{Dumusque} X.,  et~al., 2011b, \mn@doi [\aap] {10.1051/0004-6361/201117148},
  \href {http://adsabs.harvard.edu/abs/2011A%26A...535A..55D} {535, A55}

\bibitem[\protect\citeauthoryear{{Dumusque} et~al.,}{{Dumusque}
  et~al.}{2012}]{Dumusque2012b}
{Dumusque} X.,  et~al., 2012, \mn@doi [\nat] {10.1038/nature11572}, \href
  {http://adsabs.harvard.edu/abs/2012Natur.491..207D} {491, 207}

\bibitem[\protect\citeauthoryear{{Duncan}, {Quinn}  \& {Tremaine}}{{Duncan}
  et~al.}{1989}]{Duncan1989}
{Duncan} M.,  {Quinn} T.,   {Tremaine} S.,  1989, \mn@doi [\icarus]
  {10.1016/0019-1035(89)90047-X}, \href
  {http://adsabs.harvard.edu/abs/1989Icar...82..402D} {82, 402}

\bibitem[\protect\citeauthoryear{{Fischer}, {Marcy}, {Butler}, {Vogt}, {Walp}
  \& {Apps}}{{Fischer} et~al.}{2002}]{Fischer2002}
{Fischer} D.~A.,  {Marcy} G.~W.,  {Butler} R.~P.,  {Vogt} S.~S.,  {Walp} B.,
  {Apps} K.,  2002, \mn@doi [\pasp] {10.1086/341677}, \href
  {http://adsabs.harvard.edu/abs/2002PASP..114..529F} {114, 529}

\bibitem[\protect\citeauthoryear{{Fischer} et~al.,}{{Fischer}
  et~al.}{2009}]{Fischer2009}
{Fischer} D.,  et~al., 2009, \mn@doi [\apj] {10.1088/0004-637X/703/2/1545},
  \href {http://adsabs.harvard.edu/abs/2009ApJ...703.1545F} {703, 1545}

\bibitem[\protect\citeauthoryear{{Fogg} \& {Nelson}}{{Fogg} \&
  {Nelson}}{2005}]{Fogg2005a}
{Fogg} M.~J.,  {Nelson} R.~P.,  2005, \mn@doi [\aap]
  {10.1051/0004-6361:20053453}, \href
  {http://adsabs.harvard.edu/abs/2005A%26A...441..791F} {441, 791}

\bibitem[\protect\citeauthoryear{{Giguere}, {Fischer}, {Payne}, {Brewer},
  {Johnson}, {Howard}  \& {Isaacson}}{{Giguere} et~al.}{2015}]{From2015}
{Giguere} M.~J.,  {Fischer} D.~A.,  {Payne} M.~J.,  {Brewer} J.~M.,  {Johnson}
  J.~A.,  {Howard} A.~W.,   {Isaacson} H.~T.,  2015, \mn@doi [\apj]
  {10.1088/0004-637X/799/1/89}, \href
  {http://adsabs.harvard.edu/abs/2015ApJ...799...89G} {799, 89}

\bibitem[\protect\citeauthoryear{{Gillon} et~al.,}{{Gillon}
  et~al.}{2017}]{Gillon2017}
{Gillon} M.,  et~al., 2017, \mn@doi [\nat] {10.1038/nature21360}, \href
  {http://adsabs.harvard.edu/abs/2017Natur.542..456G} {542, 456}

\bibitem[\protect\citeauthoryear{{Giuppone}, {Morais}  \& {Correia}}{{Giuppone}
  et~al.}{2013}]{Giuppone2013}
{Giuppone} C.~A.,  {Morais} M.~H.~M.,   {Correia} A.~C.~M.,  2013, \mn@doi
  [\mnras] {10.1093/mnras/stt1831}, \href
  {http://adsabs.harvard.edu/abs/2013MNRAS.436.3547G} {436, 3547}

\bibitem[\protect\citeauthoryear{{Gomes}, {Levison}, {Tsiganis}  \&
  {Morbidelli}}{{Gomes} et~al.}{2005}]{Gomes2005}
{Gomes} R.,  {Levison} H.~F.,  {Tsiganis} K.,   {Morbidelli} A.,  2005, \mn@doi
  [\nat] {10.1038/nature03676}, \href
  {http://adsabs.harvard.edu/abs/2005Natur.435..466G} {435, 466}

\bibitem[\protect\citeauthoryear{{Grazier}}{{Grazier}}{2016}]{Grazier2016}
{Grazier} K.~R.,  2016, \mn@doi [Astrobiology] {10.1089/ast.2015.1321}, \href
  {http://adsabs.harvard.edu/abs/2016AsBio..16...23G} {16, 23}

\bibitem[\protect\citeauthoryear{{Han}, {Wang}, {Wright}, {Feng}, {Zhao},
  {Fakhouri}, {Brown}  \& {Hancock}}{{Han} et~al.}{2014}]{Han2014a}
{Han} E.,  {Wang} S.~X.,  {Wright} J.~T.,  {Feng} Y.~K.,  {Zhao} M.,
  {Fakhouri} O.,  {Brown} J.~I.,   {Hancock} C.,  2014, \mn@doi [\pasp]
  {10.1086/678447}, \href {http://adsabs.harvard.edu/abs/2014PASP..126..827H}
  {126, 827}

\bibitem[\protect\citeauthoryear{{Horner} \& {Jones}}{{Horner} \&
  {Jones}}{2008}]{Horner2008}
{Horner} J.,  {Jones} B.~W.,  2008, \mn@doi [International Journal of
  Astrobiology] {10.1017/S1473550408004187}, \href
  {http://adsabs.harvard.edu/abs/2008IJAsB...7..251H} {7, 251}

\bibitem[\protect\citeauthoryear{{Horner} \& {Jones}}{{Horner} \&
  {Jones}}{2009}]{Horner1908}
{Horner} J.,  {Jones} B.~W.,  2009, \mn@doi [International Journal of
  Astrobiology] {10.1017/S1473550408004357}, \href
  {http://adsabs.harvard.edu/abs/2009IJAsB...8...75H} {8, 75}

\bibitem[\protect\citeauthoryear{{Horner} \& {Jones}}{{Horner} \&
  {Jones}}{2010}]{Horner2010}
{Horner} J.,  {Jones} B.~W.,  2010, \mn@doi [International Journal of
  Astrobiology] {10.1017/S1473550410000261}, \href
  {http://adsabs.harvard.edu/abs/2010IJAsB...9..273H} {9, 273}

\bibitem[\protect\citeauthoryear{{Horner} \& {Jones}}{{Horner} \&
  {Jones}}{2012}]{Horner2012}
{Horner} J.,  {Jones} B.~W.,  2012, \mn@doi [International Journal of
  Astrobiology] {10.1017/S1473550412000043}, \href
  {http://adsabs.harvard.edu/abs/2012IJAsB..11..147H} {11, 147}

\bibitem[\protect\citeauthoryear{{Horner}, {Mousis}, {Petit}  \&
  {Jones}}{{Horner} et~al.}{2009}]{Horner2009}
{Horner} J.,  {Mousis} O.,  {Petit} J.-M.,   {Jones} B.~W.,  2009, \mn@doi
  [\planss] {10.1016/j.pss.2009.06.006}, \href
  {http://adsabs.harvard.edu/abs/2009P%26SS...57.1338H} {57, 1338}

\bibitem[\protect\citeauthoryear{{Horner}, {Gilmore}  \& {Waltham}}{{Horner}
  et~al.}{2015}]{Horner2013}
{Horner} J.,  {Gilmore} J.~B.,   {Waltham} D.,  2015, preprint, \href
  {http://adsabs.harvard.edu/abs/2015arXiv151106043H} {} (\mn@eprint {arXiv}
  {1511.06043})

\bibitem[\protect\citeauthoryear{{Howard} et~al.,}{{Howard}
  et~al.}{2010}]{W.Howard2010a}
{Howard} A.~W.,  et~al., 2010, \mn@doi [\apj] {10.1088/0004-637X/721/2/1467},
  \href {http://adsabs.harvard.edu/abs/2010ApJ...721.1467H} {721, 1467}

\bibitem[\protect\citeauthoryear{{Huang}, {Wu}  \& {Triaud}}{{Huang}
  et~al.}{2016}]{Huang2016}
{Huang} C.,  {Wu} Y.,   {Triaud} A.~H.~M.~J.,  2016, \mn@doi [\apj]
  {10.3847/0004-637X/825/2/98}, \href
  {http://adsabs.harvard.edu/abs/2016ApJ...825...98H} {825, 98}

\bibitem[\protect\citeauthoryear{{Izidoro}, {de Souza Torres}, {Winter}  \&
  {Haghighipour}}{{Izidoro} et~al.}{2013}]{Izidoro2013}
{Izidoro} A.,  {de Souza Torres} K.,  {Winter} O.~C.,   {Haghighipour} N.,
  2013, \mn@doi [\apj] {10.1088/0004-637X/767/1/54}, \href
  {http://adsabs.harvard.edu/abs/2013ApJ...767...54I} {767, 54}

\bibitem[\protect\citeauthoryear{{Izidoro}, {Haghighipour}, {Winter}  \&
  {Tsuchida}}{{Izidoro} et~al.}{2014}]{Izidoro2014}
{Izidoro} A.,  {Haghighipour} N.,  {Winter} O.~C.,   {Tsuchida} M.,  2014,
  \mn@doi [\apj] {10.1088/0004-637X/782/1/31}, \href
  {http://adsabs.harvard.edu/abs/2014ApJ...782...31I} {782, 31}

\bibitem[\protect\citeauthoryear{{Izidoro}, {Raymond}, {Morbidelli}  \&
  {Winter}}{{Izidoro} et~al.}{2015}]{Izidoro2015}
{Izidoro} A.,  {Raymond} S.~N.,  {Morbidelli} A.,   {Winter} O.~C.,  2015,
  \mn@doi [\mnras] {10.1093/mnras/stv1835}, \href
  {http://adsabs.harvard.edu/abs/2015MNRAS.453.3619I} {453, 3619}

\bibitem[\protect\citeauthoryear{{Izidoro}, {Raymond}, {Pierens}, {Morbidelli},
  {Winter}  \& {Nesvorny`}}{{Izidoro} et~al.}{2016}]{Izidoro2016}
{Izidoro} A.,  {Raymond} S.~N.,  {Pierens} A.,  {Morbidelli} A.,  {Winter}
  O.~C.,   {Nesvorny`} D.,  2016, \mn@doi [\apj] {10.3847/1538-4357/833/1/40},
  \href {http://adsabs.harvard.edu/abs/2016ApJ...833...40I} {833, 40}

\bibitem[\protect\citeauthoryear{{Jenkins} et~al.,}{{Jenkins}
  et~al.}{2015}]{Jenkins2015}
{Jenkins} J.~M.,  et~al., 2015, \mn@doi [\aj] {10.1088/0004-6256/150/2/56},
  \href {http://adsabs.harvard.edu/abs/2015AJ....150...56J} {150, 56}

\bibitem[\protect\citeauthoryear{{Jones} \& {Sleep}}{{Jones} \&
  {Sleep}}{2002}]{Jones2002}
{Jones} B.~W.,  {Sleep} P.~N.,  2002, \mn@doi [\aap]
  {10.1051/0004-6361:20021138}, \href
  {http://adsabs.harvard.edu/abs/2002A%26A...393.1015J} {393, 1015}

\bibitem[\protect\citeauthoryear{{Jones} \& {Sleep}}{{Jones} \&
  {Sleep}}{2010}]{Jones2010}
{Jones} B.~W.,  {Sleep} P.~N.,  2010, \mn@doi [\mnras]
  {10.1111/j.1365-2966.2010.16978.x}, \href
  {http://adsabs.harvard.edu/abs/2010MNRAS.407.1259J} {407, 1259}

\bibitem[\protect\citeauthoryear{{Jones}, {Sleep}  \& {Chambers}}{{Jones}
  et~al.}{2001}]{Jones2001}
{Jones} B.~W.,  {Sleep} P.~N.,   {Chambers} J.~E.,  2001, \mn@doi [\aap]
  {10.1051/0004-6361:20000078}, \href
  {http://adsabs.harvard.edu/abs/2001A%26A...366..254J} {366, 254}

\bibitem[\protect\citeauthoryear{{Jones}, {Paul Butler}, {Marcy}, {Tinney},
  {Penny}, {McCarthy}  \& {Carter}}{{Jones} et~al.}{2002}]{Jones2002a}
{Jones} H.~R.~A.,  {Paul Butler} R.,  {Marcy} G.~W.,  {Tinney} C.~G.,  {Penny}
  A.~J.,  {McCarthy} C.,   {Carter} B.~D.,  2002, \mn@doi [\mnras]
  {10.1046/j.1365-8711.2002.05787.x}, \href
  {http://adsabs.harvard.edu/abs/2002MNRAS.337.1170J} {337, 1170}

\bibitem[\protect\citeauthoryear{{Jones}, {Underwood}  \& {Sleep}}{{Jones}
  et~al.}{2005}]{Jones2005}
{Jones} B.~W.,  {Underwood} D.~R.,   {Sleep} P.~N.,  2005, \mn@doi [\apj]
  {10.1086/428108}, \href {http://adsabs.harvard.edu/abs/2005ApJ...622.1091J}
  {622, 1091}

\bibitem[\protect\citeauthoryear{{Jones}, {Butler}, {Tinney}, {Marcy},
  {Carter}, {Penny}, {McCarthy}  \& {Bailey}}{{Jones} et~al.}{2006}]{Jones2006}
{Jones} H.~R.~A.,  {Butler} R.~P.,  {Tinney} C.~G.,  {Marcy} G.~W.,  {Carter}
  B.~D.,  {Penny} A.~J.,  {McCarthy} C.,   {Bailey} J.,  2006, \mn@doi [\mnras]
  {10.1111/j.1365-2966.2006.10298.x}, \href
  {http://adsabs.harvard.edu/abs/2006MNRAS.369..249J} {369, 249}

\bibitem[\protect\citeauthoryear{{Kaib} \& {Chambers}}{{Kaib} \&
  {Chambers}}{2016}]{Kaib2016}
{Kaib} N.~A.,  {Chambers} J.~E.,  2016, \mn@doi [\mnras]
  {10.1093/mnras/stv2554}, \href
  {http://adsabs.harvard.edu/abs/2016MNRAS.455.3561K} {455, 3561}

\bibitem[\protect\citeauthoryear{{Kane}}{{Kane}}{2014}]{Kane2014}
{Kane} S.~R.,  2014, \mn@doi [\apj] {10.1088/0004-637X/782/2/111}, \href
  {http://adsabs.harvard.edu/abs/2014ApJ...782..111K} {782, 111}

\bibitem[\protect\citeauthoryear{{Kane}}{{Kane}}{2015}]{Kane2015}
{Kane} S.~R.,  2015, \mn@doi [\apjl] {10.1088/2041-8205/814/1/L9}, \href
  {http://adsabs.harvard.edu/abs/2015ApJ...814L...9K} {814, L9}

\bibitem[\protect\citeauthoryear{{Kasting}, {Whitmire}  \&
  {Reynolds}}{{Kasting} et~al.}{1993}]{Kasting1993}
{Kasting} J.~F.,  {Whitmire} D.~P.,   {Reynolds} R.~T.,  1993, \mn@doi
  [\icarus] {10.1006/icar.1993.1010}, \href
  {http://adsabs.harvard.edu/abs/1993Icar..101..108K} {101, 108}

\bibitem[\protect\citeauthoryear{{Kopparapu} et~al.,}{{Kopparapu}
  et~al.}{2013}]{Kopparapu2013}
{Kopparapu} R.~K.,  et~al., 2013, \mn@doi [\apj] {10.1088/0004-637X/770/1/82},
  \href {http://adsabs.harvard.edu/abs/2013ApJ...770...82K} {770, 82}

\bibitem[\protect\citeauthoryear{{Kopparapu}, {Ramirez}, {SchottelKotte},
  {Kasting}, {Domagal-Goldman}  \& {Eymet}}{{Kopparapu}
  et~al.}{2014}]{Kopparapu2014}
{Kopparapu} R.~K.,  {Ramirez} R.~M.,  {SchottelKotte} J.,  {Kasting} J.~F.,
  {Domagal-Goldman} S.,   {Eymet} V.,  2014, \mn@doi [\apjl]
  {10.1088/2041-8205/787/2/L29}, \href
  {http://adsabs.harvard.edu/abs/2014ApJ...787L..29K} {787, L29}

\bibitem[\protect\citeauthoryear{{Levison} \& {Duncan}}{{Levison} \&
  {Duncan}}{1994}]{Levison1994}
{Levison} H.~F.,  {Duncan} M.~J.,  1994, \mn@doi [\icarus]
  {10.1006/icar.1994.1039}, \href
  {http://adsabs.harvard.edu/abs/1994Icar..108...18L} {108, 18}

\bibitem[\protect\citeauthoryear{{Levison} \& {Duncan}}{{Levison} \&
  {Duncan}}{2000}]{Levison2000}
{Levison} H.~F.,  {Duncan} M.~J.,  2000, \mn@doi [\aj] {10.1086/301553}, \href
  {http://adsabs.harvard.edu/abs/2000AJ....120.2117L} {120, 2117}

\bibitem[\protect\citeauthoryear{{Levison}, {Kretke}, {Walsh}  \&
  {Bottke}}{{Levison} et~al.}{2015}]{Levison2015}
{Levison} H.~F.,  {Kretke} K.~A.,  {Walsh} K.~J.,   {Bottke} W.~F.,  2015,
  \mn@doi [Proceedings of the National Academy of Science]
  {10.1073/pnas.1513364112}, \href
  {http://adsabs.harvard.edu/abs/2015PNAS..11214180L} {112, 14180}

\bibitem[\protect\citeauthoryear{{Mandell} \& {Sigurdsson}}{{Mandell} \&
  {Sigurdsson}}{2003}]{Mandell2003a}
{Mandell} A.~M.,  {Sigurdsson} S.,  2003, \mn@doi [\apjl] {10.1086/381245},
  \href {http://adsabs.harvard.edu/abs/2003ApJ...599L.111M} {599, L111}

\bibitem[\protect\citeauthoryear{{Mandell}, {Raymond}  \&
  {Sigurdsson}}{{Mandell} et~al.}{2007}]{Mandell2007}
{Mandell} A.~M.,  {Raymond} S.~N.,   {Sigurdsson} S.,  2007, \mn@doi [\apj]
  {10.1086/512759}, \href {http://adsabs.harvard.edu/abs/2007ApJ...660..823M}
  {660, 823}

\bibitem[\protect\citeauthoryear{{Marcy}, {Butler}, {Vogt}, {Fischer}, {Henry},
  {Laughlin}, {Wright}  \& {Johnson}}{{Marcy} et~al.}{2005}]{Marcy2005}
{Marcy} G.~W.,  {Butler} R.~P.,  {Vogt} S.~S.,  {Fischer} D.~A.,  {Henry}
  G.~W.,  {Laughlin} G.,  {Wright} J.~T.,   {Johnson} J.~A.,  2005, \mn@doi
  [\apj] {10.1086/426384}, \href
  {http://adsabs.harvard.edu/abs/2005ApJ...619..570M} {619, 570}

\bibitem[\protect\citeauthoryear{{Martin} \& {Livio}}{{Martin} \&
  {Livio}}{2013}]{Martin2013a}
{Martin} R.~G.,  {Livio} M.,  2013, \mn@doi [\mnras] {10.1093/mnrasl/sls003},
  \href {http://adsabs.harvard.edu/abs/2013MNRAS.428L..11M} {428, L11}

\bibitem[\protect\citeauthoryear{{Matsumura}, {Brasser}  \& {Ida}}{{Matsumura}
  et~al.}{2016}]{Matsumura2016}
{Matsumura} S.,  {Brasser} R.,   {Ida} S.,  2016, \mn@doi [\apj]
  {10.3847/0004-637X/818/1/15}, \href
  {http://adsabs.harvard.edu/abs/2016ApJ...818...15M} {818, 15}

\bibitem[\protect\citeauthoryear{{Mayor} \& {Queloz}}{{Mayor} \&
  {Queloz}}{1995}]{Mayor1995}
{Mayor} M.,  {Queloz} D.,  1995, \mn@doi [\nat] {10.1038/378355a0}, \href
  {http://adsabs.harvard.edu/abs/1995Natur.378..355M} {378, 355}

\bibitem[\protect\citeauthoryear{{Mayor}, {Udry}, {Naef}, {Pepe}, {Queloz},
  {Santos}  \& {Burnet}}{{Mayor} et~al.}{2004}]{Mayor2004}
{Mayor} M.,  {Udry} S.,  {Naef} D.,  {Pepe} F.,  {Queloz} D.,  {Santos} N.~C.,
   {Burnet} M.,  2004, \mn@doi [\aap] {10.1051/0004-6361:20034250}, \href
  {http://adsabs.harvard.edu/abs/2004A%26A...415..391M} {415, 391}

\bibitem[\protect\citeauthoryear{{Minton} \& {Malhotra}}{{Minton} \&
  {Malhotra}}{2009}]{Minton2009}
{Minton} D.~A.,  {Malhotra} R.,  2009, \mn@doi [\nat] {10.1038/nature07778},
  \href {http://adsabs.harvard.edu/abs/2009Natur.457.1109M} {457, 1109}

\bibitem[\protect\citeauthoryear{{Minton} \& {Malhotra}}{{Minton} \&
  {Malhotra}}{2011}]{Minton2011}
{Minton} D.~A.,  {Malhotra} R.,  2011, \mn@doi [\apj]
  {10.1088/0004-637X/732/1/53}, \href
  {http://adsabs.harvard.edu/abs/2011ApJ...732...53M} {732, 53}

\bibitem[\protect\citeauthoryear{{Moutou} et~al.,}{{Moutou}
  et~al.}{2009}]{Mayor2009}
{Moutou} C.,  et~al., 2009, \mn@doi [\aap] {10.1051/0004-6361:200810941}, \href
  {http://adsabs.harvard.edu/abs/2009A%26A...496..513M} {496, 513}

\bibitem[\protect\citeauthoryear{{Naef} et~al.,}{{Naef}
  et~al.}{2010}]{Naef2010}
{Naef} D.,  et~al., 2010, \mn@doi [\aap] {10.1051/0004-6361/200913616}, \href
  {http://adsabs.harvard.edu/abs/2010A%26A...523A..15N} {523, A15}

\bibitem[\protect\citeauthoryear{{O'Brien}, {Walsh}, {Morbidelli}, {Raymond}
  \& {Mandell}}{{O'Brien} et~al.}{2014}]{O'Brien2014}
{O'Brien} D.~P.,  {Walsh} K.~J.,  {Morbidelli} A.,  {Raymond} S.~N.,
  {Mandell} A.~M.,  2014, \mn@doi [\icarus] {10.1016/j.icarus.2014.05.009},
  \href {http://adsabs.harvard.edu/abs/2014Icar..239...74O} {239, 74}

\bibitem[\protect\citeauthoryear{{Pasquini}, {Cristiani}, {Garcia-Lopez},
  {Haehnelt}  \& {Mayor}}{{Pasquini} et~al.}{2010}]{Pasquini2010}
{Pasquini} L.,  {Cristiani} S.,  {Garcia-Lopez} R.,  {Haehnelt} M.,   {Mayor}
  M.,  2010, The Messenger, \href
  {http://adsabs.harvard.edu/abs/2010Msngr.140...20P} {140, 20}

\bibitem[\protect\citeauthoryear{{Peek} et~al.,}{{Peek}
  et~al.}{2009}]{Peek2009}
{Peek} K.~M.~G.,  et~al., 2009, \mn@doi [\pasp] {10.1086/599862}, \href
  {http://adsabs.harvard.edu/abs/2009PASP..121..613P} {121, 613}

\bibitem[\protect\citeauthoryear{{Pepe} et~al.,}{{Pepe}
  et~al.}{2014}]{Pepe2014}
{Pepe} F.,  et~al., 2014, preprint, \href
  {http://adsabs.harvard.edu/abs/2014arXiv1401.5918P} {} (\mn@eprint {arXiv}
  {1401.5918})

\bibitem[\protect\citeauthoryear{{Perrier}, {Sivan}, {Naef}, {Beuzit}, {Mayor},
  {Queloz}  \& {Udry}}{{Perrier} et~al.}{2003}]{Perrier2003a}
{Perrier} C.,  {Sivan} J.-P.,  {Naef} D.,  {Beuzit} J.~L.,  {Mayor} M.,
  {Queloz} D.,   {Udry} S.,  2003, \mn@doi [\aap] {10.1051/0004-6361:20031340},
  \href {http://adsabs.harvard.edu/abs/2003A%26A...410.1039P} {410, 1039}

\bibitem[\protect\citeauthoryear{{Quintana} \& {Lissauer}}{{Quintana} \&
  {Lissauer}}{2014}]{Quintana2014}
{Quintana} E.~V.,  {Lissauer} J.~J.,  2014, \mn@doi [\apj]
  {10.1088/0004-637X/786/1/33}, \href
  {http://adsabs.harvard.edu/abs/2014ApJ...786...33Q} {786, 33}

\bibitem[\protect\citeauthoryear{{Raymond} \& {Morbidelli}}{{Raymond} \&
  {Morbidelli}}{2014}]{Raymond2014}
{Raymond} S.~N.,  {Morbidelli} A.,  2014, in Complex Planetary Systems,
  Proceedings of the International Astronomical Union. pp 194--203 (\mn@eprint
  {arXiv} {1409.6340}), \mn@doi{10.1017/S1743921314008254}

\bibitem[\protect\citeauthoryear{{Raymond}, {Barnes}  \& {Kaib}}{{Raymond}
  et~al.}{2006}]{Raymond2006}
{Raymond} S.~N.,  {Barnes} R.,   {Kaib} N.~A.,  2006, \mn@doi [\apj]
  {10.1086/503594}, \href {http://adsabs.harvard.edu/abs/2006ApJ...644.1223R}
  {644, 1223}

\bibitem[\protect\citeauthoryear{{Rivera} \& {Haghighipour}}{{Rivera} \&
  {Haghighipour}}{2007}]{Rivera2007}
{Rivera} E.,  {Haghighipour} N.,  2007, \mn@doi [\mnras]
  {10.1111/j.1365-2966.2006.11172.x}, \href
  {http://adsabs.harvard.edu/abs/2007MNRAS.374..599R} {374, 599}

\bibitem[\protect\citeauthoryear{{Robertson} et~al.,}{{Robertson}
  et~al.}{2012}]{Robertson2012}
{Robertson} P.,  et~al., 2012, \mn@doi [\apj] {10.1088/0004-637X/754/1/50},
  \href {http://adsabs.harvard.edu/abs/2012ApJ...754...50R} {754, 50}

\bibitem[\protect\citeauthoryear{{Robertson}, {Mahadevan}, {Endl}  \&
  {Roy}}{{Robertson} et~al.}{2014}]{Robertson2014}
{Robertson} P.,  {Mahadevan} S.,  {Endl} M.,   {Roy} A.,  2014, \mn@doi
  [Science] {10.1126/science.1253253}, \href
  {http://adsabs.harvard.edu/abs/2014Sci...345..440R} {345, 440}

\bibitem[\protect\citeauthoryear{{Sato} et~al.,}{{Sato}
  et~al.}{2009}]{Sato2009}
{Sato} B.,  et~al., 2009, \mn@doi [\apj] {10.1088/0004-637X/703/1/671}, \href
  {http://adsabs.harvard.edu/abs/2009ApJ...703..671S} {703, 671}

\bibitem[\protect\citeauthoryear{{S{\'e}gransan} et~al.,}{{S{\'e}gransan}
  et~al.}{2010}]{Segransan2009}
{S{\'e}gransan} D.,  et~al., 2010, \mn@doi [\aap]
  {10.1051/0004-6361/200912136}, \href
  {http://adsabs.harvard.edu/abs/2010A%26A...511A..45S} {511, A45}

\bibitem[\protect\citeauthoryear{{Steffen} et~al.,}{{Steffen}
  et~al.}{2012}]{Steffen2012a}
{Steffen} J.~H.,  et~al., 2012, \mn@doi [Proceedings of the National Academy of
  Science] {10.1073/pnas.1120970109}, \href
  {http://adsabs.harvard.edu/abs/2012PNAS..109.7982S} {109, 7982}

\bibitem[\protect\citeauthoryear{{Swift} et~al.,}{{Swift}
  et~al.}{2015}]{Swift2015}
{Swift} J.~J.,  et~al., 2015, \mn@doi [Journal of Astronomical Telescopes,
  Instruments, and Systems] {10.1117/1.JATIS.1.2.027002}, \href
  {http://adsabs.harvard.edu/abs/2015JATIS...1b7002S} {1, 027002}

\bibitem[\protect\citeauthoryear{{Thilliez} \& {Maddison}}{{Thilliez} \&
  {Maddison}}{2016}]{Thilliez2016}
{Thilliez} E.,  {Maddison} S.~T.,  2016, \mn@doi [\mnras]
  {10.1093/mnras/stw079}, \href
  {http://adsabs.harvard.edu/abs/2016MNRAS.457.1690T} {457, 1690}

\bibitem[\protect\citeauthoryear{{Thilliez}, {Jouvin}, {Maddison}  \&
  {Horner}}{{Thilliez} et~al.}{2014}]{Thilliez2014}
{Thilliez} E.,  {Jouvin} L.,  {Maddison} S.~T.,   {Horner} J.,  2014, preprint,
  \href {http://adsabs.harvard.edu/abs/2014arXiv1402.2728T} {} (\mn@eprint
  {arXiv} {1402.2728})

\bibitem[\protect\citeauthoryear{{Tinney}, {Butler}, {Marcy}, {Jones}, {Penny},
  {McCarthy}, {Carter}  \& {Bond}}{{Tinney} et~al.}{2003}]{Tinney2003}
{Tinney} C.~G.,  {Butler} R.~P.,  {Marcy} G.~W.,  {Jones} H.~R.~A.,  {Penny}
  A.~J.,  {McCarthy} C.,  {Carter} B.~D.,   {Bond} J.,  2003, \mn@doi [\apj]
  {10.1086/368068}, \href {http://adsabs.harvard.edu/abs/2003ApJ...587..423T}
  {587, 423}

\bibitem[\protect\citeauthoryear{{Udry}, {Mayor}, {Naef}, {Pepe}, {Queloz},
  {Santos}  \& {Burnet}}{{Udry} et~al.}{2002}]{The2002}
{Udry} S.,  {Mayor} M.,  {Naef} D.,  {Pepe} F.,  {Queloz} D.,  {Santos} N.~C.,
   {Burnet} M.,  2002, \mn@doi [\aap] {10.1051/0004-6361:20020685}, \href
  {http://adsabs.harvard.edu/abs/2002A%26A...390..267U} {390, 267}

\bibitem[\protect\citeauthoryear{{Vogt}, {Marcy}, {Butler}  \& {Apps}}{{Vogt}
  et~al.}{2000}]{American2000}
{Vogt} S.~S.,  {Marcy} G.~W.,  {Butler} R.~P.,   {Apps} K.,  2000, \mn@doi
  [\apj] {10.1086/308981}, \href
  {http://adsabs.harvard.edu/abs/2000ApJ...536..902V} {536, 902}

\bibitem[\protect\citeauthoryear{{Vogt} et~al.,}{{Vogt}
  et~al.}{2015}]{Vogt2015}
{Vogt} S.~S.,  et~al., 2015, \mn@doi [\apj] {10.1088/0004-637X/814/1/12}, \href
  {http://adsabs.harvard.edu/abs/2015ApJ...814...12V} {814, 12}

\bibitem[\protect\citeauthoryear{{Walsh}, {Morbidelli}, {Raymond}, {O'Brien}
  \& {Mandell}}{{Walsh} et~al.}{2011}]{Walsh2011}
{Walsh} K.~J.,  {Morbidelli} A.,  {Raymond} S.~N.,  {O'Brien} D.~P.,
  {Mandell} A.~M.,  2011, \mn@doi [\nat] {10.1038/nature10201}, \href
  {http://adsabs.harvard.edu/abs/2011Natur.475..206W} {475, 206}

\bibitem[\protect\citeauthoryear{{Ward} \& {Brownlee}}{{Ward} \&
  {Brownlee}}{2000}]{Ward2000}
{Ward} P.,  {Brownlee} D.,  2000, {Rare earth : why complex life is uncommon in
  the universe}

\bibitem[\protect\citeauthoryear{{Wetherill}}{{Wetherill}}{1994}]{Wetherill1994}
{Wetherill} G.~W.,  1994, \mn@doi [\apss] {10.1007/BF00984505}, \href
  {http://adsabs.harvard.edu/abs/1994Ap%26SS.212...23W} {212, 23}

\bibitem[\protect\citeauthoryear{{Wisdom}}{{Wisdom}}{1980}]{Wisdom1980}
{Wisdom} J.,  1980, \mn@doi [\aj] {10.1086/112778}, \href
  {http://adsabs.harvard.edu/abs/1980AJ.....85.1122W} {85, 1122}

\bibitem[\protect\citeauthoryear{{Wittenmyer}, {Tinney}, {Butler}, {O'Toole},
  {Jones}, {Carter}, {Bailey}  \& {Horner}}{{Wittenmyer}
  et~al.}{2011}]{Wittenmyer2011a}
{Wittenmyer} R.~A.,  {Tinney} C.~G.,  {Butler} R.~P.,  {O'Toole} S.~J.,
  {Jones} H.~R.~A.,  {Carter} B.~D.,  {Bailey} J.,   {Horner} J.,  2011,
  \mn@doi [\apj] {10.1088/0004-637X/738/1/81}, \href
  {http://adsabs.harvard.edu/abs/2011ApJ...738...81W} {738, 81}

\bibitem[\protect\citeauthoryear{{Wittenmyer}, {Horner}  \&
  {Tinney}}{{Wittenmyer} et~al.}{2012}]{Wittenmyer2012}
{Wittenmyer} R.~A.,  {Horner} J.,   {Tinney} C.~G.,  2012, \mn@doi [\apj]
  {10.1088/0004-637X/761/2/165}, \href
  {http://adsabs.harvard.edu/abs/2012ApJ...761..165W} {761, 165}

\bibitem[\protect\citeauthoryear{{Wittenmyer} et~al.,}{{Wittenmyer}
  et~al.}{2013}]{Wittenmyer2013}
{Wittenmyer} R.~A.,  et~al., 2013, \mn@doi [\apjs] {10.1088/0067-0049/208/1/2},
  \href {http://adsabs.harvard.edu/abs/2013ApJS..208....2W} {208, 2}

\bibitem[\protect\citeauthoryear{{Wittenmyer} et~al.,}{{Wittenmyer}
  et~al.}{2016}]{Wittenmyer2016}
{Wittenmyer} R.~A.,  et~al., 2016, \mn@doi [\apj] {10.3847/0004-637X/819/1/28},
  \href {http://adsabs.harvard.edu/abs/2016ApJ...819...28W} {819, 28}

\bibitem[\protect\citeauthoryear{{Wright}, {Wittenmyer}, {Tinney}, {Bentley}
  \& {Zhao}}{{Wright} et~al.}{2016}]{Wright2015}
{Wright} D.~J.,  {Wittenmyer} R.~A.,  {Tinney} C.~G.,  {Bentley} J.~S.,
  {Zhao} J.,  2016, \mn@doi [\apjl] {10.3847/2041-8205/817/2/L20}, \href
  {http://adsabs.harvard.edu/abs/2016ApJ...817L..20W} {817, L20}

\bibitem[\protect\citeauthoryear{{da Silva} et~al.,}{{da Silva}
  et~al.}{2007}]{DaSilva2007}
{da Silva} R.,  et~al., 2007, \mn@doi [\aap] {10.1051/0004-6361:20077314},
  \href {http://adsabs.harvard.edu/abs/2007A%26A...473..323D} {473, 323}

\makeatother
\end{thebibliography}

%\begin{thebibliography}{99}
%\end{thebibliography}

%%%%%%%%%%%%%%%%%%%%%%%%%%%%%%%%%%%%%%%%%%%%%%%%%%

%%%%%%%%%%%%%%%%% APPENDICES %%%%%%%%%%%%%%%%%%%%%

 \appendix
 \section{Resonant Argument Plots}
 \begin{figure*}
 \centering
 	\begin{subfigure}{0.24\textwidth}
   		\centering
   		\includegraphics[width=\textwidth]{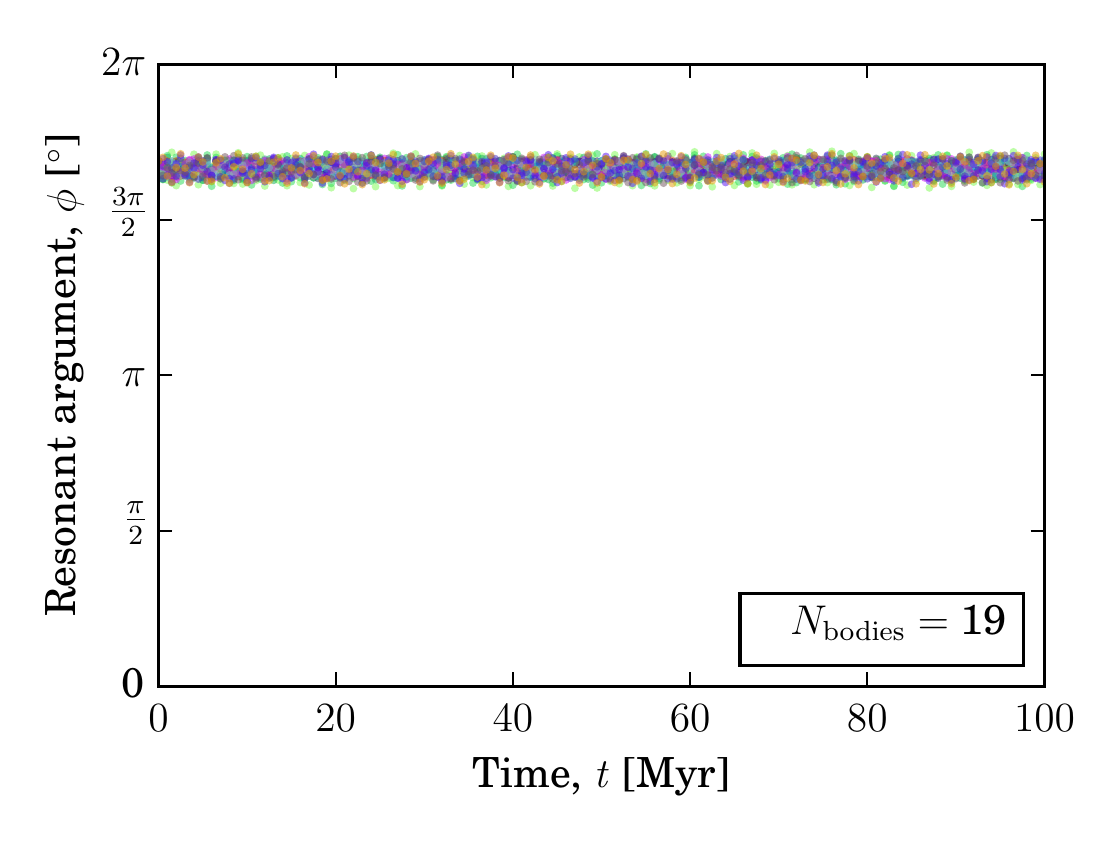}
   		\caption{HD 16760 ($1:1$)}
 	\end{subfigure}
 	\begin{subfigure}{0.24\textwidth}
   		\centering
   		\includegraphics[width=\textwidth]{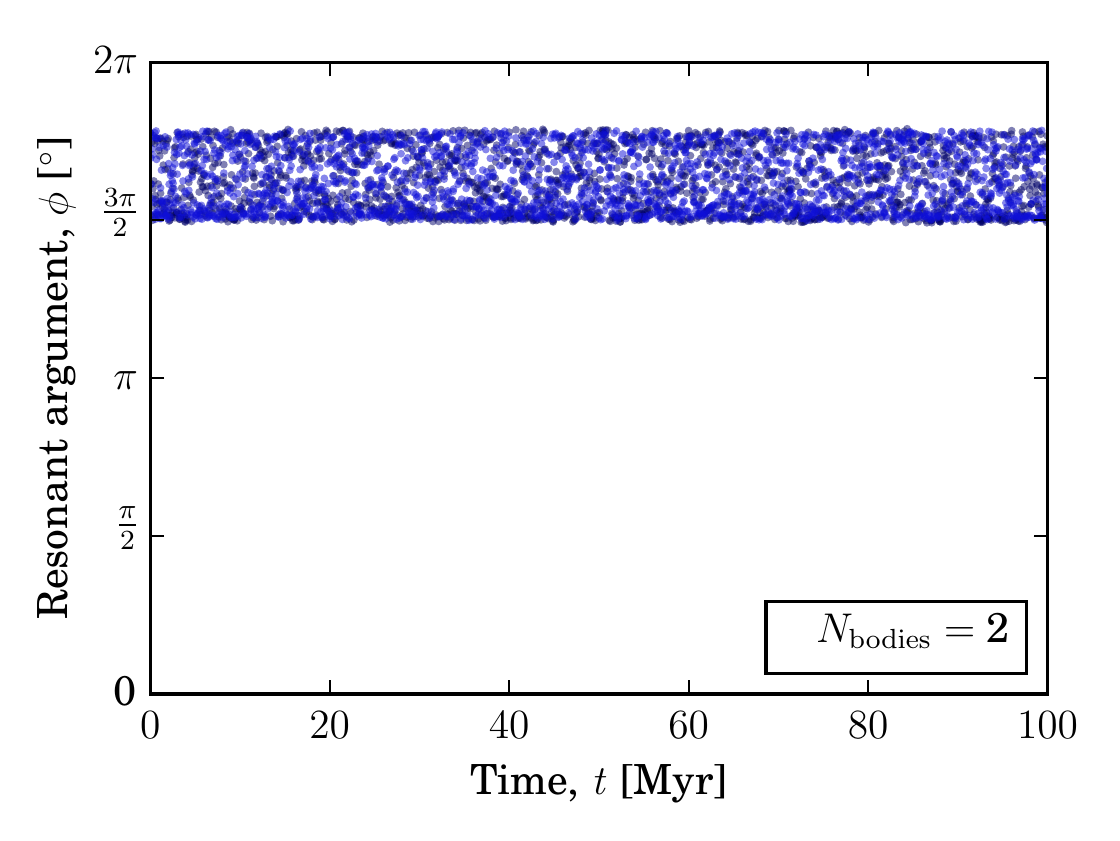}
   		\caption{HD 34445 ($1:1$)}
 	\end{subfigure}
 	\begin{subfigure}{0.24\textwidth}
   		\centering
   		\includegraphics[width=\textwidth]{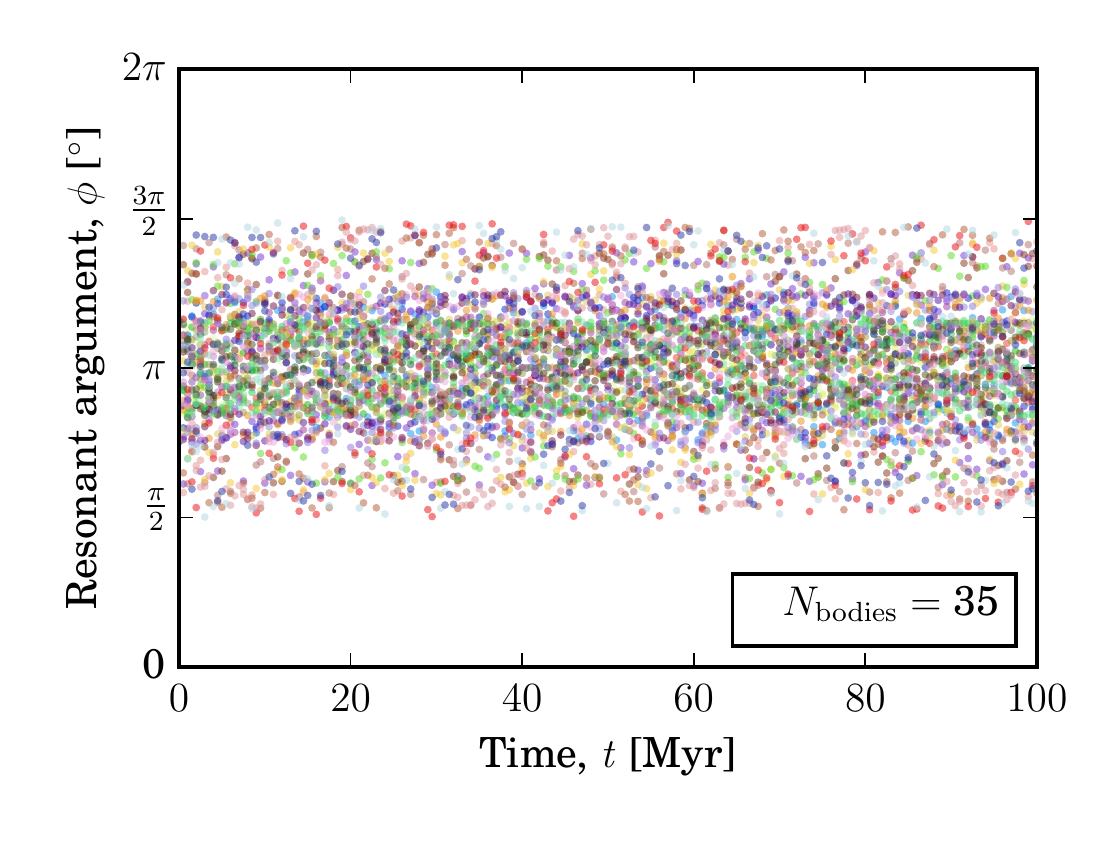}
   		\caption{HD 34445 ($3:2$)}
 	\end{subfigure}
 	\begin{subfigure}{0.24\textwidth}
   		\centering
   		\includegraphics[width=\textwidth]{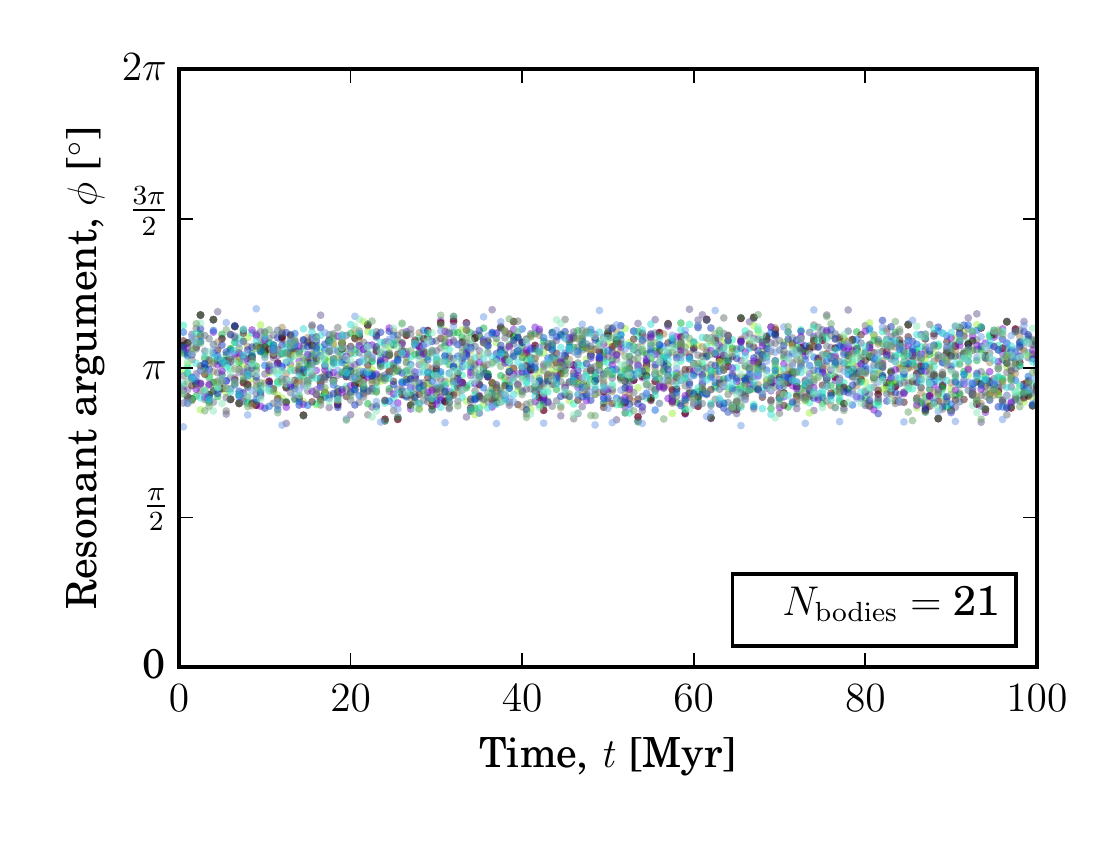}
   		\caption{HD 34445 ($4:3$)}
 	\end{subfigure}
	
 	\begin{subfigure}{0.24\textwidth}
   		\centering
   		\includegraphics[width=\textwidth]{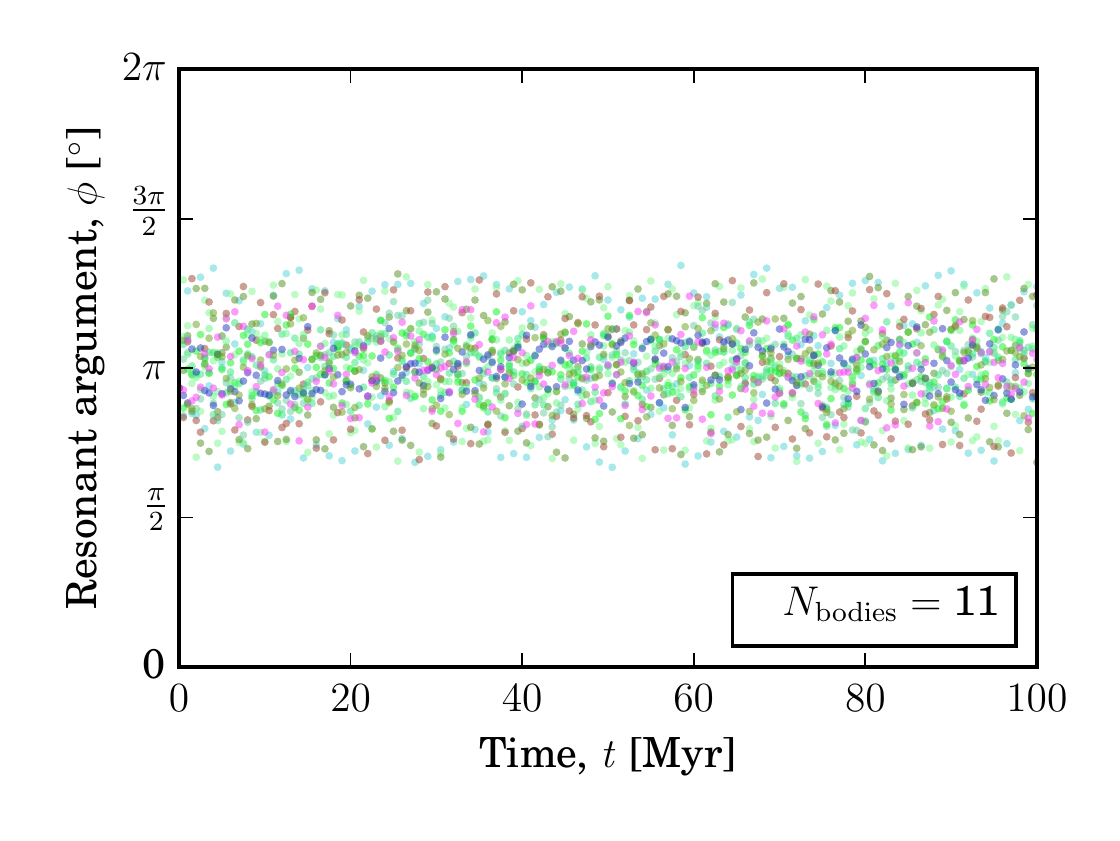}
   		\caption{HD 34445 ($5:3$)}
 	\end{subfigure}
 	\begin{subfigure}{0.24\textwidth}
   		\centering
   		\includegraphics[width=\textwidth]{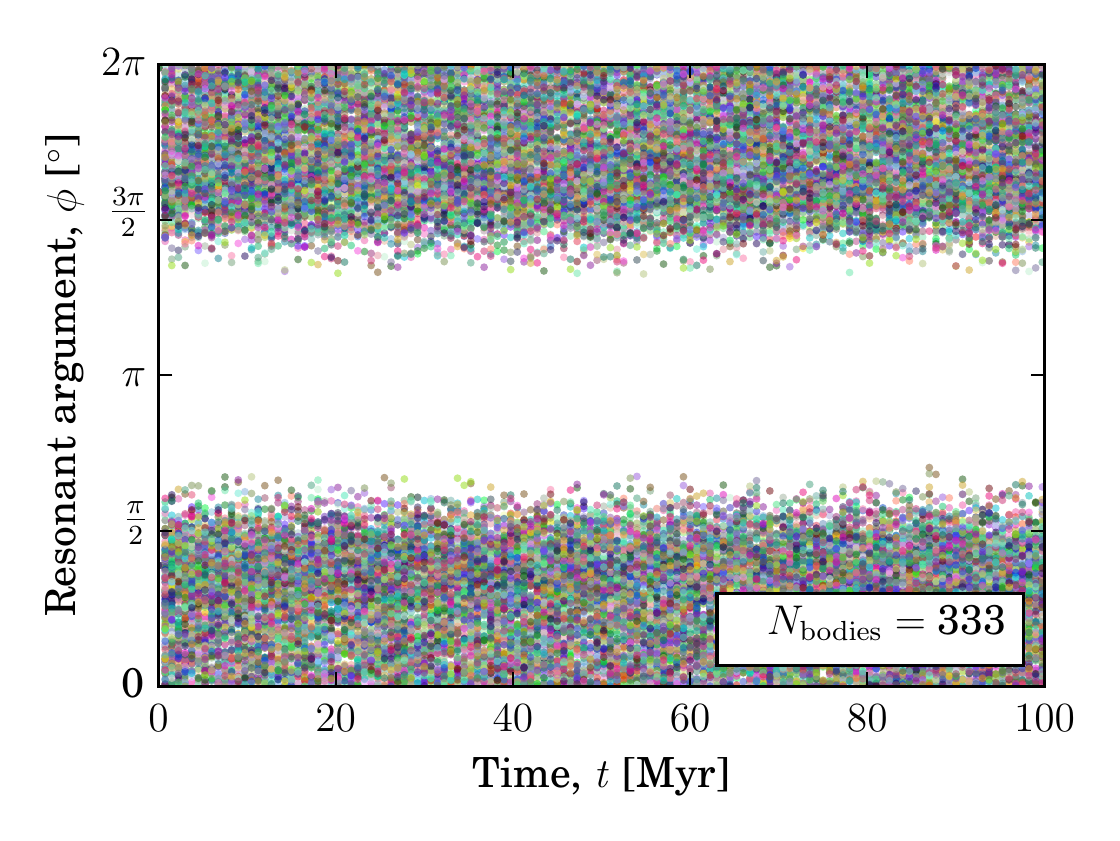}
   		\caption{HD 43197 ($1:2$)}
 	\end{subfigure}
 	\begin{subfigure}{0.24\textwidth}
   		\centering
   		\includegraphics[width=\textwidth]{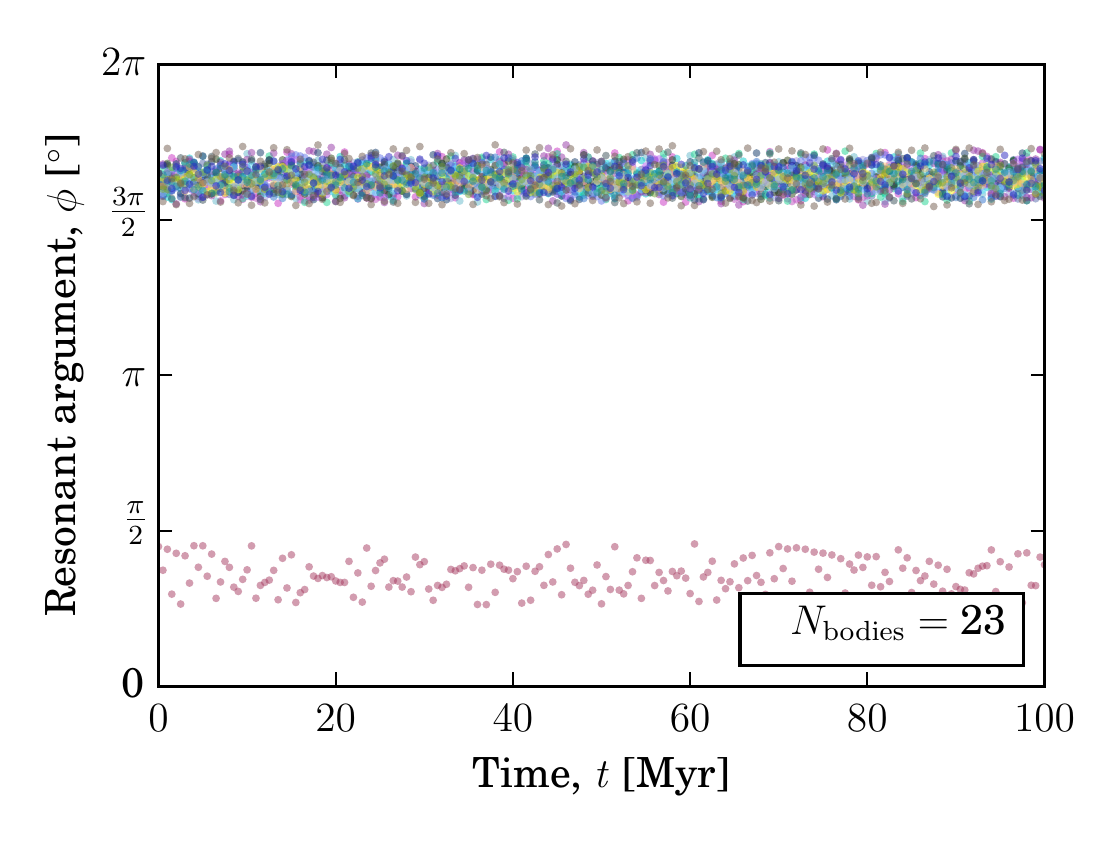}
   		\caption{HD 66428 ($2:1$)}
 	\end{subfigure}
 	\begin{subfigure}{0.24\textwidth}
   		\centering
   		\includegraphics[width=\textwidth]{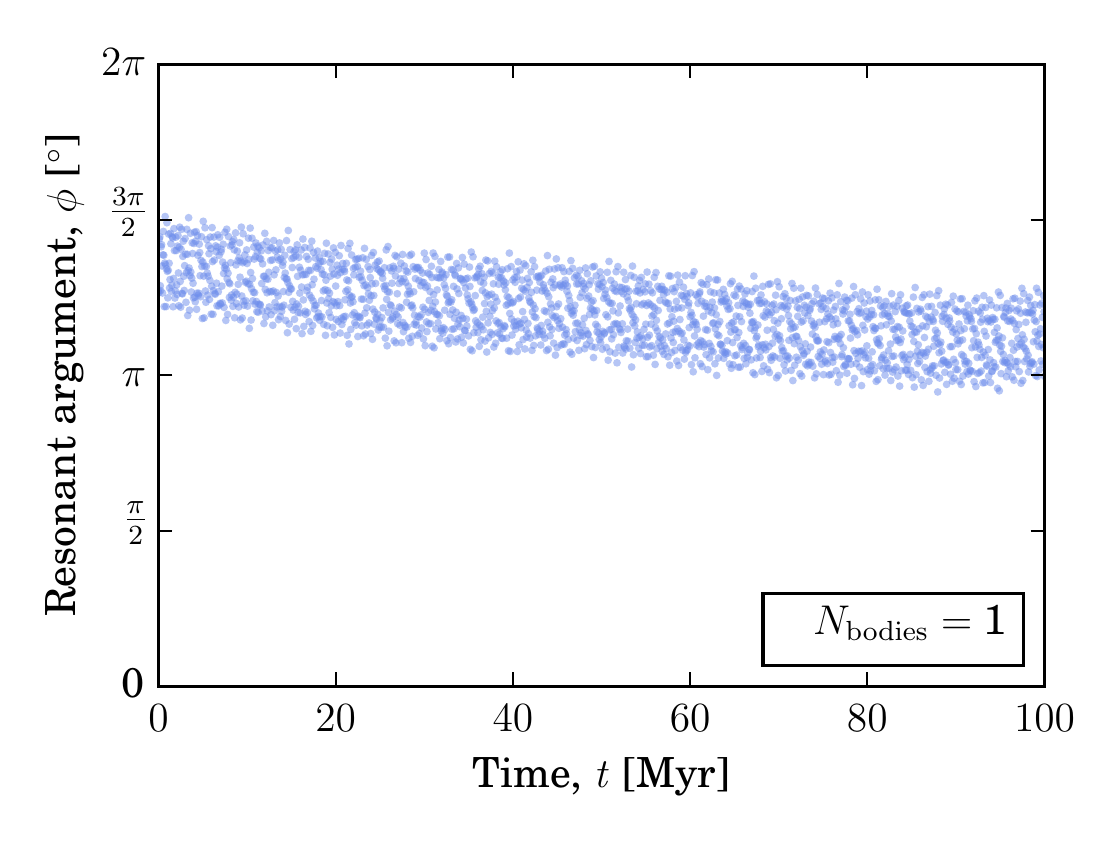}
   		\caption{HD 75784 ($3:1$)}
 	\end{subfigure}
	
 	\begin{subfigure}{0.24\textwidth}
   		\centering
   		\includegraphics[width=\textwidth]{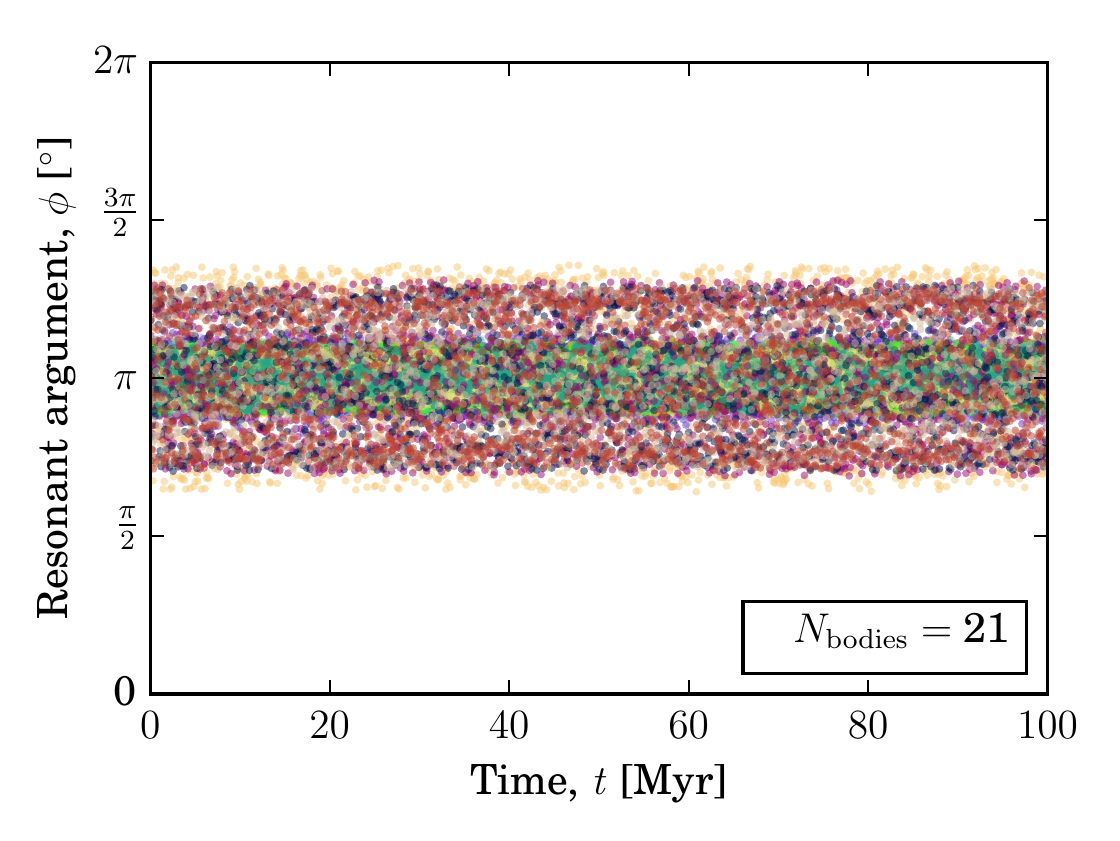}
   		\caption{HD 132563 B ($3:2$)}
 	\end{subfigure}
 		\begin{subfigure}{0.24\textwidth}
   		\centering
   		\includegraphics[width=\textwidth]{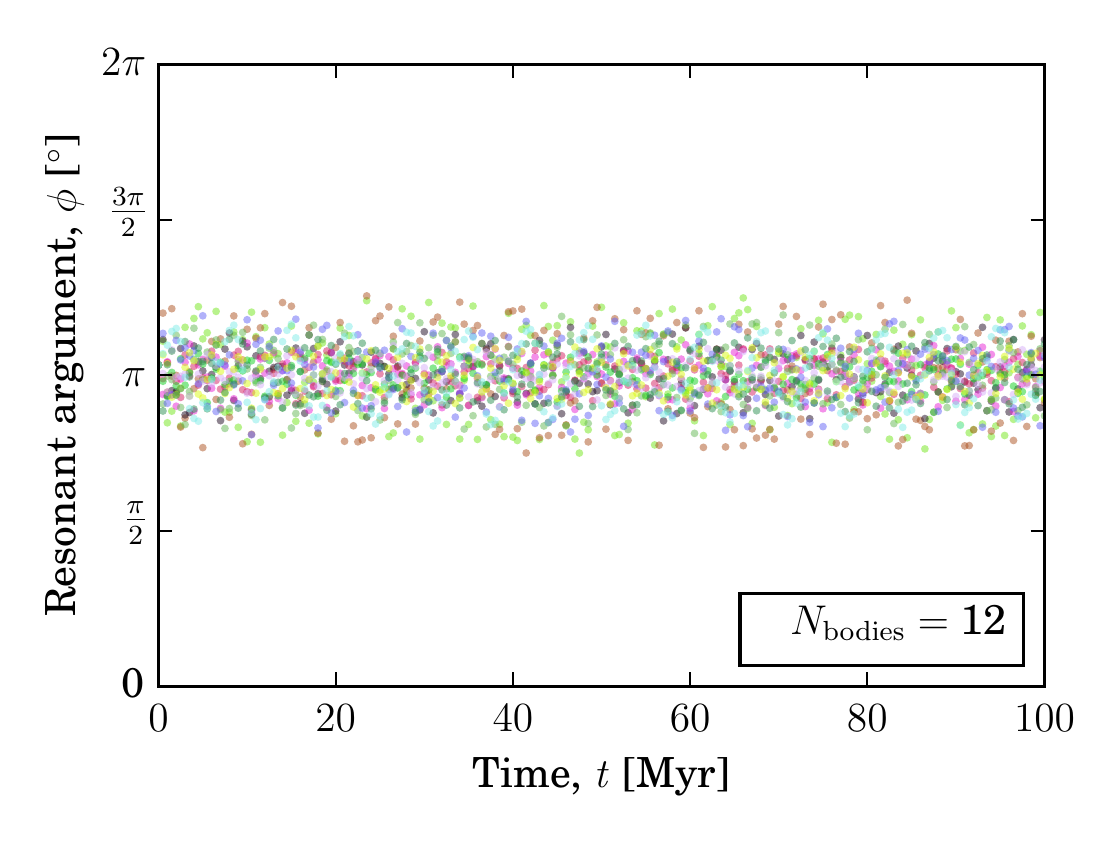}
   		\caption{HD 132563 B ($4:3$)}
 	\end{subfigure}
 	\begin{subfigure}{0.24\textwidth}
   		\centering
   		\includegraphics[width=\textwidth]{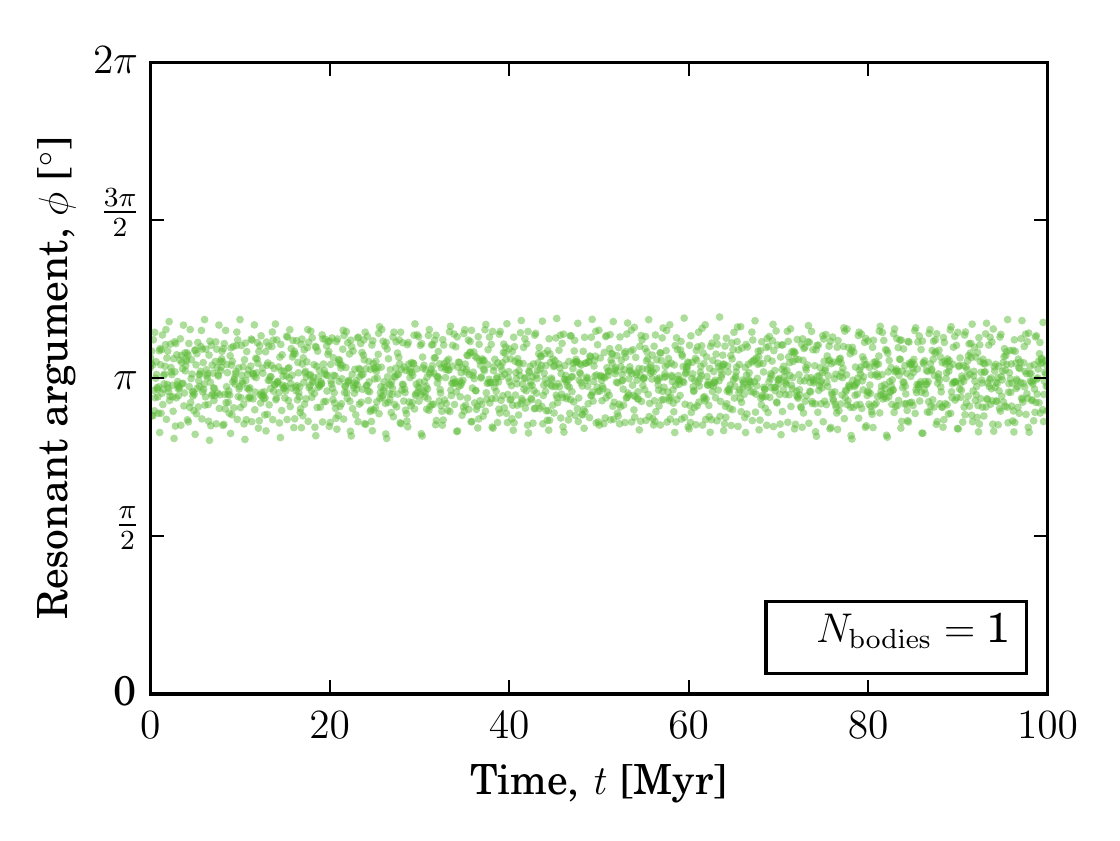}
   		\caption{HD 132563 B ($5:3$)}
 	\end{subfigure}
 	\begin{subfigure}{0.24\textwidth}
   		\centering
   		\includegraphics[width=\textwidth]{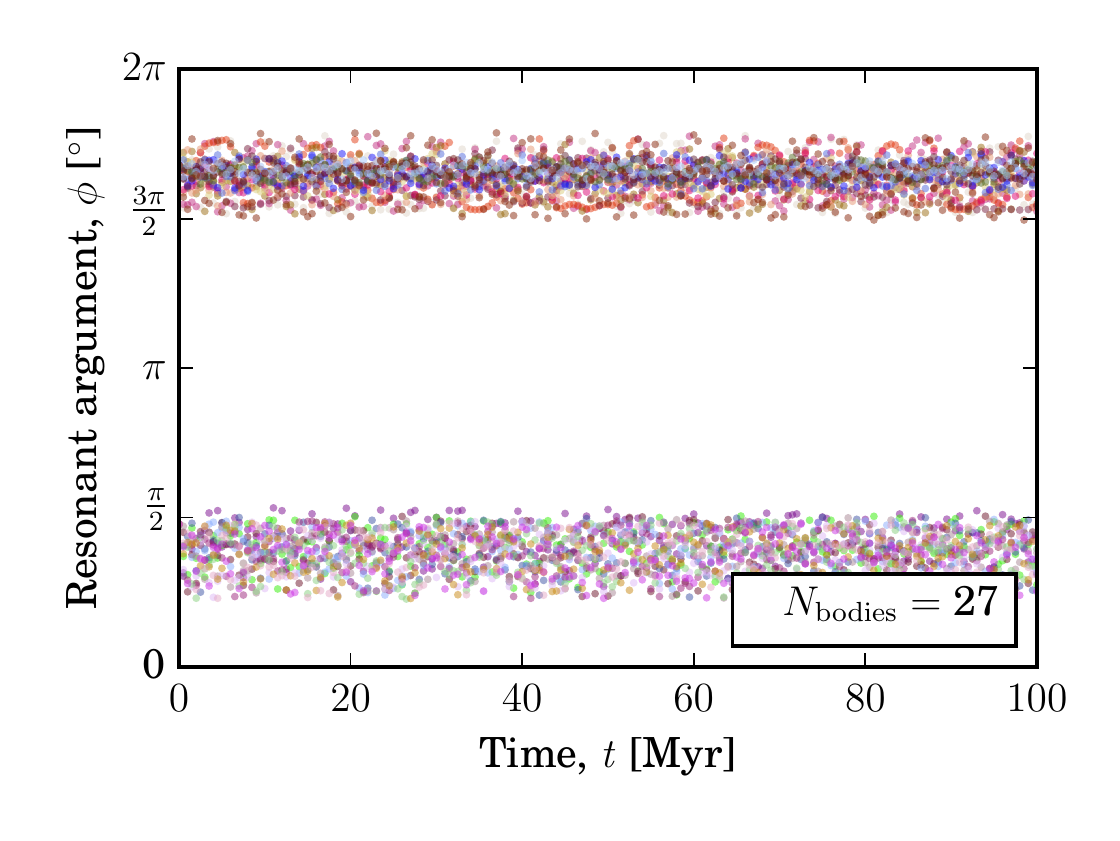}
   		\caption{HD 137388 ($1:1$)}
 	\end{subfigure}
	
 	\begin{subfigure}{0.24\textwidth}
   		\centering
   		\includegraphics[width=\textwidth]{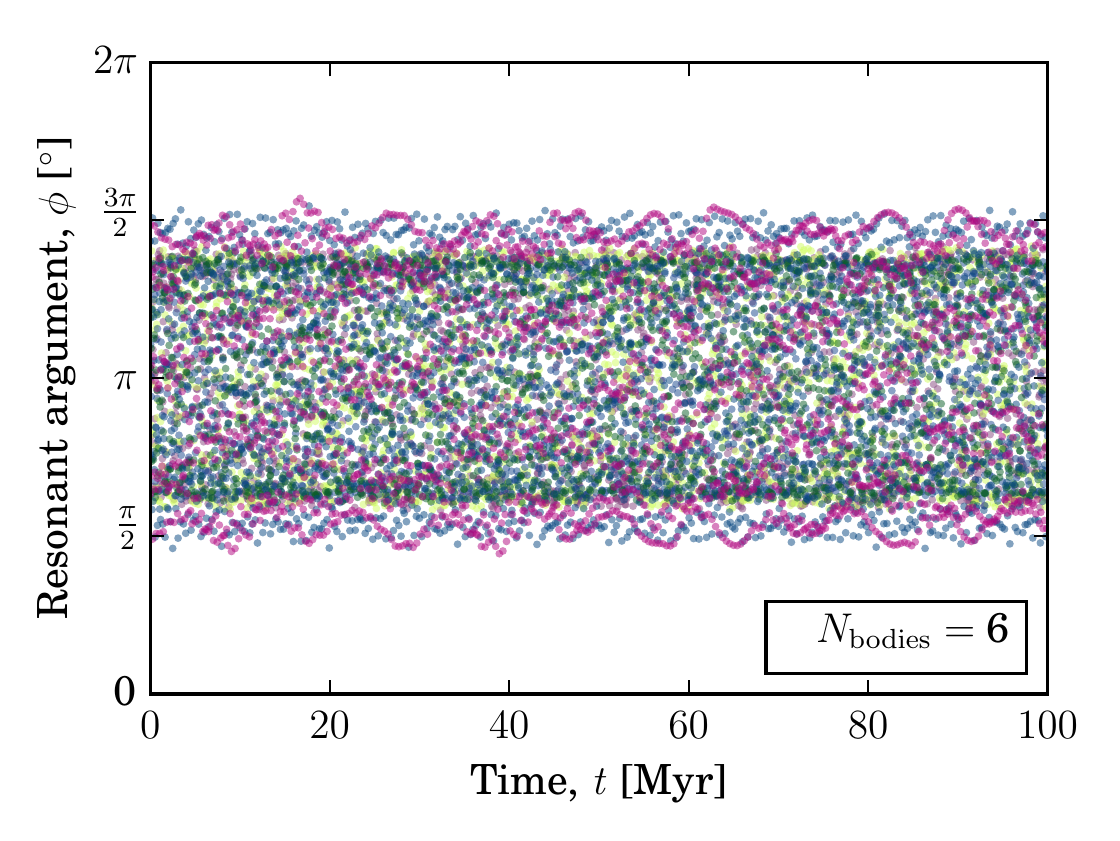}
   		\caption{HD 137388 ($4:3$)}
 	\end{subfigure}
 	\begin{subfigure}{0.24\textwidth}
   		\centering
   		\includegraphics[width=\textwidth]{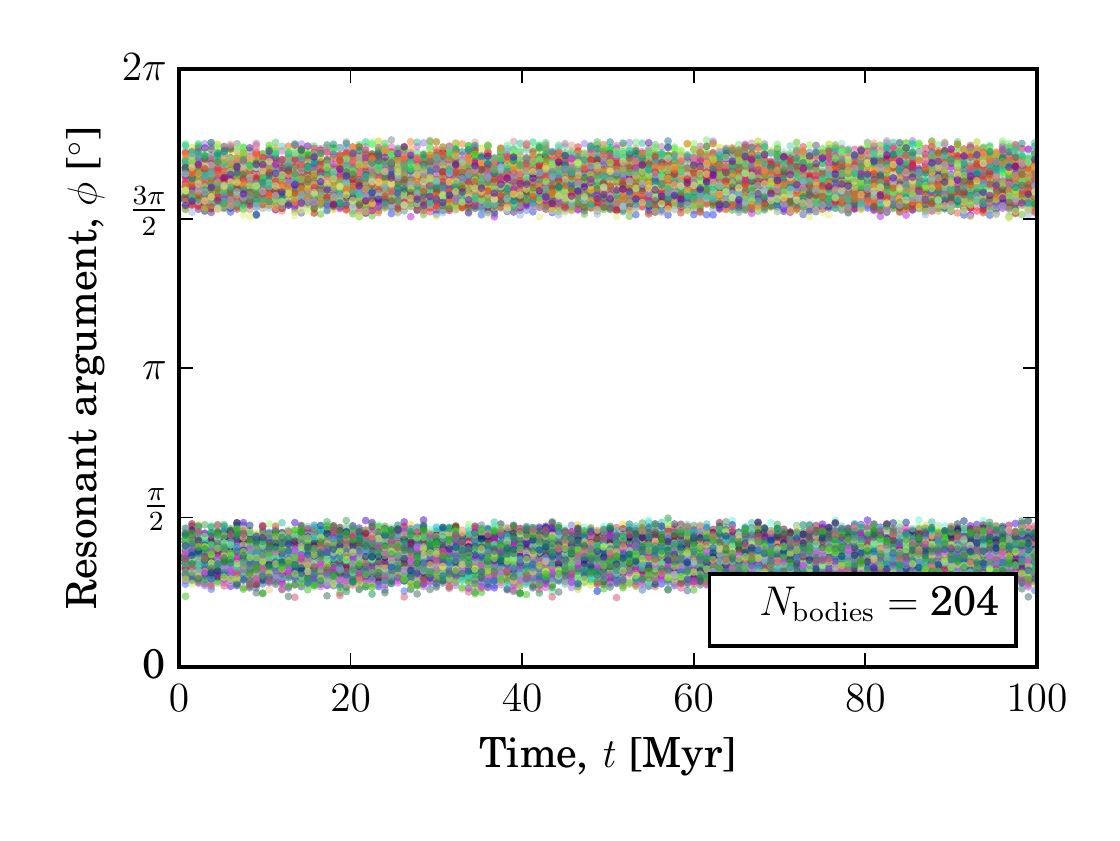}
   		\caption{HD 147513 ($1:1$)}
 	\end{subfigure}
 	\begin{subfigure}{0.24\textwidth}
   		\centering
   		\includegraphics[width=\textwidth]{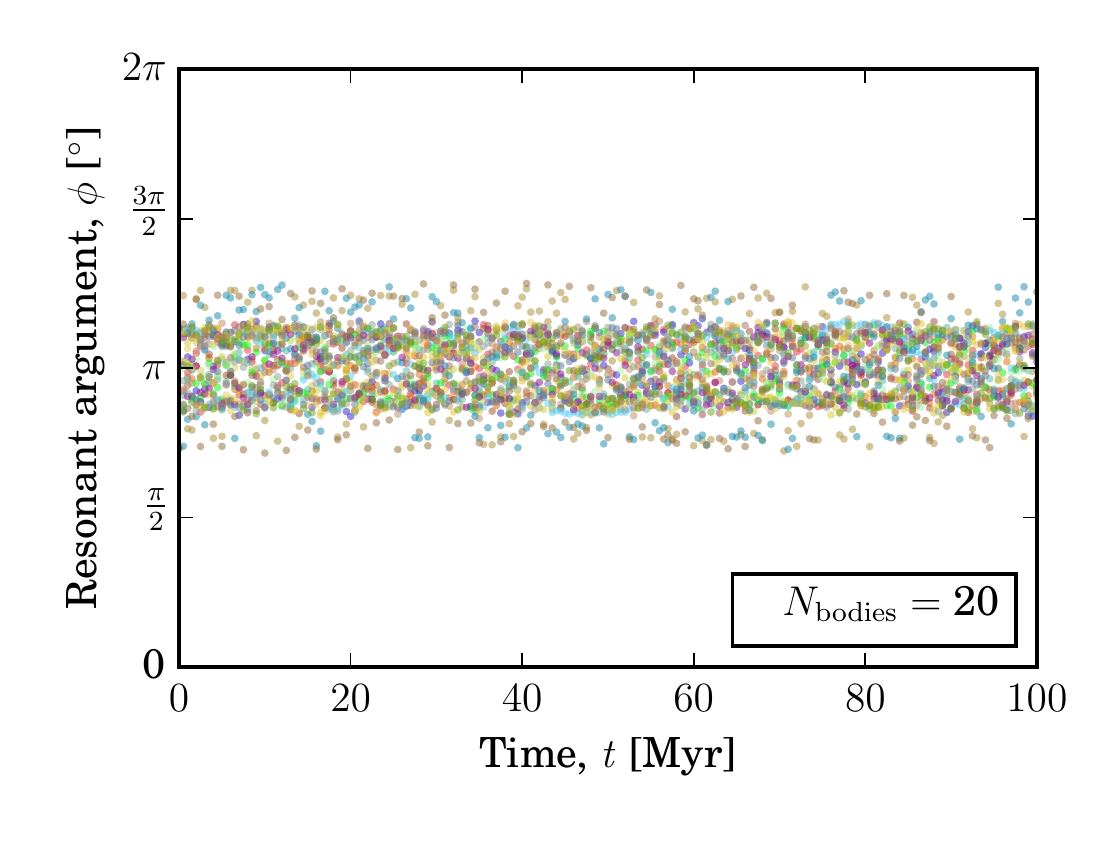}
   		\caption{HD 147513 ($3:2$)}
 	\end{subfigure}
 	\begin{subfigure}{0.24\textwidth}
   		\centering
   		\includegraphics[width=\textwidth]{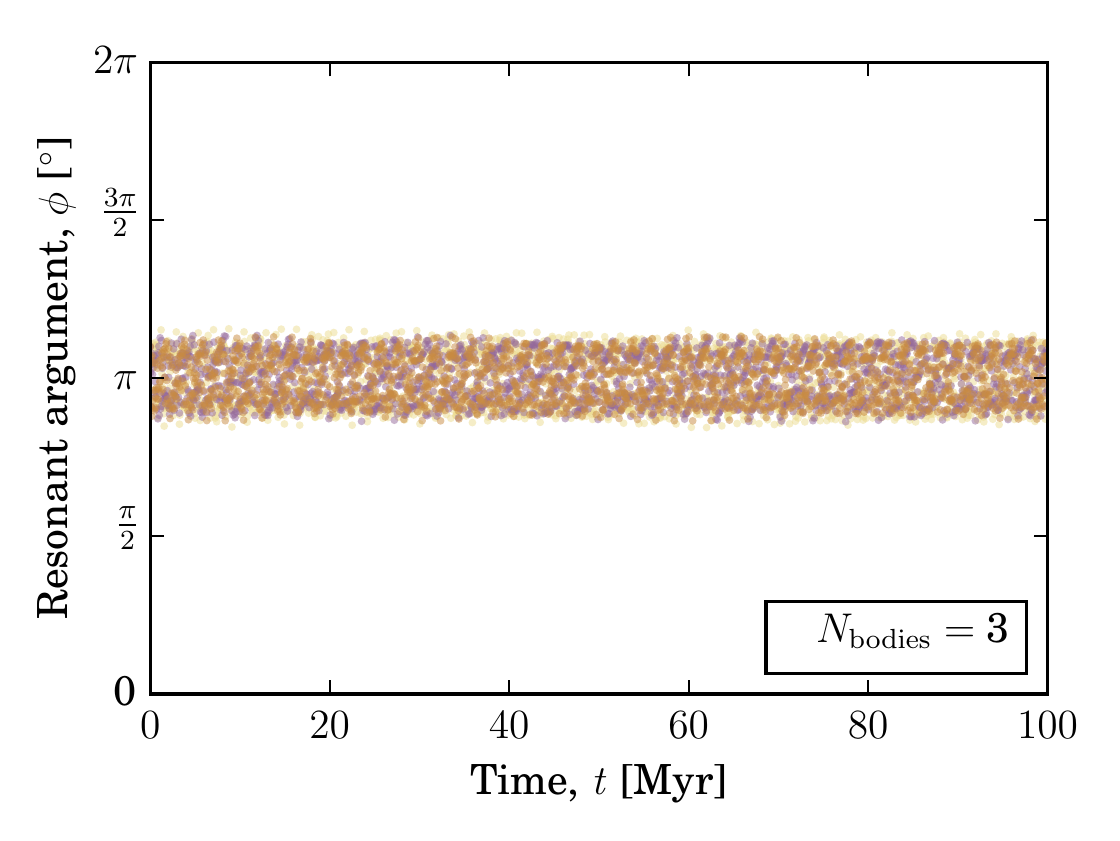}
   		\caption{HD 147513 ($4:3$)}
 	\end{subfigure}
	
	\begin{subfigure}{0.24\textwidth}
   		\centering
   		\includegraphics[width=\textwidth]{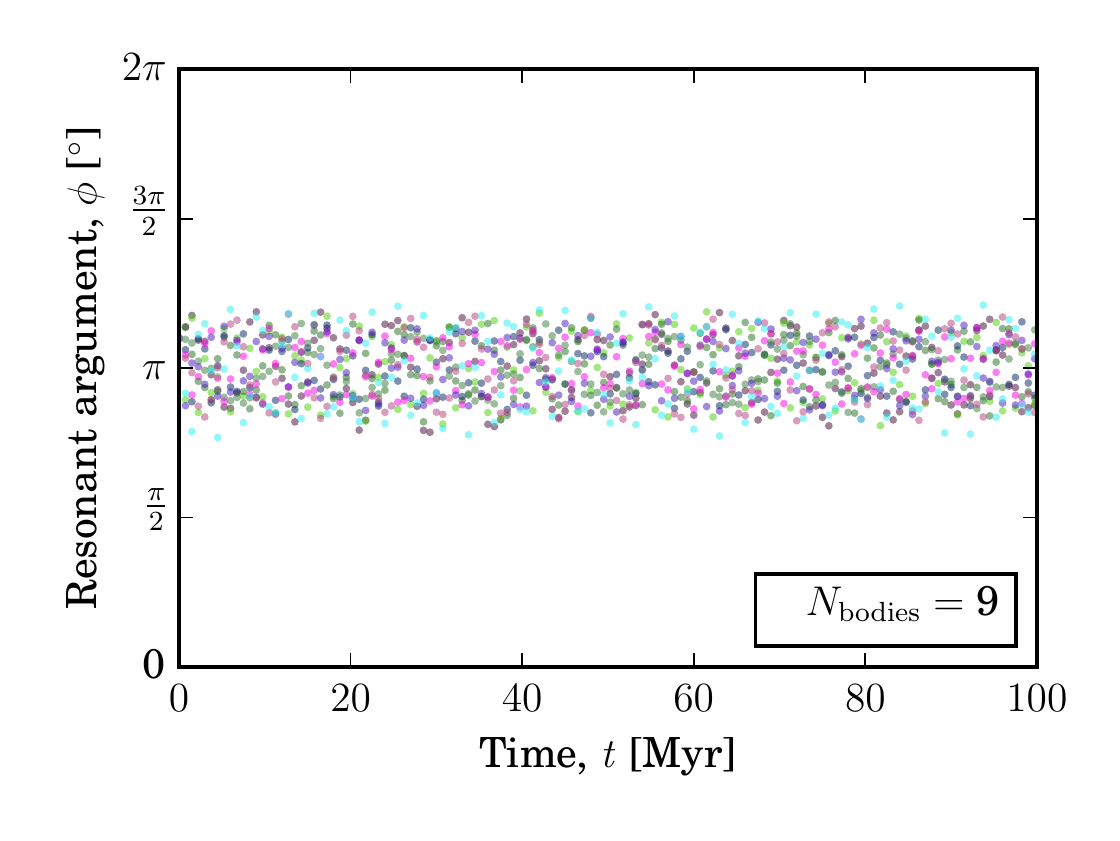}
   		\caption{HD 147513 ($5:3$)}
 	\end{subfigure}
 	\begin{subfigure}{0.24\textwidth}
   		\centering
   		\includegraphics[width=\textwidth]{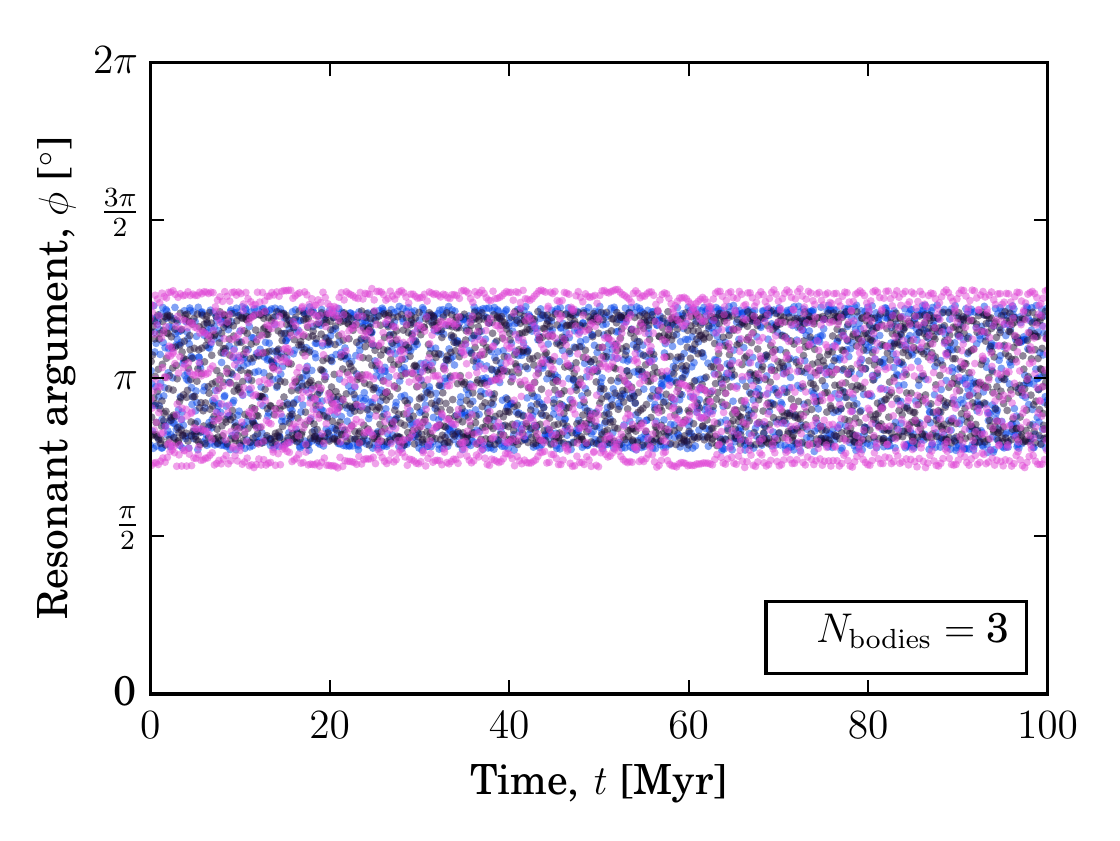}
   		\caption{HD 148156 ($3:2$)}
 	\end{subfigure}
 	\begin{subfigure}{0.24\textwidth}
   		\centering
   		\includegraphics[width=\textwidth]{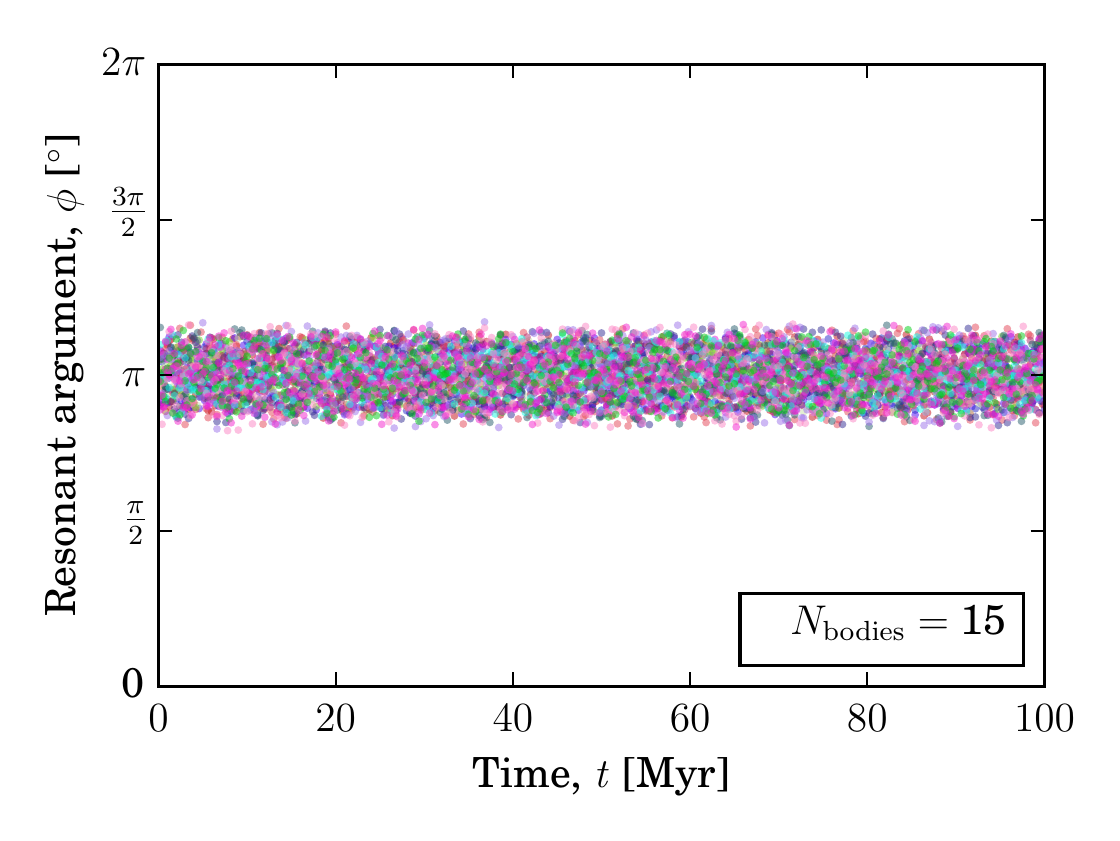}
   		\caption{HD 148156 ($4:3$)}
 	\end{subfigure}
 	\begin{subfigure}{0.24\textwidth}
   		\centering
   		\includegraphics[width=\textwidth]{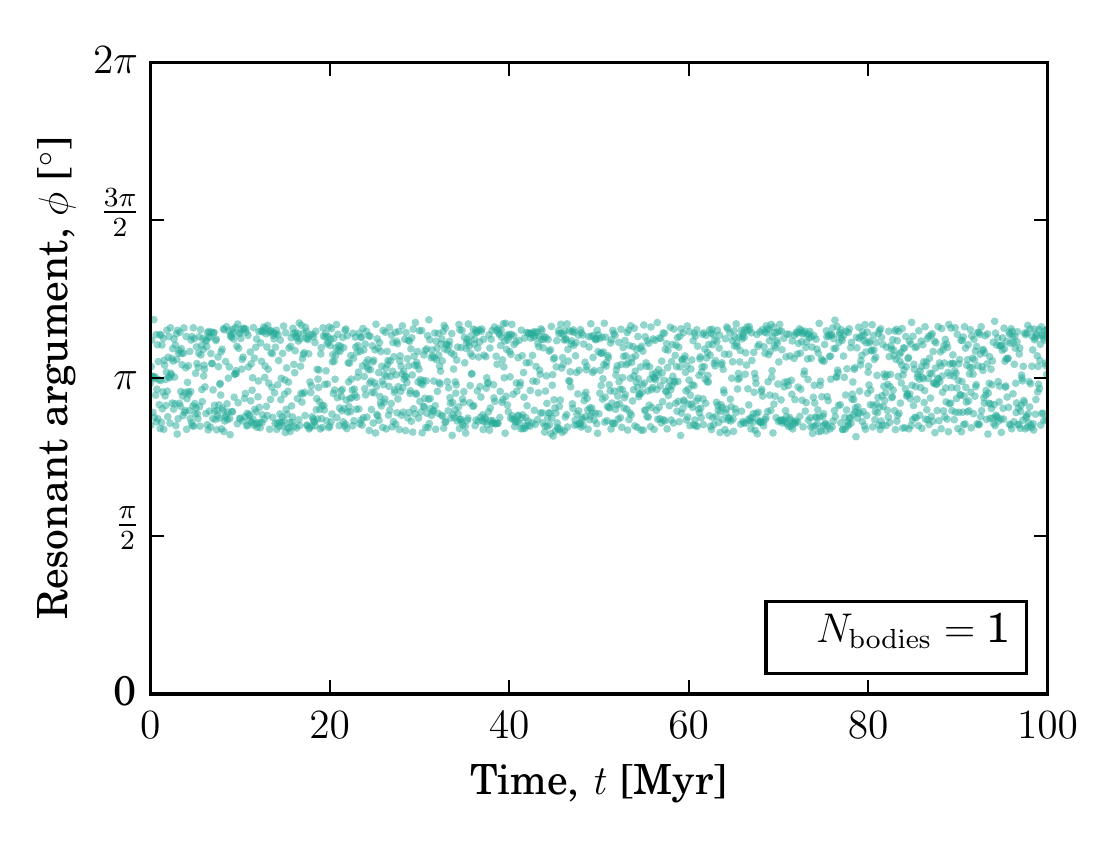}
   		\caption{HD 148156 ($5:3$)}
 	\end{subfigure}
	
 	\begin{subfigure}{0.24\textwidth}
   		\centering
   		\includegraphics[width=\textwidth]{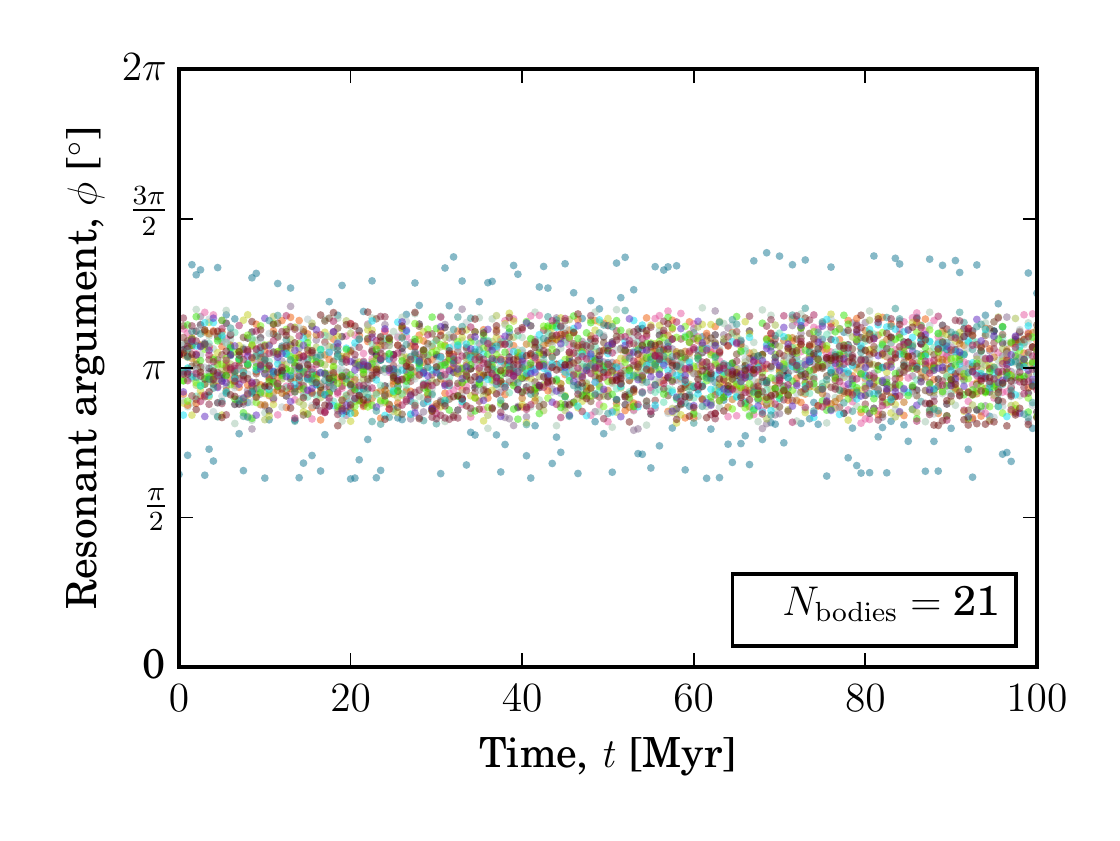}
   		\caption{HD 187085 ($3:2$)}
 	\end{subfigure}
 	\begin{subfigure}{0.24\textwidth}
   		\centering
   		\includegraphics[width=\textwidth]{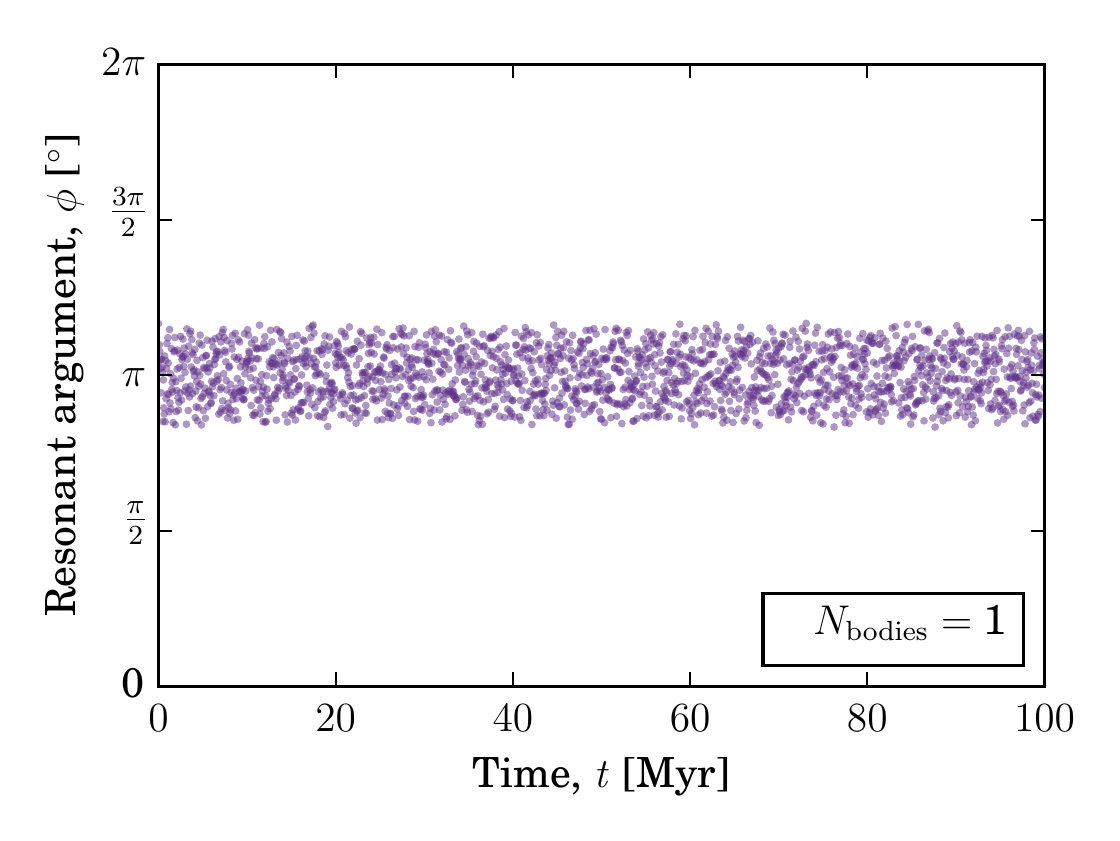}
   		\caption{HD 187085 ($4:3$)}
 	\end{subfigure}
 	\caption{The librating resonant angles $\phi=(p+q)\lambda'-p\lambda-q\omega'$ versus time for all the stable bodies of the $(p+q):p$ MMR for the Jovians in the red completely overlapping systems with stable regions. Note that each body is run in its own simulation, just the resonant angle plots are stacked.}
 	\label{fig:res_plots}
 \end{figure*}

%%%%%%%%%%%%%%%%%%%%%%%%%%%%%%%%%%%%%%%%%%%%%%%%%%

% Don't change these lines
\bsp	% typesetting comment
\label{lastpage}
\end{document}